\documentclass[english,floatfix,onecolumn,nofootinbib,superscriptaddress,aps,prd,12pt]{revtex4-1}
\usepackage[utf8]{inputenc}
\usepackage{float}
\usepackage{array}
\usepackage{lipsum}
\usepackage{graphicx}
\usepackage{rotating}
\usepackage{amsmath,amsthm,amsfonts,amssymb}
\usepackage{tikz}
\usepackage{multirow}
\newcommand{\be}{\begin{equation}}
\newcommand{\ee}{\end{equation}}
\newcommand{\Be}{\begin{eqnarray}}
\newcommand{\Ee}{\end{eqnarray}}

\newcommand{\mincir}{\raise
-3.truept\hbox{\rlap{\hbox{$\sim$}}\raise4.truept\hbox{$<$}\ }}
\newcommand{\magcir}{\raise
-3.truept\hbox{\rlap{\hbox{$\sim$}}\raise4.truept\hbox{$>$}\ }}

\newcolumntype{Y}{>{\centering\arraybackslash}X}
\providecommand{\U}[1]

 \usetikzlibrary{quotes,angles}
 \usetikzlibrary{arrows} 
\usetikzlibrary{decorations.markings}
\usepackage{graphicx}
\usepackage[english]{babel}
\usepackage{color}
\usepackage{subfig}
\usepackage{caption}
\usepackage{tensor}
\usepackage{esint}
\usepackage[dvips]{epsfig}
\usepackage[dvips]{graphicx}
\usepackage{float}
\usepackage{units}
\usepackage{textcomp}
\usepackage{mathrsfs}
\usepackage{amsmath}
\usepackage[makeroom]{cancel}
\usepackage{amssymb}
\usepackage{amsbsy}
\usepackage{amsfonts}
\usepackage{amssymb,mathrsfs,xcolor}
\usepackage{esint}
\usepackage{array}
\usepackage{graphicx}

\usepackage{hyperref}
\hypersetup{
    colorlinks,
    citecolor=blue,
    filecolor=green,
    linkcolor=purple,
    urlcolor=red,
}

\usepackage{slashed}

\newcommand{\ie}{\begin{equation}}
\newcommand{\fe}{\end{equation}}
\newcommand{\se}{\begin{eqnarray}}
\newcommand{\ff}{\end{eqnarray}}

\usepackage{hyperref}
\hypersetup{colorlinks,breaklinks,
			citecolor=[rgb]{0,0.0,1.0},
            urlcolor=[rgb]{0.0,0.0,1.0},
            linkcolor=[rgb]{0,0.5,0.9}}

\begin{document}

    \title{Axisymmetric black hole in a non--commutative gauge theory: classical and quantum gravity effects}

\author{A. A. Ara\'{u}jo Filho}
\email{dilto@fisica.ufc.br}

\affiliation{Departamento de Física, Universidade Federal da Paraíba, Caixa Postal 5008, 58051--970, João Pessoa, Paraíba,  Brazil.}
\affiliation{Departamento de Física, Universidade Federal de Campina Grande Caixa Postal 10071, 58429-900 Campina Grande, Paraíba, Brazil.}
\affiliation{Center for Theoretical Physics, Khazar University, 41 Mehseti Street, Baku, AZ-1096, Azerbaijan.}


\author{N. Heidari}
\email{heidari.n@gmail.com}

\affiliation{Center for Theoretical Physics, Khazar University, 41 Mehseti Street, Baku, AZ-1096, Azerbaijan.}
\affiliation{School of Physics, Damghan University, Damghan, 3671641167, Iran.}

\author{Ali \"Ovg\"un}
\email{ali.ovgun@emu.edu.tr}
\affiliation{Physics Department, Eastern Mediterranean University, Famagusta, 99628 North
Cyprus via Mersin 10, Turkiye.}


\date{\today}

\begin{abstract}

This work explores both classical and quantum aspects of an axisymmetric black hole within a non--commutative gauge theory. The rotating solution is derived using a modified Newman--Janis procedure. The analysis begins with the horizon structure, ergospheres, and angular velocity. The thermodynamic properties are examined through surface gravity, focusing on the Hawking temperature, entropy, and heat capacity. In addition, the remnant mass is calculated. The Hawking radiation is treated as a tunneling process for bosonic and fermionic particles, along with the corresponding particle creation density. Geodesic motion is explored, emphasizing null geodesics, radial accelerations, the photon sphere, and black hole shadows. Finally, the gravitational lensing in the strong deflection limit is investigated.

\end{abstract}


\maketitle

\pagebreak

\tableofcontents

\section{Introduction}
\label{introduction}

Within general relativity, the geometry of spacetime does not impose a definitive lower limit on measurable distances. The Planck length, however, is commonly regarded as a fundamental constraint. To explore the implications of this limitation, non-commutative spacetimes have been proposed as a theoretical framework. This concept, which is connected to string theory and other quantum gravity models, has become increasingly relevant in the study of supersymmetric field theories, especially through the lens of superfield formalism \cite{ferrari2003finiteness,szabo2003quantum,ferrari2004superfield,3,szabo2006symmetry,ferrari2004towards}. In addition, a prominent technique for incorporating non--commutative structures into gravitational models is the Seiberg--Witten map, which facilitates the gauging of specific symmetry groups \cite{chamseddine2001deforming}. Such a formalism has been used to study black hole thermodynamics, evaporation, and thermal behavior, including emission spectra and thermal state quantites \cite{sharif2011thermodynamics,banerjee2008noncommutative,AraujoFilho:2024rss,nozari2007thermodynamics,t29,nozari2006reissner,myung2007thermodynamics,lopez2006towards}.

The exploration of non--commutative effects in gravitational systems has seen significant progress with the alteration of matter sources in Einstein's field equations, maintaining the form of the Einstein tensor unchanged \cite{nicolini2006noncommutative}. This approach substitutes the conventional point--mass density with smoothly distributed profiles, such as Gaussian \cite{ghosh2018noncommutative} and Lorentzian functions \cite{nicolini2009noncommutative}. Such modifications have fueled extensive research into black hole thermodynamics, including analyses of thermal radiation and quantum tunneling processes \cite{nozari2008hawking,banerjee2008noncommutative,sharif2011thermodynamics}. They have also been applied to topological studies within Gauss--Bonnet gravity \cite{lekbich2024optical}, black hole shadow formation \cite{wei2015shadow,sharif2016shadow,ovgun2020shadow}, and phenomena like matter accretion and gravitational lensing \cite{ding2011strong,saleem2023observable,ding2011probing}. Recent efforts have explored non--commutativity as a perturbative correction in gravitational frameworks, expanding its theoretical implications \cite{newcommutativity}.

Non–commutativity can emerge through different formulations, each introducing specific geometric deformations. These include: (i) approaches based on smeared matter sources, where point–like distributions are replaced by Gaussian or Lorentzian profiles that regularize curvature singularities \cite{araujo2025particle,campos2022quasinormal,Anacleto:2022shk,Anacleto:2019tdj,AraujoFilho:2025rzh}; (ii) formulations where the non–commutative structure is implemented directly in the action through Moyal–type $\star$–products, modifying the coordinate algebra and leading to non–local and higher–derivative corrections in the field equations \cite{Pozar:2025yoj}; and (iii) gauge–theoretic realizations grounded on the Seiberg–Witten map \cite{araujo2025non,araujo2025gedddodesics,heidari2025non,Juric:2025kjl}, such as the one employed in this work, in which the deformation is introduced perturbatively in the tetrads and spin connections while maintaining gauge covariance. As we shall see in the forthcoming sections, our choice of framework is therefore motivated by its consistency with gauge principles, its analytical tractability, and its suitability for investigating how non–commutative effects influence measurable gravitational properties such as horizons, thermodynamics, shadows, and lensing phenomena.

Gravitational waves are essential for understanding a range of physical phenomena, from early universe dynamics to stellar processes such as oscillations and binary interactions \cite{yakut2005evolution,pretorius2005evolution,kjeldsen1994amplitudes,dziembowski1992effects,unno1979nonradial,hurley2002evolution,heuvel2011compact}. The intensity and mode of these waves fluctuate, while their spectral features convey details about their sources \cite{riles2017recent}. Black holes stand out as prominent emitters, generating radiation during their formation, with distinct frequencies known as \textit{quasinormal} modes \cite{kokkotas1999quasi,roy2020revisiting,oliveira2019quasinormal, berti2009quasinormal, horowitz2000quasinormal, heidari2024impact,Hamil:2024ppj,nollert1999quasinormal,ferrari1984new,santos2016quasinormal,ovgun2018quasinormal,jusufi2024charged, rincon2020greybody,araujo2024dark,london2014modeling,maggiore2008physical, flachi2013quasinormal,blazquez2018scalar,konoplya2011quasinormal}. The analysis of these modes in black hole physics often employs the weak field approximation, a method utilized not only within the scope of general relativity but also in various alternative gravitational frameworks, including those exploring Lorentz symmetry violations and extended theoretical models  \cite{araujo2024exploring,lee2020quasi, jawad2020quasinormal,Fernando:2012yw,jcap5,maluf2014einstein,jcap4,JCAP1,Daghigh:2008jz,JCAP2,Gogoi:2023kjt,JCAP3,Daghigh:2020fmw,mmm2,Daghigh:2022uws,Daghigh:2020jyk,mmm1,Fernando:2016ftj,Heidari:2023bww,Daghigh:2020mog,liu2022quasinormal,yang2023probing,Gogoi:2023fow,maluf2013matter,kim2018quasi,Daghigh:2005ph,lambiase2023investigating,hassanabadi2023gravitational,Yang:2022ifo,gogoi2023quasinormal}

The imaging of the supermassive black hole at the core of the M87 galaxy by the Event Horizon Telescope has drawn considerable scientific focus \cite{023,024,025,026,027,028,029,pantig2023testing,ccimdiker2021black,lambiase2023investigating,pantig2022shadow}. Virbhadra and Ellis were among the first to propose a simplified lensing formula for supermassive black holes in asymptotically flat spacetimes, revealing how strong gravitational fields create multiple images symmetrically distributed along the optical axis \cite{030,031}. Analytical techniques for examining strong gravitational lensing were later expanded by Fritelli et al. \cite{032}, Bozza et al. \cite{033}, and Tsukamoto \cite{035}, leading to more comprehensive studies of light deflection. Investigations have since extended to various frameworks, such as Reissner--Nordström backgrounds \cite{036,aa2024remarks,036.2,aa2024static,036.1}, axisymmetric black holes \cite{37.5,37.2,37.4,37.3,37.6,37.1}, exotic objects like wormholes \cite{38.4,38.5,38.3,38.2,38.1}, and spacetime modifications arising from alternative gravity theories \cite{40,Donmez:2024lfi,Koyuncu:2014nga,Donmez:2023egk}.

Hawking introduced a fundamental concept linking quantum mechanics with gravity, contributing to the foundation of quantum gravity research \cite{o11,o1,o111}. He demonstrated that black holes can emit thermal radiation, causing a gradual loss of mass, a phenomenon now known as Hawking radiation \cite{eeeOvgun:2019ygw,gibbons1977cosmological,eeeKuang:2018goo,eeeKuang:2017sqa,eeeOvgun:2015box,eeeOvgun:2019jdo}. This result, derived from quantum field theory in curved spacetime near the event horizon, has significantly influenced the study of black hole thermodynamics and quantum effects \cite{sedaghatnia2023thermodynamical,araujo2024dark,o3,o9,o6,araujo2023analysis,o8,aa2024implications,o4,o7}. Kraus and Wilczek \cite{o10}, later expanded by Parikh and Wilczek \cite{013,o11,o12}, reinterpreted \textit{Hawking} radiation as a quantum tunneling effect within a semi--classical framework. This perspective has since been extensively explored in the context of various black hole configurations \cite{anacleto2015quantum,medved2002radiation,mirekhtiary2024tunneling,silva2013quantum,del2024tunneling,calmet2023quantum,johnson2020hawking,vanzo2011tunnelling,mitra2007hawking,zhang2005new,touati2024quantum,senjaya2024bocharova}.

The Newman--Janis method, designed to incorporate angular momentum into static spacetimes, offers a widely applied approach for generating rotating black hole solutions from spherically symmetric metrics \cite{n57i,n57}. This technique relies on a complex coordinate transformation and has been instrumental in deriving the Kerr--like solutions from simpler static models, such as the Schwarzschild--like spacetimes. Recognizing the limitations of the original formulation, recent works have been used modified versions to extend its applicability to a broader spectrum of gravitational scenarios \cite{afrim}.

An alternative to the standard Newman--Janis procedure, known as the modified Newman--Janis algorithm \cite{n59,n58}, avoids the conventional complexification process, which has long been a topic of debate. This revised method instead employs a more geometrically motivated transformation, providing a clearer interpretation of the process. It has been particularly effective in deriving rotating metrics for spacetimes with imperfect fluids, broadening the scope of the original approach, which primarily addressed perfect fluid models. By applying this modified technique, rotating solutions can be obtained from static, spherically symmetric metrics in a wider range of physical contexts, enhancing its versatility across gravitational models \cite{n63,afrim,n62,n60,n64,n65,n61}.

This work examines an axisymmetric black hole within a non--commutative gauge theory, emphasizing both its classical structure and quantum properties. The rotating solution is obtained through a modified Newman--Janis procedure, with the horizon structure, ergospheres, and angular velocity studied as fundamental aspects of the spacetime geometry. The thermodynamic behavior is characterized by surface gravity, from which the Hawking temperature, entropy, and heat capacity are determined. Furthermore, the remnant mass is also addressed. Quantum effects are addressed by modeling Hawking radiation as a tunneling process for bosonic and fermionic particles. The motion of test particles is explored through a detailed geodesic analysis, including null trajectories, radial accelerations, the photon sphere, and the formation of black hole shadows. Finally, the gravitational lensing is examined in the strong deflection regime.


\section{Black hole in a non--comutative gauge theory}

This section outlines the essential framework for examining the non--commutative gauge theory of gravity. As previously noted, the gauge group involved is the de Sitter group, $\mathrm{SO}(4,1)$. To set the stage, we will first describe the formulation of the $\mathrm{SO}(4,1)$ gauge theory in a commutative $(3+1)$--dimensional Minkowski spacetime, where the metric in spherical coordinates takes the form
\ie
\mathrm{d}s^{2}=\mathrm{d}r^{2}+r^{2}\mathrm{d}\Omega^{2}_{2}-c^{2}\mathrm{d}t^{2},
\fe
where $\mathrm{d}\Omega^{2}_{2} = \mathrm{d}\theta^{2} + \sin^{2}\theta \,\mathrm{d}\varphi^{2}$. The group $\mathrm{SO}(4,1)$ consists of ten generators, denoted by $\mathcal{M}_{\mathcal{A}\mathcal{B}}$, which satisfy the antisymmetry relation $\mathcal{M}_{\mathcal{A}\mathcal{B}} = -\mathcal{M}_{\mathcal{B}\mathcal{A}}$. The indices involved are $\mathcal{A}, \mathcal{B} = a, 5$ with $a, b = 0, 1, 2, 3$. These generators decompose into two distinct sets: the components $\mathcal{M}_{ab} = -\mathcal{M}_{ba}$, responsible for rotations, and the components $\mathcal{P}_{a} = \mathcal{M}_{a5}$, which describe translations.

The undeformed gauge potentials are denoted by $\omega^{\mathcal{A}\mathcal{B}}_{\mu}(x)$, which satisfy the antisymmetric condition $\omega^{\mathcal{A}\mathcal{B}}_{\mu}(x) = -\omega^{\mathcal{B}\mathcal{A}}_{\mu}(x)$. These potentials are distinct from the spin connection, expressed as $\omega^{ab}_{\mu}(x) = -\omega^{ba}_{\mu}(x)$, and the tetrad fields, denoted by $e^{a}_{\mu}(x)$. The components $\hat{\omega}^{a5}_{\mu}(x)$ are related to the tetrad fields through the expression $\hat{\omega}^{a5}_{\mu}(x) = \mathcal{K} \hat{e}^{a}_{\mu}(x)$, where $\mathcal{K}$ represents a contraction parameter. Additionally, another gauge field can be defined as $\hat{\omega}^{55}_{\mu}(x) = \mathcal{K} \hat{\phi}_{\mu}(x,\Theta)$, where the field $\hat{\phi}_{\mu}(x,\Theta)$ vanishes when the limit $\mathcal{K} \to 0$ is taken. This limiting case effectively reduces the gauge group to the Poincaré group $\mathrm{ISO}(3,1)$ \cite{1,2}. The corresponding field strength associated with the gauge potential $\omega^{\mathcal{A}\mathcal{B}}_{\mu}(x)$ is given by
\ie
	F^{\mathcal{A}\mathcal{B}}_{\mu} = \partial_{\mu}\omega^{\mathcal{A}\mathcal{B}}_{\nu} - \partial_{\nu}\omega^{\mathcal{A}\mathcal{B}}_{\mu} + \left(\omega^{\mathcal{A}\mathcal{C}}_{\mu}\omega^{\mathcal{D}\mathcal{B}}_{\nu}-\omega^{\mathcal{A}\mathcal{C}}_{\nu}\omega^{\mathcal{D}\mathcal{B}}_{\mu}\right)\eta_{\mathcal{C}\mathcal{D}}
\fe
where $\mu, \nu = 0, 1, 2, 3$ and the metric tensor is defined as $\eta_{\mathcal{A}\mathcal{B}} = \mathrm{diag}(+,+,+,-,+)$. Furthermore, the expression can be rewritten as
\begin{subequations}
	\begin{align}
&F^{a5}_{\mu\nu}=\mathcal{K}\left[\partial_{\mu}e^{a}_{\nu}-\partial_{\nu}e^{a}_{\mu}+\left(\omega^{ab}_{\mu}e^{a}_{\nu}-\omega^{ab}_{\nu}e^{c}_{\mu}\right)\eta_{bc}\right]=\mathcal{K}T^{a}_{\mu\nu},\label{torsion}\\
		&F^{ab}_{\mu\nu} = \partial_{\mu} \omega^{ab}_{\nu}-\partial_{\nu}\omega^{ab}_{\mu}+\left(\omega^{ac}_{\mu}\omega^{db}_{\nu}-\omega^{ac}_{\nu}\omega^{db}_{\mu}\right)\eta_{cd}+\mathcal{K}\left(e^{a}_{\mu}e^{b}_{\nu}-e^{a}_{\nu}e^{b}_{\mu}\right)=R^{ab}_{\mu\nu},
	\end{align}
\end{subequations}
where $\eta_{ab} = \mathrm{diag}(+,+,+,-)$. It is important to highlight that the Poincaré gauge group being considered here is linked to the geometric structure of Riemann--Cartan spacetime, which incorporates both curvature and torsion fields \cite{1,6}. The torsion tensor, defined as $T^{a}_{\mu\nu} \equiv F^{a5}_{\mu\nu}/\mathcal{K}$, and the curvature tensor, expressed as $R^{ab}_{\mu\nu} \equiv F^{ab}_{\mu\nu}$, are described in terms of the tetrad fields $e^{a}_{\mu}(x)$ and the spin connection $\omega^{ab}_{\mu}(x)$. When the torsion field vanishes, as indicated by Eq. \eqref{torsion}, the spin connection can be determined solely from the tetrad fields.

Next, we consider a possible configuration for spherically symmetric gauge fields associated with the $\mathrm{SO}(4,1)$ group \cite{1,6}:
\ie
\label{tetrads1}
	e^{1}_{\mu} = \left(\frac{1}{\mathcal{A}}, 0,0,0\right), \quad e^{2}_{\mu} = \left(0, r,0,0\right), \quad e^{3}_{\mu} = \left(0,0,r\, \mathrm{sin}\theta,0\right), \quad e^{0}_{\mu} = \left(,0,0,0, \mathcal{A}\right),
\fe
and 
\ie
\label{omega}
	\begin{split}
		& \omega^{12}_{\mu} = \left(0, \mathcal{W},0,0\right), \quad 
		\omega^{13}_{\mu} = \left(0,0, \mathcal{Z}\, \mathrm{sin}\theta,0\right),  \quad \omega^{10}_{\mu} = \left(0,0,0,\mathcal{U}\right),\\
		& \omega ^{23}_{\mu} = \left(0,0,-\mathrm{cos}\theta, \mathcal{V}\right), \quad \omega^{20}_{\mu} = \omega^{30}_{\mu} = \left(0,0,0,0\right),
	\end{split}
\fe
with $\mathcal{A}$, $\mathcal{U}$, $\mathcal{V}$, $\mathcal{W}$, and $\mathcal{Z}$ are functions dependent solely on the three--dimensional radial coordinate. Moreover, the non--vanishing components of the torsion tensor can be expressed as \cite{2}
\ie
\label{torsion2}
	\begin{split}
		&T^{0}_{01} = -\frac{\mathcal{A}\mathcal{A}'+U}{\mathcal{A}}, \qquad 
		T^{2}_{03} = r\, \mathcal{V} \mathrm{sin}\theta \,T^{2}_{12} = \frac{\mathcal{A}+\mathcal{W}}{\mathcal{A}},\\
		& T^{3}_{02} = -r\, \mathcal{V}, \,\,\,\qquad\qquad T^{3}_{13} = \frac{\left(\mathcal{A}+\mathcal{Z}\right)\mathrm{sin}\theta}{\mathcal{A}},
	\end{split}
\fe
and, therefore, the curvature tensor reads \cite{2}
\ie
\label{curvature}
	\begin{split}
		&R^{01}_{01} = \mathcal{U}', \quad R^{23}_{01} = -\mathcal{V}', \quad R^{13}_{23} = \left(\mathcal{Z}-\mathcal{W}\right) \mathrm{cos} \theta,\quad R^{01}_{01} = -\mathcal{U}\mathcal{W}, \quad R^{13}_{01} = - \mathcal{V}\mathcal{W},\\
		& R^{03}_{03} = -\mathcal{U} \mathcal{Z} \mathrm{sin} \theta, \quad R^{12}_{03} = \mathcal{V} \mathcal{Z} \mathrm{sin}\theta \, R^{12}_{12} = \mathrm{W}',\quad R^{23}_{23} = \left(1-\mathcal{Z}\mathcal{W}\right)\mathrm{sin}\theta, \quad R^{13}_{13} = \mathcal{Z}' \mathrm{sin}\theta.
	\end{split}
\fe
The symbols $\mathcal{A}'$, $\mathcal{U}'$, $\mathcal{V}'$, $\mathcal{W}'$, and $\mathcal{Z}'$ indicate derivatives taken with respect to the radial coordinate. To ensure the torsion field is absent, as outlined in Eq. \eqref{torsion2}, the following conditions are proposed: 
\ie
\mathcal{V} = 0, \qquad \mathcal{U} = - \mathcal{A}\mathcal{A}', \qquad \mathcal{W} =  -\mathcal{A} = \mathcal{Z}.
\fe
Taking into account the field equation 
\ie
\label{fieldequation}
	R^{a}_{\mu} - \frac{1}{2} R\, e^{a}_{\mu} = 0,
\fe
expressed through the tetrad fields $e^{a}_{\mu}(x)$, with the definitions $R^{a}_{\mu} = R^{ab}_{\mu\nu} e^{\nu}_{b}$ and $R = R^{ab}_{\mu\nu} e^{\mu}_{a} e^{\nu}_{b}$, the resulting solution can be written as
\ie
\mathcal{A}(r) = \sqrt{1-\frac{\alpha}{r}},
\fe
where the constant $\alpha$ is defined as $\alpha = 2GM/c^{2}$, with $G$ denoting the gravitational constant, $M$ the black hole mass, and $c$ the speed of light. To obtain the modified metric $\mathrm{d}s^{2} = \hat{g}_{\mu\nu}(x,\Theta)\mathrm{d}x^{\mu}\mathrm{d}x^{\nu}$, expressed in spherical coordinates $x^{\mu} = (r, \theta, \varphi, c t)$ for a $(3+1)$--dimensional non--commutative Schwarzschild spacetime, it is necessary to determine the deformed tetrad fields $\hat{e}^{a}_{\mu}(x,\Theta)$. These tetrads emerge from a contraction between the non--commutative gauge group $\mathrm{SO}(4,1)$ and the Poincaré group $\mathrm{ISO}(3,1)$, employing the Seiberg--Witten map formalism \cite{3,4,5}. The structure of the non--commutative spacetime can then be defined under the following conditions:
\ie
\label{NonCommSTcond1}
\left[x^{\mu},x^{\nu}\right]=i\Theta^{\mu\nu}.
\fe
The constants $\Theta^{\mu\nu}$ are assumed to be real and satisfy the antisymmetric relation $\Theta^{\mu\nu} = -\Theta^{\nu\mu}$. As a result, the gravitational fields, specifically the modified tetrad fields $\hat{e}^{a}_{\mu}(x,\Theta)$ and the gauge connection $\hat{\omega}^{\mathcal{A}\mathcal{B}}_{\mu}(x,\Theta)$, within a non-commutative spacetime, can be expanded as a power series in the parameter $\Theta$ \cite{1,2,3,4}.
\ie
\begin{split}
		&\hat{e}^{a}_{\mu}(x,\Theta) = e^{a}_{\mu}(x)-i \Theta^{\nu\rho}e^{a}_{\mu\nu\rho}(x)+\Theta^{\nu\rho}\Theta^{\lambda\tau}e^{a}_{\mu\nu\rho\lambda\tau}(x)\dots,\\	
		&\hat{\omega}^{\mathcal{A}\mathcal{B}}_{\mu}(x,\Theta) = \omega^{\mathcal{A}\mathcal{B}}_{\mu}(x)-i \Theta^{\nu\rho} \omega^{\mathcal{A}\mathcal{B}}_{\mu\nu\rho}(x)+\Theta^{\nu\rho}\Theta^{\lambda\tau}\omega^{\mathcal{A}\mathcal{B}}_{\mu\nu\rho\lambda\tau}(x)\dots  \,\,. \label{omeganoncom}
	\end{split}
\fe

The tetrad fields $\hat{e}^{a}_{\mu}(x,\Theta)$ result from the expansion of the non--commutative modifications applied to the gauge connection $\hat{\omega}^{\mathcal{A}\mathcal{B}}_{\mu}(x,\Theta)$, presented in Eq. \eqref{omeganoncom}, truncated to second--order terms in the parameter $\Theta$  
\begin{subequations}
\label{noncommcorr}
	\begin{align}
		\omega^{\mathcal{A}\mathcal{B}}_{\mu\nu\rho} (x) &= \frac{1}{4} \left\{\omega_{\nu},\partial_{\rho}\omega_{\mu}+R_{\rho\mu}\right\}^{\mathcal{A}\mathcal{B}},\label{noncommcorr-tetrad}\\
		\nonumber\omega^{\mathcal{A}\mathcal{B}}_{\mu\nu\rho\lambda\tau} (x) &= \frac{1}{32}\left(-\left\{\omega_{\lambda},\partial_{\tau}\left\{\omega_{\nu},\partial_{\rho}\omega_{\mu}+R_{\rho\mu}\right\}\right\}+2\left\{\omega_{\lambda},\left\{R_{\tau\nu},R_{\mu\rho}\right\}\right\}\right.\\
		\nonumber&\left.-\left\{\omega_{\lambda},\left\{\omega_{\nu},D_{\rho}R_{\tau\mu}+\partial_{\rho}R_{\tau\mu}\right\}\right\}-\left\{\left\{\omega_{\nu},\partial_{\rho}\omega_{\lambda}+R_{\rho\lambda}\right\},\left(\partial_{\tau}\omega_{\mu}+R_{\tau\mu}\right)\right\}\right.\\
		&\left.+2\left[\partial_{\nu}\omega_{\lambda},\partial_{\rho}\left(\partial_{\tau}\omega_{\mu}+R_{\tau\mu}\right)\right]\right)^{\mathcal{A}\mathcal{B}}.\label{noncommcorr-omega}
	\end{align}
\end{subequations}
Obtained through the application of the Seiberg--Witten map, Eqs. \eqref{noncommcorr-tetrad} and \eqref{noncommcorr-omega} obey to the following conditions:
\ie
\left[\alpha,\beta\right]^{\mathcal{A}\mathcal{B}} =  \alpha^{\mathcal{A}\mathcal{C}}\beta^{\mathcal{B}}_{\mathcal{C}}-\beta^{\mathcal{A}\mathcal{C}}\alpha^{\mathcal{B}}_{\mathcal{C}},\qquad
	\left\{\alpha,\beta\right\}^{\mathcal{A}\mathcal{B}} = \alpha^{\mathcal{A}\mathcal{C}}\beta^{\mathcal{B}}_{\mathcal{C}}+\beta^{\mathcal{A}\mathcal{C}}\alpha^{\mathcal{B}}_{\mathcal{C}},
\fe
and
\ie
	D_{\mu}R^{\mathcal{A}\mathcal{B}}_{\rho\sigma} = \partial_{\mu}R^{\mathcal{A}\mathcal{B}}_{\rho\sigma} +\left(\omega^{\mathcal{A}\mathcal{C}}_{\mu}R^{\mathcal{D}\mathcal{B}}_{\rho\sigma}+\omega^{\mathcal{B}\mathcal{C}}_{\mu}R^{\mathcal{D}\mathcal{A}}_{\rho\sigma}\right)\eta_{\mathcal{C}\mathcal{D}}.
\fe
It is essential to emphasize certain constraints associated with the gauge connection $\hat{\omega}^{\mathcal{A}\mathcal{B}}_{\mu}(x,\Theta)$: 
\ie
\label{CondDefOmega}
	\hat{\omega}^{\mathcal{A}\mathcal{B}\star}_{\mu}(x,\Theta) = -\hat{\omega}^{\mathcal{A}\mathcal{B}}_{\mu}(x,\Theta), \quad 
	\hat{\omega}^{\mathcal{A}\mathcal{B}}_{\mu}(x,\Theta) ^{r} \equiv \hat{\omega}^{\mathcal{A}\mathcal{B}}_{\mu}(x,-\Theta) = -\hat{\omega}^{\mathcal{B}\mathcal{A}}_{\mu}(x,\Theta),
\fe
The superscript ${}^\star$ denotes the complex conjugate operation. Additionally, the non-commutative corrections arising from the constraints given in Eq. \eqref{CondDefOmega} can be formulated as
\ie
	\omega^{\mathcal{A}\mathcal{B}}_{\mu} (x) = - \omega^{\mathcal{B}\mathcal{A}}_{\mu} (x), \quad \omega^{\mathcal{A}\mathcal{B}}_{\mu\nu\rho} (x) = \omega^{\mathcal{B}\mathcal{A}}_{\mu\nu\rho} (x), \quad \omega^{\mathcal{A}\mathcal{B}}_{\mu\nu\rho\lambda\tau} (x) = -\omega^{\mathcal{B}\mathcal{A}}_{\mu\nu\rho\lambda\tau} (x).
\fe
Above equations are derived by applying Eqs. \eqref{noncommcorr-tetrad} and \eqref{noncommcorr-omega} under the conditions of vanishing torsion field $T^{a}_{\mu\nu}$ and the limit $\mathcal{K}\rightarrow 0$. For this expression, the complex conjugate of the deformed tetrad fields can be represented as
\ie
\label{ComConjDefTetrads}
	\hat{e}^{a\star}_{\mu}(x,\Theta) = e^{a}_{\mu}(x)+i \Theta^{\nu\rho}e^{a}_{\mu\nu\rho}(x)+\Theta^{\nu\rho}\Theta^{\lambda\tau}e^{a}_{\mu\nu\rho\lambda\tau}(x)\dots,
\fe
in which
\ie
	\begin{split}
		e^{a}_{\mu\nu\rho} &= \frac14	\left[\omega^{ac}_{\nu}\partial_{\rho} e^{d}_{\mu}+\left(\partial_{\rho}\omega^{ac}_{\mu}+R^{ac}_{\rho\mu}\right)e^{d}_{\nu}\right]\eta_{ad},
	\end{split}
\fe
and
\ie
	\begin{split}
		e^{a}_{\mu\nu\rho\lambda\tau}(x) &= \frac{1}{32}\left[2\left\{R_{\tau\nu},R_{\mu\rho}\right\}^{ab}e^{c}_{\lambda} - \omega^{ab}_{\lambda}\left(D_{\rho}+\partial_{\rho}\right)R^{cd}_{\tau\mu}e^{m}_{\nu}\eta_{dm}\right.\\
		&\left.-\left\{\omega_{\nu},\left(D_{\rho}+\partial_{\rho}\right)R_{\tau\mu}\right\}^{ab}e^{c}_{\lambda}-\partial_{\tau}\left\{\omega_{\nu},\left(\partial_{\rho}\omega_{\mu}+R_{\rho\mu}\right)\right\}^{ab}e^{c}_{\lambda}\right.\\
		&\left.-\omega^{ab}_{\lambda}\partial_{\tau}\left(\omega^{cd}_{\nu}\partial_{\rho}e^{m}_{\mu}+\left(\partial_{\rho}\omega^{cd}_{\mu}+R^{cd}_{\rho\mu}\right)e^{m}_{\nu}\right)
		\eta_{dm}+2\partial_{\nu}\omega^{ab}_{\lambda}\partial_{\rho}\partial_{\tau}e^{c}_{\mu}\right.\\
		&\left. -2\partial_{\rho}\left(\partial_{\tau}\omega^{ab}_{\mu}+R^{ab}_{\tau\mu}\right)\partial_{\nu}e^{c}_{\lambda}-\left\{\omega_{\nu},\left(\partial_{\rho}\omega_{\lambda}+R_{\rho\lambda}\right)\right\}^{ab}\partial_{\tau}e^{c}_{\mu}\right.\\
		&\left.-\left(\partial_{\tau}\omega^{ab}_{\mu}+R^{ab}_{\tau\mu}\right)\left(\omega^{cd}_{\nu}\partial_{\rho}e^{m}_{\lambda}+\left(\partial_{\rho}\omega^{cd}_{\lambda}+R^{cd}_{\rho\lambda}\right)e^{m}_{\nu}\eta_{dm}\right)
		\right]\eta_{bc}.
	\end{split}
\fe

Consequently, the deformed metric tensor can be expressed as
\ie
\label{DefMetTensor}
	g^{\Theta}_{\mu\nu}\left(x,\Theta\right) = \frac12 \eta_{ab}\left(\hat{e}^{a}_{\mu}(x,\Theta)\ast\hat{e}^{b\star}_{\nu}(x,\Theta)+\hat{e}^{b}_{\mu}(x,\Theta)\ast\hat{e}^{a\star}_{\nu}(x,\Theta)\right),
\fe
in which the symbol $\ast$ represents the standard star product. For the remainder of the calculations, we shall adopt natural units, setting $\hbar = c = G = 1$. Thereby, we have
\ie
\label{met}
\begin{array}{l}
 g_{tt}^{\Theta} = g_{tt} - \frac{{\alpha (8r - 11\alpha )}}{{16{r^4}}}{\Theta ^2} + \mathcal{O}({\Theta ^4}),\\
g^{\Theta}_{rr} = g_{rr} - \frac{{\alpha (4r - 3\alpha )}}{{16{r^2}{{(r - \alpha )}^2}}}{\Theta ^2} + \mathcal{O}({\Theta ^4}),\\
 g^{\Theta}_{\theta\theta} = g_{\theta\theta} - \frac{{2{r^2} - 17\alpha r + 17{\alpha ^2}}}{{32r(r - \alpha )}}{\Theta ^2} + \mathcal{O}({\Theta ^4}),\\
g^{\Theta}_{\varphi\varphi} = {g_{\varphi\varphi}} - \frac{{({r^2} + \alpha r - {\alpha ^2})\cos \theta  - \alpha (2r - \alpha )}}{{16r(r - \alpha )}}{\Theta ^2} + \mathcal{O}({\Theta ^4}).
\end{array}
\fe
To determine the radius of the deformed Schwarzschild event horizon, the condition $1/g^{\Theta}_{rr} = 0$ is imposed, i.e., up to the second order of $\Theta$, consistent with the method described in Ref. \cite{t29}. Applying this condition yields:
\ie
\label{rad}
{r_{s\Theta }} = 2M - \frac{{{\Theta ^2}}}{{32M}}.
\fe
The radius \(r_{s\Theta} = 2M_\Theta\) represents the modified event horizon associated with the deformed non--commutative mass of the Schwarzschild black hole. This correction introduces a redefined mass parameter expressed as \cite{heidari2023gravitational,heidari2024quantum}:
\begin{equation}\label{mass}
M_\Theta = M - \frac{1}{64M}\Theta^2.
\end{equation}
Throughout this analysis, the Schwarzschild metric is utilized, incorporating the corrected non--commutative mass defined in Eq. (\ref{mass}). In addition, all results presented in this work, including the figures, are expressed in natural (geometrized) units, where $\hbar = c = G = 1$.


\section{A corrected Newman--Janis technique \label{cnj}}

A widely utilized method for constructing rotating black hole solutions from spherically symmetric spacetimes is the Newman--Janis procedure \cite{n57i,n57}. In this work, a modified approach called the non-complexification Newman--Janis algorithm \cite{n59,n58} is applied. This variation eliminates the complexification step, making it particularly effective for obtaining rotating metrics in systems involving imperfect fluids, starting from static, spherically symmetric configurations \cite{n64,n61,n65,n60,n63,n62}. In our case, we consider
\ie
\mathrm{d}s^2 = -f_{\Theta}(r)\mathrm{d}t^2 + \frac{\mathrm{d}r^2}{f_{\Theta}(r)} + r^2 (\mathrm{d}\theta^2 + \sin^2\theta \, \mathrm{d}\varphi^2),
\fe
in which $f_{\Theta}(r) \equiv 1 - 2M_{\Theta}/r$.

The method begins by converting the metric into advanced null coordinates, specifically the Eddington--Finkelstein coordinates $(u, r, \theta, \phi)$. This is achieved through the coordinate transformation:
\ie
\mathrm{d}u = \mathrm{d}t - \frac{\mathrm{d}r}{f_{\Theta}(r)},
\fe
which allows the static metric to be re--expressed as:
\ie
\mathrm{d} s^{2} = - f_{\Theta}(r) \mathrm{d}u^{2} - 2\mathrm{d}u \mathrm{d}r + r^{2}\left(  \mathrm{d}\theta^{2} + \sin^{2}\theta \mathrm{d}\phi^{2}   \right).
\fe
A basis of null tetrad vectors is introduced, defined as $Z^\alpha_\mu = (l_\mu, n_\mu, m_\mu, \bar{m}_\mu)$. The inverse metric tensor $g^{\mu\nu}$ can then be expressed in terms of these vectors as
\ie
g^{\mu\nu} = -l^\mu n^\nu - l^\nu n^\mu + m^\mu \bar{m}^\nu + m^\nu \bar{m}^\mu.
\fe
The tetrad vector components are explicitly defined as
\ie 
l^{\mu} = \delta^{\mu}_{r}, \quad n^\mu = \delta^u_\mu - \frac{1}{2} f_{\Theta}(r) \delta^\mu_r, \quad m^\mu = \frac{1}{\sqrt{2}r} \left(\delta_\theta^\mu + \frac{i} {\sin\theta} \, \delta_\phi^\mu\right).
\fe
Within this framework, $\bar{m}_\mu$ represents the complex conjugate of $m_\mu$. The null tetrad vectors form an orthonormal basis and satisfy the following conditions $
l^\mu l_\mu = n^\mu n_\mu = m^\mu m_\mu = l^\mu m_\mu = n^\mu m_\mu = 0$,
and
$ l^\mu n_\mu = -m^\mu \bar{m}_\mu = -1$.

A complex coordinate transformation is then implemented, redefining $\delta^\nu_\mu$ according to the following relation \cite{afrim}
\ie
\delta^r_\mu \rightarrow \delta^r_\mu, \quad \delta^u_\mu \rightarrow \delta^u_\mu, \quad \delta^\theta_\mu \rightarrow \delta^r_\mu + i a \sin \theta (\delta^u_\mu - \delta^r_\mu), \quad \delta^\phi_\mu \rightarrow \delta^\phi_\mu.
\fe
In this formulation, the spin parameter of the black hole is denoted by $a$. The challenge of complexifying the radial coordinate is addressed using the modified Newman--Janis algorithm developed in \cite{azreg2014generating,afrim}. Rather than directly complexifying $r$, this method transforms the radial function $f(r)$ into a more general expression $F(r, a, \theta)$ and replaces $r^2$ with a redefined term $H(r, a, \theta)$. By following the steps detailed in \cite{azreg2014generating,afrim}, the null tetrad vectors after this transformation are given by:
\ie
l'_\mu = \delta_r^\mu, \quad n'^\mu = \delta_u^\mu - F(r, a, \theta) \delta_r^\mu, \quad m'^\mu = \frac{1}{\sqrt{2H(r, a, \theta)}} \left[i a \sin\theta (\delta_u^\mu - \delta_r^\mu) + \delta_\theta^\mu + \sin\theta \, \delta_\phi^\mu \right],
\fe
With the necessary preliminaries established, the resulting expression is obtained below
\ie
g^{\mu\nu} = -l'^\mu n'^\nu - l'^\nu n'^\mu + m'^\mu \bar{m}'^\nu + m'^\nu \bar{m}'^\mu.
\fe

Following this method, a rotating black hole solution can be constructed in Eddington--Finkelstein coordinates, as presented as follows \cite{azreg2014generating}
\ie
\begin{split}
\mathrm{d}s^2 = &  -F(r, a, \theta) \mathrm{d}u^2 - 2 \mathrm{d}u \mathrm{d}r + 2 a \sin^2\theta \left(F(r, a, \theta) - 1\right) \mathrm{d}u \mathrm{d}\phi + 2 a \sin^2\theta \mathrm{d}r \mathrm{d}\phi \\ + & H(r, a, \theta) \mathrm{d}\theta^2 
+ \sin^2\theta \left[H(r, a, \theta) + a^2 \sin^2\theta \left(2 - F(r, a, \theta)\right)\right] \mathrm{d}\phi^2.
\end{split}
\fe

The function $F(r, a, \theta)$ originates from the deformation of the radial component $f_{\Theta}(r)$. This modified approach provides a versatile framework applicable to generating rotating spacetimes from any spherically symmetric black hole solution \cite{kumar2020rotating,brahma2021testing,islam2023investigating,afrin2022testing}. To express the metric in Boyer–Lindquist coordinates, a global transformation is performed \cite{azreg2014generating}:
\ie
\mathrm{d}u = \mathrm{d}t' + \lambda(r) \mathrm{d}r, \quad \mathrm{d}\phi = \mathrm{d}\phi' + \chi(r) \mathrm{d}r.
\fe
The functions $\lambda(r)$ and $\chi(r)$ are entirely radial dependent and can be expressed as \cite{azreg2014generating}:
\ie
\lambda(r) = -\frac{r^2 + a^2}{f_{\Theta}(r)(r^2 + a^2)}, \quad \chi(r) = -\frac{a}{f_{\Theta}(r)(r^2 + a^2)}.
\fe

\begin{figure}
    \centering
     \includegraphics[scale=0.53]{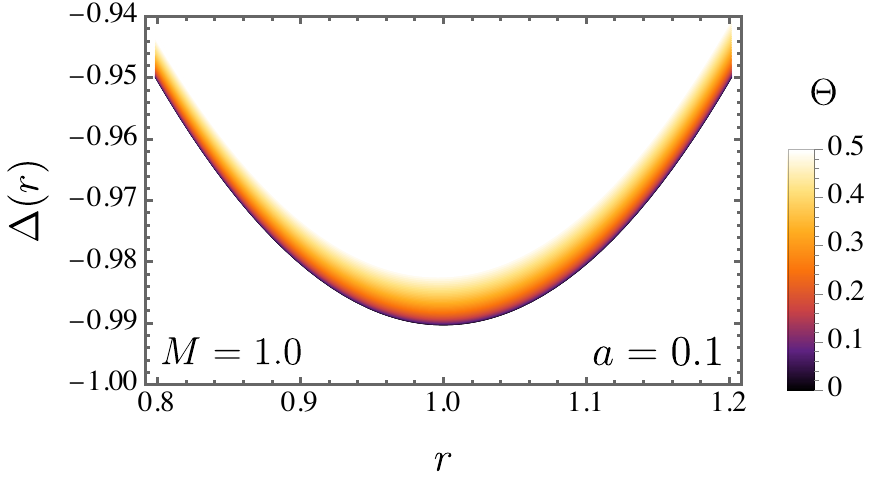}
     \includegraphics[scale=0.53]{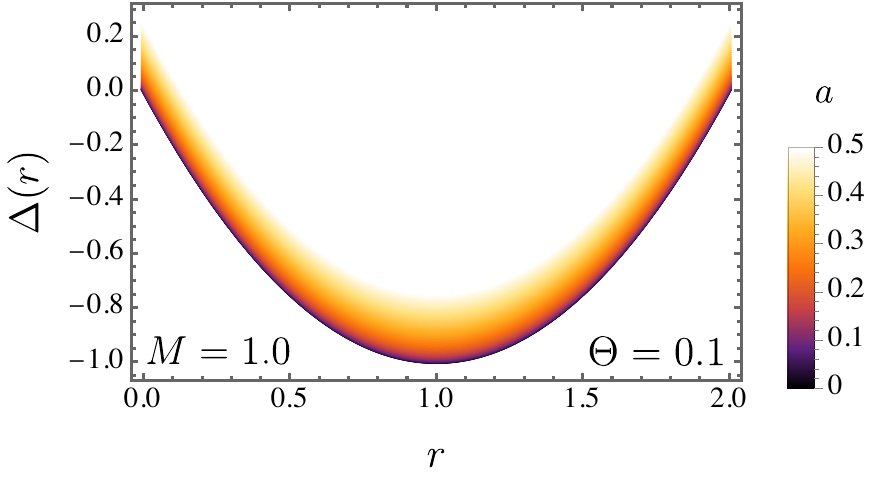}
    \caption{The parameter $\Delta(r)$ is shown as a function of $r$ for different combinations of $\Theta$ and $a$.}
    \label{noncommutativedeltas}
\end{figure}

To eliminate the cross--term $\mathrm{d}t \mathrm{d}r$ from the metric, the following condition is imposed:
\ie
F(r,a,\theta) = \frac{f_{\Theta}(r)r^2 + a^2\cos^2\theta}{H(r,a,\theta)}.
\fe
Additionally, the requirement for the Einstein tensor component $G_{r\theta}$ to vanish leads to the constraint:
\ie
H(r,a,\theta) = r^2 + a^2\cos^2\theta.
\fe
These conditions, when applied and followed by a series of algebraic adjustments, yield the final expression for the rotating black hole metric in Boyer--Lindquist coordinates \cite{azreg2014generating, afrim}
\ie
\begin{split}
\label{rotatingmetric}
\mathrm{d} s^{2} = & \left[  \frac{\Delta(r) - a^{2} \sin^{2}\theta       }{\Sigma}   \right] \mathrm{d}t^{2} + \frac{\Sigma}{\Delta(r)} \mathrm{d}r^{2} + \Sigma \,\mathrm{d}\theta^{2} - 2 a \sin^{2} \theta \left[ 1 - \frac{\Delta(r) - a^{2}\sin^{2}\theta}{\Sigma}      \right] \mathrm{d}t \mathrm{d}\phi \\
& + \frac{\sin^{2}\theta}{\Sigma} \left[ (r^{2} + a^{2})^{2}  - \Delta(r) a^{2} \sin^{2} \theta   \right] \mathrm{d}\phi^{2},
\end{split}
\fe
in which $\Delta(r) = a^2 + r^2 f_{\Theta}(r)$ and $\Sigma = r^2 + a^2 \cos^2\theta$. These relations define the rotating modification of the static metric. Since $\Delta(r)$ governs key geometric aspects of the black hole, its behavior is analyzed in Fig. \ref{noncommutativedeltas}, where it is plotted against the radial coordinate $r$ for different values of the non--commutative parameter $\Theta$ and the spin parameter $a$.

The subsequent sections will explore the fundamental properties and features of the derived rotating metric.


\section{The general remarks and thermal behavior \label{generalfeatures}}

\subsection{The horizons and the corresponding ergosheres}

\begin{figure}
    \centering
     \includegraphics[scale=0.545]{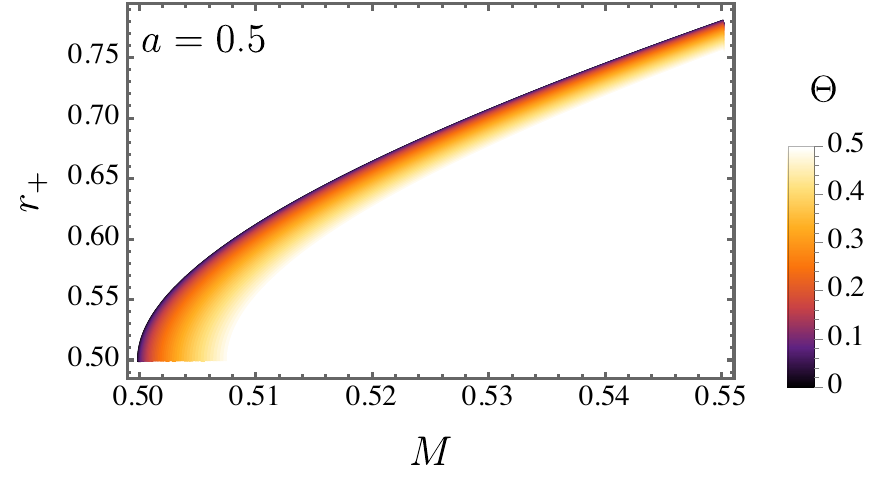}
     \includegraphics[scale=0.545]{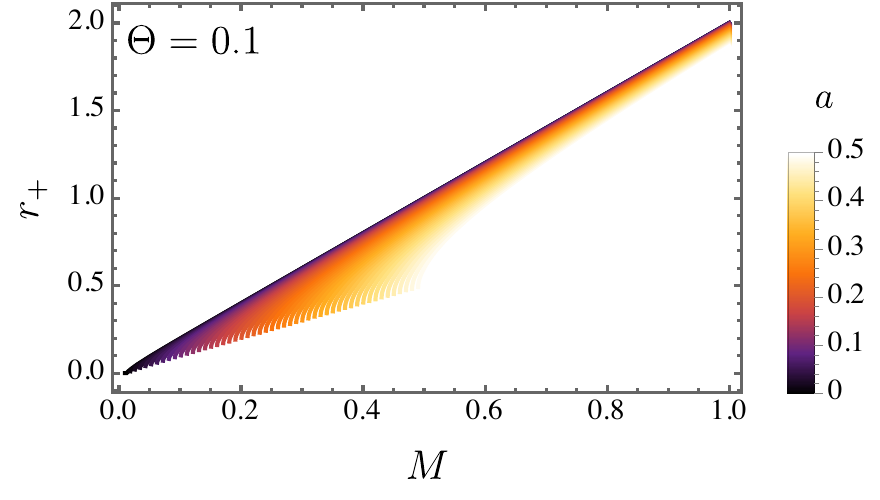}
     \includegraphics[scale=0.64]{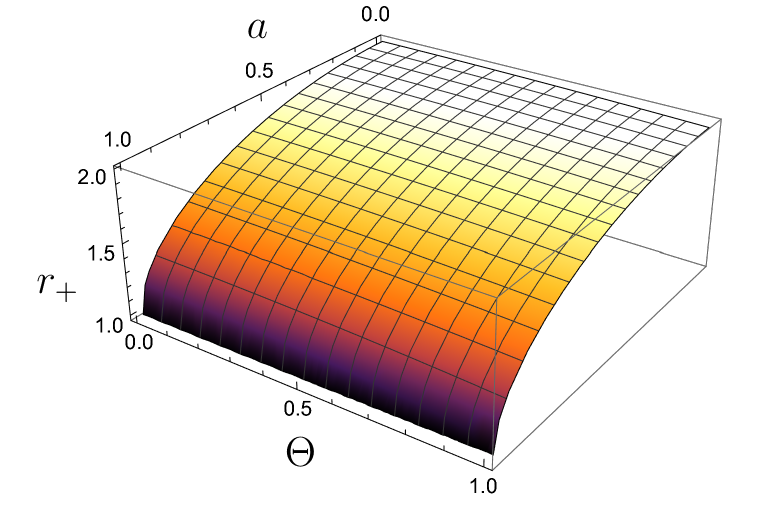}
    \caption{The top left panel displays the behavior of $r_{+}$ as a function of the mass parameter \(M\) for various values of $\Theta$ while keeping the spin parameter fixed at $a = 0.5$. The top right panel illustrates the dependence of $r_{+}$ on $M$ for different spin parameter values, with $\Theta$ held constant at $0.1$. The bottom panel presents a three--dimensional plot of $r_{+}$ as a function of both $a$ and $\Theta$, keeping the mass parameter fixed at $M = 1$.}
    \label{eventhor}
\end{figure}

This section focuses on examining both the thermodynamic properties of the metric in Eq. (\ref{rotatingmetric}) and the geometric structures of the event horizon and ergosphere. To identify the physical horizons, the condition $1/g_{rr} = 0$ is imposed, which simplifies to solving $\Delta(r) = 0$
\ie
r_{+} = \frac{\sqrt{\left(\Theta ^2-64 M^2\right)^2-4096 a^2 M^2}-\Theta ^2+64 M^2}{64 M},
\fe
where $r_{+}$ denotes the outer event horizon and
\ie
r_{-} = -\frac{\sqrt{\left(\Theta ^2-64 M^2\right)^2-4096 a^2 M^2}+\Theta ^2-64 M^2}{64 M},
\fe
represents the inner horizon. As the parameter $\Theta$ approaches zero, the expressions for both horizons reduce to those of the standard Kerr black hole. To illustrate the behavior of the outer event horizon more effectively, Fig. \ref{eventhor} provides a detailed visualization. The top left panel shows the variation of $r_{+}$ with the mass parameter $M$ for different values of $\Theta$, keeping the spin parameter fixed at $a = 0.5$. In the top right panel, the dependence of $r_{+}$ on $M$ is plotted for varying spin parameter values while $\Theta$ is held constant at $0.1$. The bottom panel further explores this relation with a three--dimensional plot of $r_{+}$ as a function of both $a$ and $\Theta$ with a fixed mass value of $M = 2$.

The analysis now moves to the ergosphere structure, defined by the condition $g_{tt} = 0$. Solving this equation, it leads to the following expression:
\ie
r_{e_{\pm}} =  \frac{\pm \sqrt{\left(\Theta ^2-64 M^2\right)^2-4096 a^2 M^2 \cos ^2(\theta )}-\Theta ^2+64 M^2}{64 M}.
\fe
The ergosphere is characterized by the surfaces defined by $r_{e_{\pm}}$, which depend on the spin parameter $a$, mass $M$, non--commutative parameter $\Theta$, and the angular coordinate $\theta$. As $\Theta \to 0$, this structure reduces to the familiar Kerr black hole ergosphere. The presence of the $\cos(2\theta)$ term introduces angular dependence, causing the ergosphere to deviate from a perfect sphere, appearing flattened near the rotation axis at $\theta = 0$ and $\theta = \pi$. The region lies between the two surfaces described by $r_{e_{\pm}}$, where the temporal component of the metric, $g_{tt}$, becomes negative and behaves like a spatial coordinate. Consequently, any particle within this region must rotate in the same direction as the central mass to maintain a time--like trajectory, ensuring positive proper time as it moves through spacetime.

\begin{figure}
    \centering
     \includegraphics[scale=0.545]{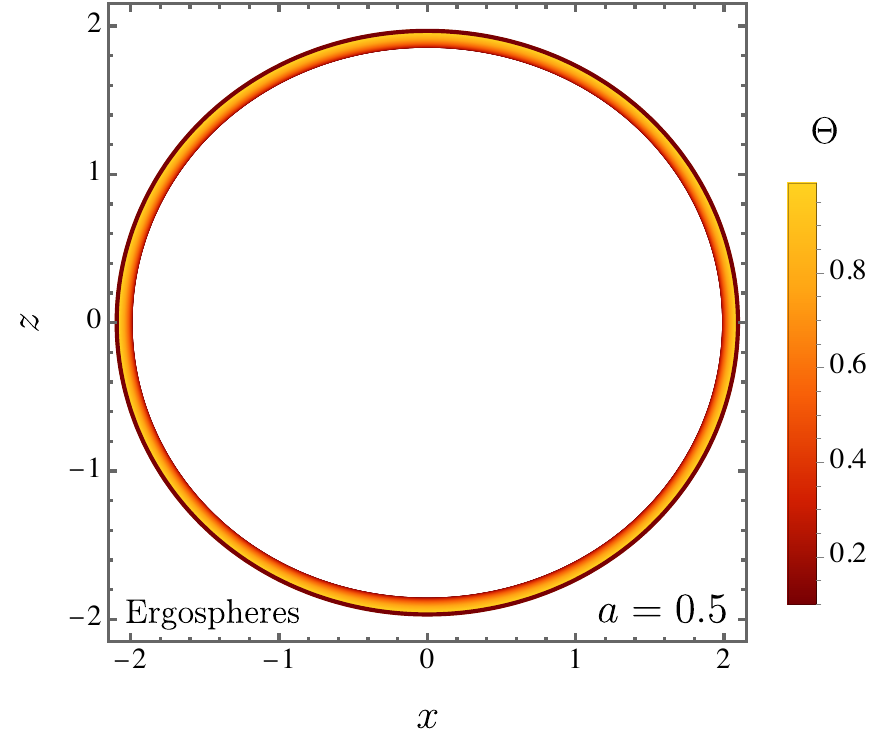}
     \includegraphics[scale=0.545]{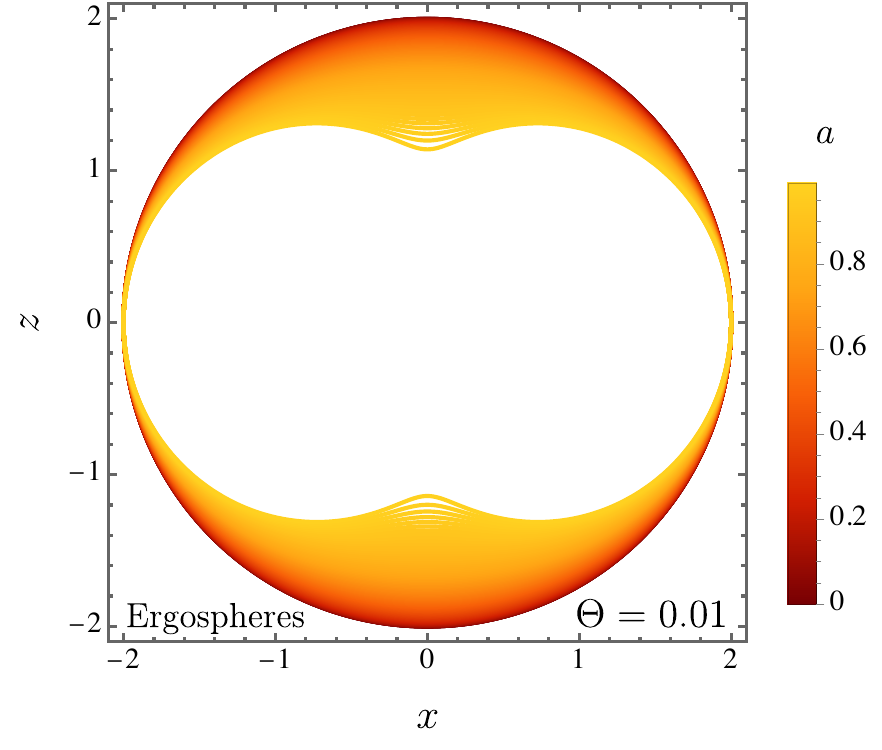}
    \caption{The parametric plot of the ergosphere is shown for different combinations of the non--commutative parameter $\Theta$ and the spin parameter $a$.}
    \label{ergos}
\end{figure}

Fig. \ref{ergos} illustrates the inner and outer ergosphere boundaries through a parametric plot for varying spin parameter values $a$, with the non--commutative parameter fixed at $\Theta = 0.01$ and mass set to $M = 1$. A significant effect of non--commutativity, compared to the standard Kerr black hole, is the noticeable contraction of the ergosphere along the \(z\)--axis, resulting in a more compressed structure in that direction.

\subsection{The angular velocity}

Analogous to the Kerr black hole, the presented solution displays symmetries related to time invariance and rotational symmetry, characterized by the existence of Killing vectors. The \textit{Lense--Thirring} effect causes a stationary observer positioned outside the event horizon, with zero angular momentum relative to an observer at infinity, to be set into rotational motion. In the non--commutative black hole spacetime, this frame--dragging originates, among other factors, from the non--zero off--diagonal metric component \(g_{t\phi}\). Therefore, the observer's angular velocity is determined by \cite{visser2007kerr,grumiller2022black}
\ie
\begin{split}
\omega(r) & = - \frac{g_{t\phi}}{g_{\phi\phi}} = \frac{a[a^{2} + r^{2} - \Delta(r)  ]}{(r^{2} + a^{2} )^{2} - a^{2} \Delta(r) \sin \theta } \\
& =\frac{a r^2 \left(32 a^2 M-r \left(\Theta ^2+32 M (r-2 M)\right)\right)}{a^2 \sin (\theta ) \left(32 a^2 M \left(r^2-1\right)+r^2 \left(\Theta ^2 (-r)-32 M \left(-2 M r+r^2+1\right)\right)\right)+32 M \left(a^2+r^2\right)^2}.
\end{split}
\fe

\begin{figure}
    \centering
     \includegraphics[scale=0.545]{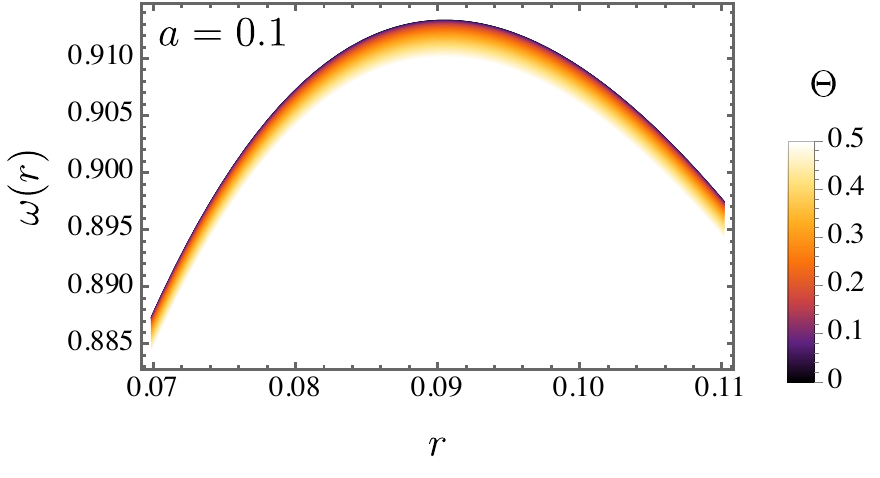}
     \includegraphics[scale=0.545]{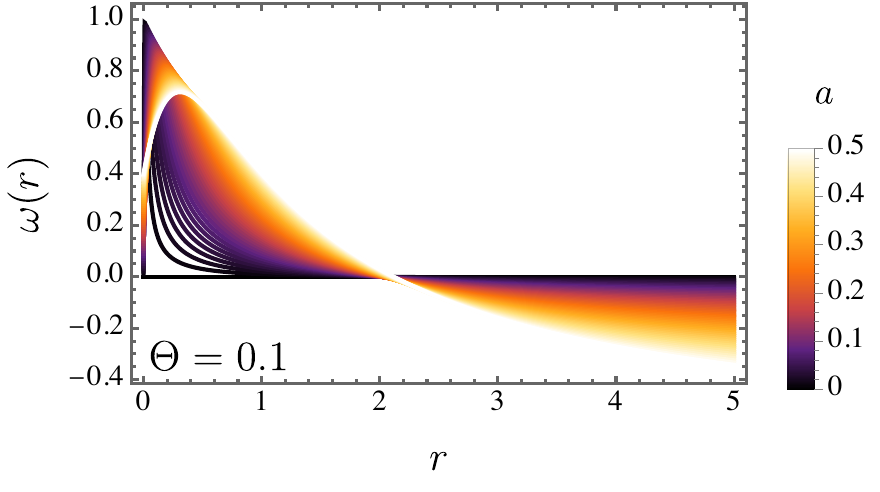}
    \caption{The angular velocity $\omega$ is shown in two--dimensional plots for different values of the spin parameter $a$ and the non--commutative parameter $\Theta$.}
    \label{angularvelocity2d}
\end{figure}

\begin{figure}
    \centering
     \includegraphics[scale=0.62]{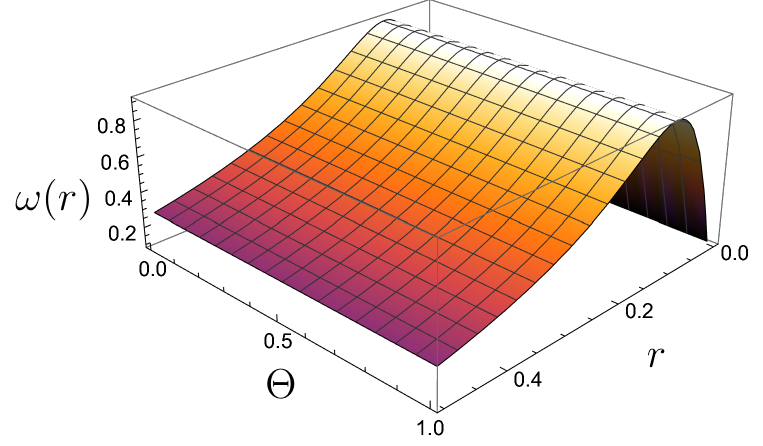}
     \includegraphics[scale=0.62]{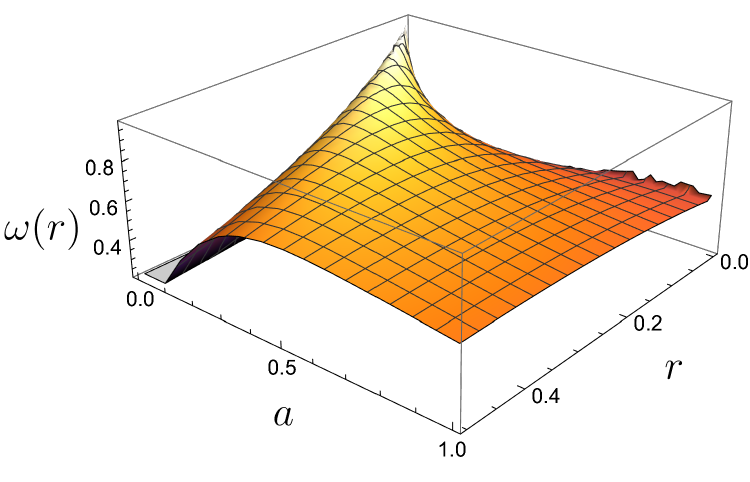}
    \caption{The angular velocity $\omega$ is depicted using three--dimensional plots for various values of the spin parameter $a$ and the non--commutative parameter $\Theta$.}
    \label{angularvelocity3d}
\end{figure}

As we can naturally expect, by setting the non--commutative parameter $\Theta$ to zero, we recover the standard Kerr expression for the angular velocity. Fig. \ref{angularvelocity2d} depicts the angular velocity $\omega$ for different values of the spin parameter $a$ and the non--commutative parameter $\Theta$, highlighting the comparison with the classical Kerr case ($\Theta = 0$). As $\Theta$ increases, $\omega(r)$ also tends to decrease, whereas larger values of $a$ lead to an increase in the magnitude of $\omega(r)$ (until $r \approx 2 $). Fig. \ref{angularvelocity3d} provides a three--dimensional representation of this behavior: the left panel illustrates $\omega(r)$ as a function of $\Theta$ and $r$, while the right panel displays the dependence of the angular velocity on both $a$ and $r$.

\subsection{The surface gravity}

The calculation of the surface gravity involves a coordinate transformation from the form given in Eq. \eqref{rotatingmetric} to the Eddington--Finkelstein coordinates \cite{christodoulou1971reversible,ruiz2019thermodynamic}, which remove the coordinate singularities occurring when $\Delta(r) = 0$. This transformation introduces modified time and angular coordinates, $\tau$ and $\chi$, respectively, defined as follows:
\ie
\mathrm{d} \tau  = \mathrm{d}t + \frac{r^{2}+a^{2}}{\Delta(r)}\mathrm{d}r,
\fe
and, also, 
\ie
\mathrm{d}\chi = \mathrm{d} \varphi + \frac{a}{\Delta(r)} \mathrm{d}r,
\fe
so that
\ie
\begin{split}
\mathrm{d}s^{2}  = & \frac{\Delta(r) - a^{2} \sin^{2}\theta }{\Sigma} \mathrm{d}\tau^{2} + 2 \mathrm{d}\tau \mathrm{d}r - \frac{2a(r^{2}+ a^{2} - \Delta(r) )\sin^{2}\theta  }{\Sigma} \mathrm{d}\tau\mathrm{d}\chi \\
&-2a \sin^{2}\theta \mathrm{d}r \mathrm{d}\chi - \frac{(r^{2} + a^{2} )^{2} - \Delta(r) a^{2} \sin^{2}\theta  }{\Sigma}  \mathrm{d}\chi^{2}.
\end{split}
\fe

By performing this change of variables, it turns out to be  evident that on the hypersurfaces where $r = r_\pm$, the Killing vectors associated with the coordinates can be expressed as:
\ie
\psi_{\pm} = \frac{\partial}{\partial \tau} + \frac{\Delta(r)}{r_{\pm}^{2} + a^{2}} \frac{\partial}{\partial \chi}.
\fe

Additionally, the Killing vectors can be rewritten using Boyer--Lindquist coordinates. Consider a Killing horizon $K$, defined by the normal Killing vector $\xi$. The surface gravity $\kappa$ is described as the constant of proportionality between the vectors $\xi^\nu \nabla_\nu \xi^\mu$ and $\xi^\mu$ \cite{wald2010general}. In this manner, we have 
\ie
\begin{split}
k_{+} = \frac{\Delta^{\prime}(r_{+})}{2(r^{2}_{+} + a^{2})}   = \frac{32 M \sqrt{\left(\Theta ^2-64 M^2\right)^2-4096 a^2 M^2}}{\left(64 M^2-\Theta ^2\right) \left(\sqrt{\left(\Theta ^2-64 M^2\right)^2-4096 a^2 M^2}-\Theta ^2+64 M^2\right)}.
\end{split}
\fe

The surface gravity, as will be examined in the subsequent subsections, plays a fundamental role in the analysis of the thermodynamic properties associated with the black hole under study.


\subsection{The Hawking temperature}

In 1973, the foundational principles of black hole mechanics were outlined by Bardeen, Carter, and Hawking, highlighting a close correspondence with the classical laws of thermodynamics \cite{bardeen1973four}. The zeroth law asserts that surface gravity remains constant across the event horizon, resembling the uniform temperature of a system in thermal equilibrium \cite{page2005hawking}. The first law connects variations in a black hole’s mass to changes in its surface area, angular momentum, and possibly electric charge, analogous to how internal energy shifts in thermodynamics relate to heat and work \cite{carlip2014black}. The second law establishes that the total area of the event horizon never decreases, reflecting the thermodynamic rule that entropy in an isolated system cannot decrease \cite{davies1978thermodynamics}. Lastly, the third law states that surface gravity cannot be reduced to zero through physical processes, mirroring the thermodynamic principle that absolute zero cannot be achieved \cite{hawking1976black}.

The understanding of black hole mechanics was further developed through the contributions of Christodoulou, who investigated irreversible processes in black hole dynamics \cite{christodoulou1970reversible}, and Bekenstein, who introduced the concept of black hole entropy. Bekenstein proposed that a black hole's entropy is directly proportional to the area of its event horizon \cite{1bekenstein2020black,2bekenstein1974generalized}, leading to the formulation of the Bekenstein--Hawking entropy, which formally connects black hole mechanics to thermodynamic principles. Throughout this analysis, unless otherwise specified, all thermodynamic quantities will be evaluated under the assumption that $\Theta, a < M$. The expression for the Hawking temperature is given by:
\ie
\begin{split}
T(\Theta,a,M)  = \frac{k_{+}}{2\pi} = \frac{16 M \sqrt{\left(\Theta ^2-64 M^2\right)^2-4096 a^2 M^2}}{\pi  \left(64 M^2-\Theta ^2\right) \left(\sqrt{\left(\Theta ^2-64 M^2\right)^2-4096 a^2 M^2}-\Theta ^2+64 M^2\right)}.
\end{split}
\fe

\begin{figure}
    \centering
     \includegraphics[scale=0.545]{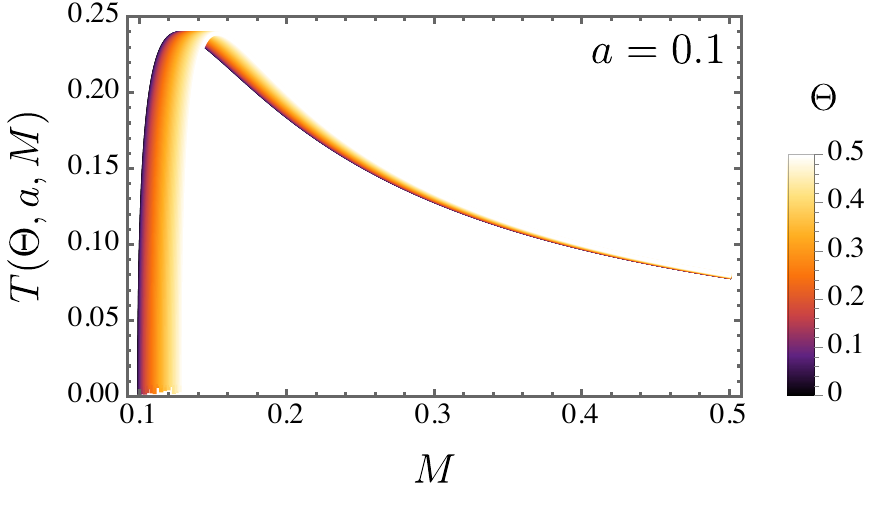}
     \includegraphics[scale=0.545]{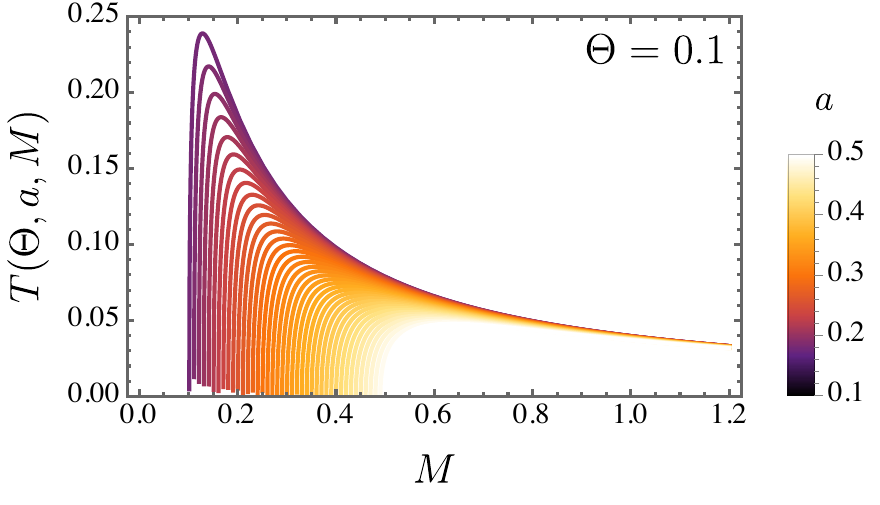}
    \caption{The behavior of the Hawking temperature is represented as a function of $M$ for different choices of the non--commutative parameter $\Theta$, while keeping the rotational parameter fixed at $a = 0.1$.}
    \label{Hawkingtemperature}
\end{figure}

Fig. \ref{Hawkingtemperature} illustrates how the thermodynamic quantity varies under the influence of the non--commutative parameter $\Theta$ and the rotational parameter $a$. As shown in the left panel, an increase in $\Theta$ results in a lower temperature magnitude (for small values of $M$), while a similar effect is observed when $a$ increases, further suppressing the temperature. To emphasize the impact of these modifications, the behavior is compared with the standard Kerr black hole, showing how the additional parameters alter the thermal properties of the system.

\subsection{The remnant mass}

\begin{figure}
    \centering
     \includegraphics[scale=0.7]{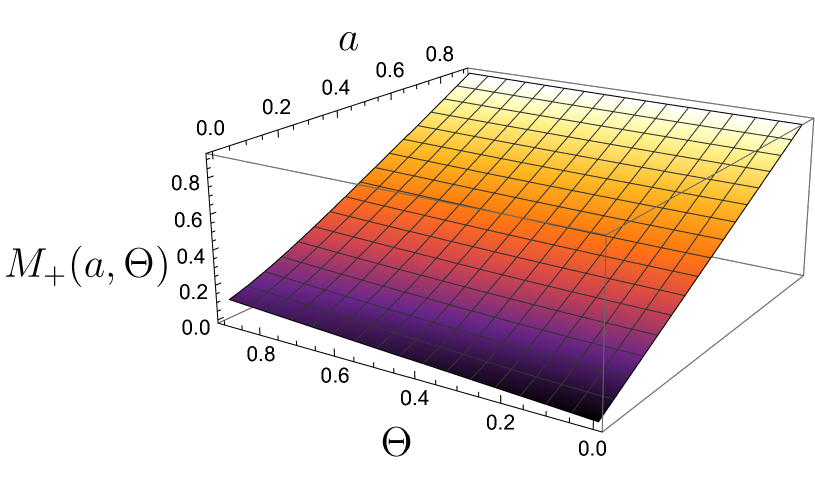}
    \caption{The remnant mass $M_{+}(a,\Theta)$ is depicted in three--dimensional plots for different values of the spin parameter $a$ and the non--commutative parameter $\Theta$.}
    \label{m3d}
\end{figure}

By examining the extreme condition where the black hole reaches its final state, i.e., $T(\Theta, a, M) \to 0$, the corresponding remnant mass is obtained
\ie
M_{\pm}(a,\Theta) = \frac{1}{8} \left(\sqrt{16 a^2+\Theta ^2} \pm 4 a\right).
\fe
As expected, this parameter depends on $\Theta$ and $a$. Additionally, in the limit $\Theta \to 0$, the remnant mass of the Kerr black hole is recovered. Fig. \ref{m3d} presents a three--dimensional plot of the remnant mass $M_{+}(a, \Theta)$ for different values of $a$ and $\Theta$. It is important to emphasize that Eq. (49) yields two branches associated with the extremal configuration of the black hole. The physically admissible solution corresponds to the positive branch $M_{+}(a,\Theta)$, which ensures a real and positive remnant mass for all $a,\Theta>0$. The negative branch $M_{-}(a,\Theta)$ instead leads to a very small physical mass, depending on the intensity of the parameters $a$ and $\Theta$. In the commutative and non–rotating limits, $\Theta\to0$ and $a\to0$, one recovers $M_{+}\to0$, consistent with the vanishing remnant of the Schwarzschild case. Conversely, finite values of $\Theta$ or $a$ prevent complete evaporation, yielding a non–zero minimal mass that characterizes the stable remnant.

\subsection{The entropy}

The expression for entropy is directly given by:
\ie
\begin{split}
& S(\Theta,a,M) = \,\pi (r_{+}^{2} + a^{2}) \\
& = \pi  \left(\frac{\left(\sqrt{\left(\Theta ^2-64 M^2\right)^2-4096 a^2 M^2}-\Theta ^2+64 M^2\right)^2}{4096 M^2}+a^2\right).
\end{split}
\fe

\begin{figure}
    \centering
     \includegraphics[scale=0.545]{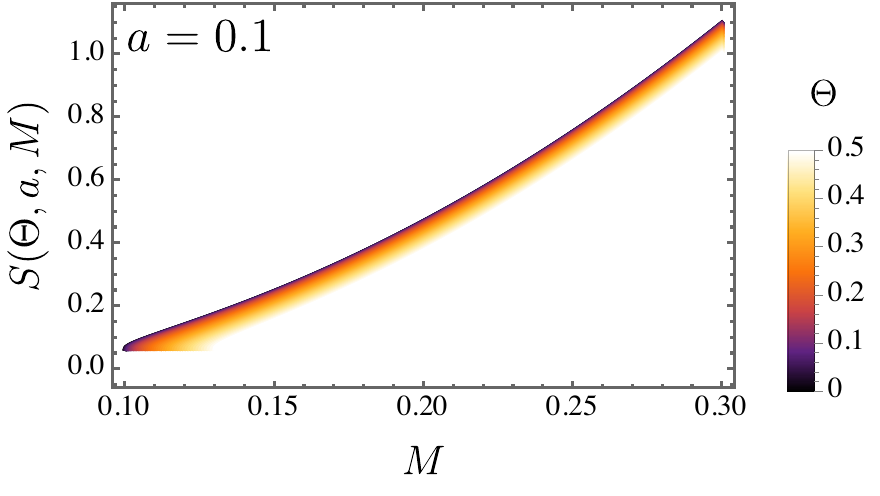}
     \includegraphics[scale=0.545]{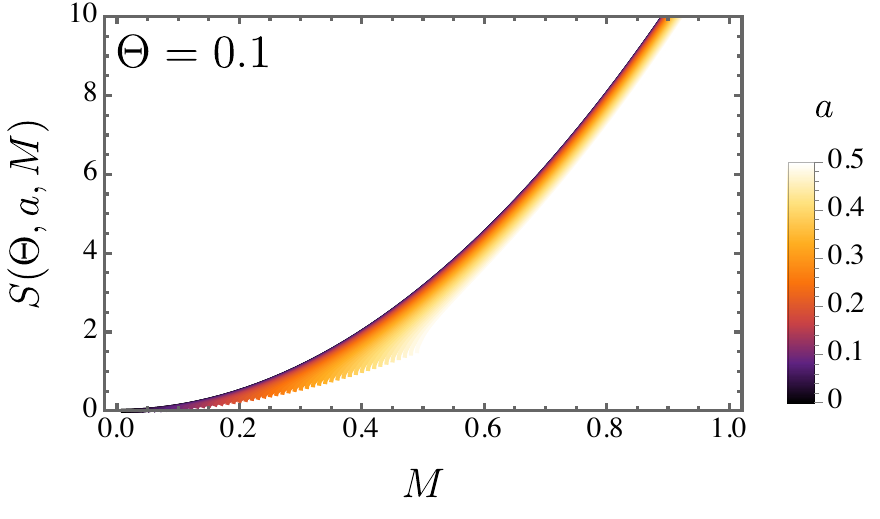}
    \caption{The entropy is shown as a function of $M$ for different values of the non--commutative parameter $\Theta$, with the rotational parameter fixed at $a = 0.1$.}
    \label{entropyrot}
\end{figure}

Fig. \ref{entropyrot} illustrates the entropy behavior for a fixed rotational parameter $a = 0.1$, showing the influence of the non--commutative parameter $\Theta$. To highlight the modifications introduced by $\Theta$, the results are compared with the standard Kerr black hole ($\Theta =  0$), revealing how the entropy is altered in the presence of non--commutativity. In general, an increase in both $\Theta$ and $a$ leads to a reduction in the magnitude of $S(\Theta, a, M)$.

\subsection{Heat capacity}

To ensure a comprehensive analysis, the heat capacity is also examined. Its expression is given by:
\ie
\begin{split}
& C_{V}(\Theta,a,M) = T\frac{\partial S}{\partial T} = T \frac{\partial S/ \partial M }{\partial T/ \partial M} \\
&= \frac{\pi  \Xi ^3 \left(\Theta ^2-64 M^2\right) \sqrt{\left(\Theta ^2-64 M^2\right)^2-4096 a^2 M^2}}{2048 M^2 \left(\Xi  \left(\Theta ^2-64 M^2\right)^2-4096 a^2 M^2 \left(\sqrt{\left(\Theta ^2-64 M^2\right)^2-4096 a^2 M^2}-2 \Theta ^2+128 M^2\right)\right)},
\end{split}
\fe
in which 
\ie
\begin{split}
\Xi =\sqrt{\left(\Theta ^2-64 M^2\right)^2-4096 a^2 M^2}-\Theta ^2+64 M^2.
\end{split}
\fe

\begin{figure}
    \centering
     \includegraphics[scale=0.545]{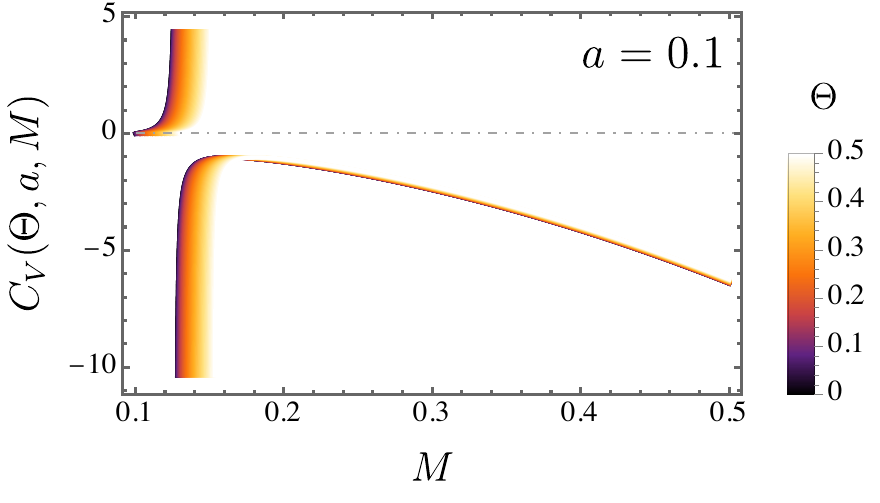}
     \includegraphics[scale=0.545]{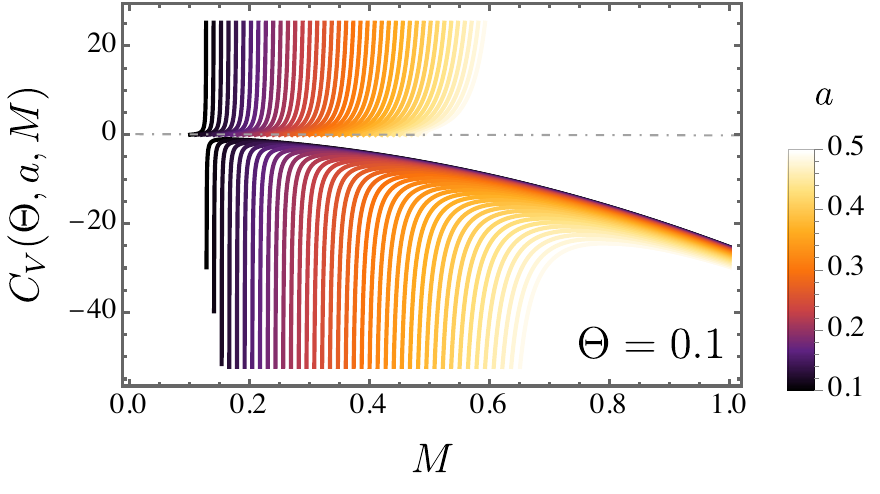}
    \caption{The heat capacity is represented as a function of $M$, considering different choices of the non--commutative parameter $\Theta$ and the rotational parameter $a$.}
    \label{heatrot}
\end{figure}

Fig. \ref{heatrot} displays the behavior of the heat capacity for different values of the non--commutative parameter $\Theta$, with the rotational parameter held constant at $a = 0.1$. To emphasize the effects introduced by non--commutativity, the results are compared with the standard Kerr black hole, analogous to the previous thermodynamic functions. Furthermore, non--commutative gravity has been widely explored in various scenarios, including modifications of general relativity \cite{hrelja2024entropy}, scalar field models, Reissner--Nordström black holes \cite{ciric2018noncommutative,ciric2024noncommutative,dimitrijevic2020noncommutative}, and BTZ black holes \cite{juric2023noncommutative}.

Finally, in Tab. \ref{compppasd}, we present the black hole parameters that are compared with those of another recently proposed axisymmetric black hole in the literature \cite{AraujoFilho:2024rss}. In addition, $\hat{\xi} \equiv \pi ^{3/4} M \sqrt{\sqrt{\pi } \left(M^2-a^2\right)-8 \sqrt{\Theta } M}-4 \sqrt[4]{\pi } \sqrt{\Theta } \sqrt{\sqrt{\pi } \left(M^2-a^2\right)-8 \sqrt{\Theta } M}+8 \Theta +\pi  M^2-8 \sqrt{\pi } \sqrt{\Theta } M$ \cite{AraujoFilho:2024rss}. Furthermore, to avoid repeating the lengthy definitions of $\Gamma$ and $\tilde{\Gamma}$ in this manuscript, readers are referred to Ref. \cite{AraujoFilho:2024rss} for their explicit forms.

\begin{table}[t]
  \centering
  \caption{\label{compppasd} The black hole parameters are compared with those of another recently proposed axisymmetric black hole in the literature \cite{AraujoFilho:2024rss}.}

  \rotatebox{90}{
    \begin{minipage}{\textheight} 
      \centering
      \setlength{\arrayrulewidth}{0.3mm}
      \setlength{\tabcolsep}{12pt}
      \renewcommand{\arraystretch}{1}
      \begin{tabular}{c c c}
        \hline \hline
        Parameters & This work & Lorentzian NC black hole \cite{AraujoFilho:2024rss} \\ \hline \\
        $\Delta(r)$ & $a^2+r^2 \left[1-\frac{2 \left(M-\frac{ \Theta ^2}{64 M}\right)}{r}\right]$ & $a^2+r^2 \left[\frac{8 \sqrt{\Theta } M}{\sqrt{\pi } r^2}-\frac{2 M}{r}+1\right]$ \\ \\
        $r_{\pm}$ & $  \frac{\pm \sqrt{\left(\Theta ^2-64 M^2\right)^2-4096 a^2 M^2}-\Theta ^2+64 M^2}{64 M}$ & $\frac{\pm \sqrt{\left(3 \Theta ^2+64 M^2\right)^2-4096 a^2 M^2}+3 \Theta ^2+64 M^2}{64 M}$ \\ \\
        $r_{e_{+}}$ & $ \frac{\pm \sqrt{\left(\Theta ^2-64 M^2\right)^2-4096 a^2 M^2 \cos ^2(\theta )}-\Theta ^2+64 M^2}{64 M}$ & $\frac{\pm \sqrt{2} \sqrt{-\pi  a^2 \cos (2 \theta )-\pi  a^2+2 \pi  M^2-16 \sqrt{\pi } \sqrt{\Theta } M}+2 \sqrt{\pi } M}{2 \sqrt{\pi }}$ \\ \\
        $\omega(r)$ & $\frac{a r^2 \left(32 a^2 M-r \left(\Theta ^2+32 M (r-2 M)\right)\right)}{a^2 \sin (\theta ) \left(32 a^2 M \left(r^2-1\right)+r^2 \left(\Theta ^2 (-r)-32 M \left(-2 M r+r^2+1\right)\right)\right)+32 M \left(a^2+r^2\right)^2}$ & $\frac{2 a M \left(\sqrt{\pi } r-4 \sqrt{\Theta }\right)}{\sqrt{\pi } \left(a^2+r^2\right)^2-a^2 \sin (\theta ) \left(\sqrt{\pi } \left(a^2+r (r-2 M)\right)+8 \sqrt{\Theta } M\right)}$ \\ \\
        $k_{+}$ & $ \frac{32 M \sqrt{\left(\Theta ^2-64 M^2\right)^2-4096 a^2 M^2}}{\left(64 M^2-\Theta ^2\right) \left(\sqrt{\left(\Theta ^2-64 M^2\right)^2-4096 a^2 M^2}-\Theta ^2+64 M^2\right)}$ & $\frac{\sqrt{M \left(M-\frac{8 \sqrt{\Theta }}{\sqrt{\pi }}\right)-a^2}}{\left(\sqrt{M \left(M-\frac{8 \sqrt{\Theta }}{\sqrt{\pi }}\right)-a^2}+M\right)^2+a^2}$ \\ \\ 
        $T(\Theta,a,M)$ & $\frac{16 M \sqrt{\left(\Theta ^2-64 M^2\right)^2-4096 a^2 M^2}}{\pi\left(64 M^2-\Theta ^2\right) \left(\sqrt{\left(\Theta ^2-64 M^2\right)^2-4096 a^2 M^2}-\Theta ^2+64 M^2\right)}$ & $\frac{\sqrt{M \left(M-\frac{8 \sqrt{\Theta }}{\sqrt{\pi }}\right)-a^2}}{2 \pi  \left(\left(\sqrt{M \left(M-\frac{8 \sqrt{\Theta }}{\sqrt{\pi }}\right)-a^2}+M\right)^2+a^2\right)}$ \\ \\
        $S(\Theta,a,M)$ & $ \pi  \left(\frac{\left(\sqrt{\left(\Theta ^2-64 M^2\right)^2-4096 a^2 M^2}-\Theta ^2+64 M^2\right)^2}{4096 M^2}+a^2\right)$ & $\frac{2 \sqrt{M^2 \left(2 \hat{\xi} -\pi  a^2\right)}}{\sqrt{\pi }}$ \\ \\
        $C_{V}(\Theta,a,M)$ &  $\frac{\pi  \Xi ^3 \left(\Theta ^2-64 M^2\right) \sqrt{\left(\Theta ^2-64 M^2\right)^2-4096 a^2 M^2}}{2048 M^2 \left(\Xi  \left(\Theta ^2-64 M^2\right)^2-4096 a^2 M^2 \left(\sqrt{\left(\Theta ^2-64 M^2\right)^2-4096 a^2 M^2}-2 \Theta ^2+128 M^2\right)\right)}$ & $ - \frac{\Gamma}{\tilde{\Gamma}}$ \\ \\
        \hline\hline
      \end{tabular}
    \end{minipage}
  }
\end{table}


\section{Hawking radiation as a tunneling process}

\subsection{Bosonic particle modes}

To investigate the tunneling of massless particles from a rotating black hole, we adopt a metric where the worldlines describe photons with energy $\omega$ propagating at infinity with constant $\theta$, and with the angular momentum component projected along the black hole's rotation axis given by $L_z = a\omega\sin^2\theta$. These coordinates, referred to as the \textit{Kerr ingoing coordinates} \cite{kerr1963gravitational,vanzo2011tunnelling}, are derived through the transformation
\ie
 \mathrm{d} v = \mathrm{d} t + \frac{(r^2+a^2)}{\Delta(r)} \mathrm{d} r ,\ \ \ \mathrm{d}\phi = \mathrm{d}\Tilde{\phi} + \frac{a}{\Delta(r)} \mathrm{d} r.
\fe
Additionally, the line element described in Eq. (\ref{rotatingmetric}) takes the form
\begin{eqnarray}
\label{kerr}
 \mathrm{d} s^2 =& -\left( 1-\frac{2 M_{\Theta} r}{\Sigma} \right) \mathrm{d} v^2 + 2 \mathrm{d} v \,\mathrm{d} r + \Sigma \,\mathrm{d}\theta^2 - \frac{4 a M_{\Theta} r \sin^2\theta}{\Sigma}\, \mathrm{d}\phi\, \mathrm{d} v  \nonumber
\\
 &- 2 a \sin^2\theta \,\mathrm{d}\phi\,\mathrm{d} r + \frac{(r^2+a^2)^2 - a^2 \Delta(r)\,\sin^2\theta}{\Sigma}\sin^2\theta\,\mathrm{d}\phi^2  \;.
\end{eqnarray}

In rotating black hole spacetimes, the \textit{static limit surface} $r_{\text{st}}$, defined by the condition $g_{tt}=0$, represents the outer boundary of the ergosphere and does not coincide with the event horizon. The applicability of the semiclassical approach in this scenario has been debated \cite{jiang2006hawking,zhang2005hawking,li2008hawking,agullo2010hawking,murata2006hawking,corda2013effective,xu2007hawking,umetsu2010hawking,arbey2020evolution,wang2024entanglement,mcmaken2024hawking,senjaya2024kerr}, as the geometric optics approximation remains more accurate near $r_{\text{st}}$ than at the event horizon, where tunneling processes are typically analyzed. To address this issue, some works \cite{5vanzzo,jiang2006hawking} have implemented a co-rotating coordinate transformation, $\phi \rightarrow \phi - \Omega^H t$, where $\Omega^H = a/(r_+^2 + a^2)$ denotes the angular velocity at the event horizon. However, it will be demonstrated that such a transformation is unnecessary, as the standard formulation already yields the correct result.

The semiclassical approach relies on the classical action $I$ to determine the transition probability across the event horizon, governed by the relativistic Hamilton--Jacobi equation. Given that the metric in Eq. (\ref{kerr}) does not explicitly depend on the coordinates $v$ and $\phi$, the action can be expressed in the form $I = -\omega \, v + J\phi + \mathcal{F}(\theta, r)$. It will be shown that the imaginary contribution arises from the $r$--dependent component, which generates a pole at the horizon. Substituting this expression for $I$ into the Hamilton--Jacobi equation leads to the following result
\begin{eqnarray}
 &a^2 (m \csc\theta - \omega  \sin\theta)^2+2\left[a^2 m-\left(a^2+r^2\right) \omega \right]\mathcal F_r +\nonumber
\\ 
 &+ \left[a^2+r (-2 m+r)\right]\mathcal F_r^2 + \mathcal F_\theta^2 =0,
\end{eqnarray}
where the subscripts $r$ and $\theta$ denote partial derivatives with respect to $r$ and $\theta$, respectively. When solving for $\mathcal{F}_r(\theta, r)$, it becomes evident that the integrand also depends on the coordinate $\theta$:
\ie
\mathcal F_r(\theta,r)=-\frac{X(r)}{ \Delta(r)}\pm\frac{\sqrt{ X(r)^2- \Delta(r) \left[a^2 (M_{\Theta} \csc\theta-\omega  \sin\theta)^2+ \mathcal F_\theta(\theta ,r)^2\right]}}{\Delta(r)}
\fe
where $X(r) \equiv a^2M_{\Theta} - (a^2 + r^2)\omega$. A straightforward approach to address the $\theta$ dependence is to set $\theta = \theta_0$ as a constant and verify that the final expression does not depend on this choice \cite{07femions}. However, this step can be avoided since the tunneling process is examined near the event horizon, where $\Delta(r_+) = 0$. In this regime, the $\theta$--dependent contribution vanishes, reducing the expression to a function of $r$ alone. This simplification arises from the complete separability of the Hamilton--Jacobi equation in Kerr spacetime. Expressing all terms in relation to $r_+$ and $r_-$, the expression for $X(r)$ becomes $X(r) = \frac{(r_+ + r_-)a^2}{2} - \omega(r_+^2 + a^2)$, leading to
\ie
\text{Im}\,\mathcal F(\theta,r)=  - \text{Im}\,\int \mathrm{d} r \, \frac{X(r) + \sqrt{
X^2(r) - (r-r_+)(r-r_-)\ (\dots)}}{(r-r_-) (r-r_+)}  
\fe
where $(\dots)$ represents all terms involving the $\theta$ dependence. Applying Feynman's prescription to regularize the integral,
\ie
\mbox{ Im}\,I = - 2 \pi\, \frac{X(r_+)}{r_+-r_-} =\ \pi \left[ 2\omega\frac{(r_+^2+a^2)}{(r_+-r_-)} - \frac{a^2 (r_++r_-)}{(r_+-r_-)} \right] .
\fe

By expressing $J$ as $J = \frac{a(r_+ + r_-)}{2}$ and incorporating the horizon's angular velocity $\Omega^H$, the terms can be reorganized, resulting in the final form of the tunneling probability rate
\ie
\bar{\Gamma} = \exp\left[ -2\, \mbox{ Im}\, I \right] = \exp\left[ -\beta (\omega - \Omega^{ H} J) \right],
\fe
where the inverse temperature is given by $\beta = \frac{4\pi(r_+^2 + a^2)}{r_+ - r_-}$. In the special case where both $a \rightarrow 0$ and $\Theta \rightarrow 0$, the expression reduces to the well--known result for the Schwarzschild black hole. Moreover, the particle density for bosons can be expressed as
\ie
n_{b}(\omega,\Theta,a,M) = \frac{ \bar{\Gamma}}{1 -  \bar{\Gamma}} = \frac{1}{\exp \left(\frac{\pi  \left(64 M^2-\Theta ^2\right) \left(\omega  \left(\sqrt{\left(\Theta ^2-64 M^2\right)^2-4096 a^2 M^2}-\Theta ^2+64 M^2\right)-32 a^2 M\right)}{16 M \sqrt{\left(\Theta ^2-64 M^2\right)^2-4096 a^2 M^2}}\right)-1}.
\fe

\begin{figure}
    \centering
     \includegraphics[scale=0.545]{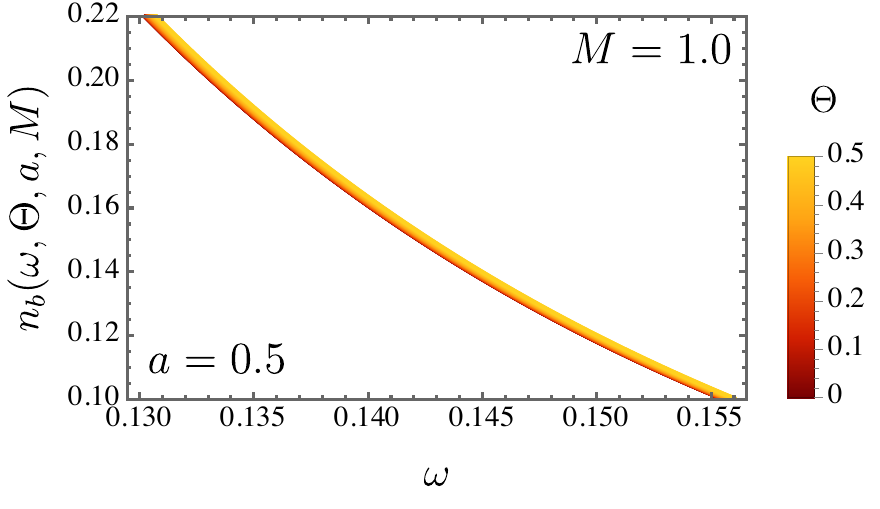}
     \includegraphics[scale=0.545]{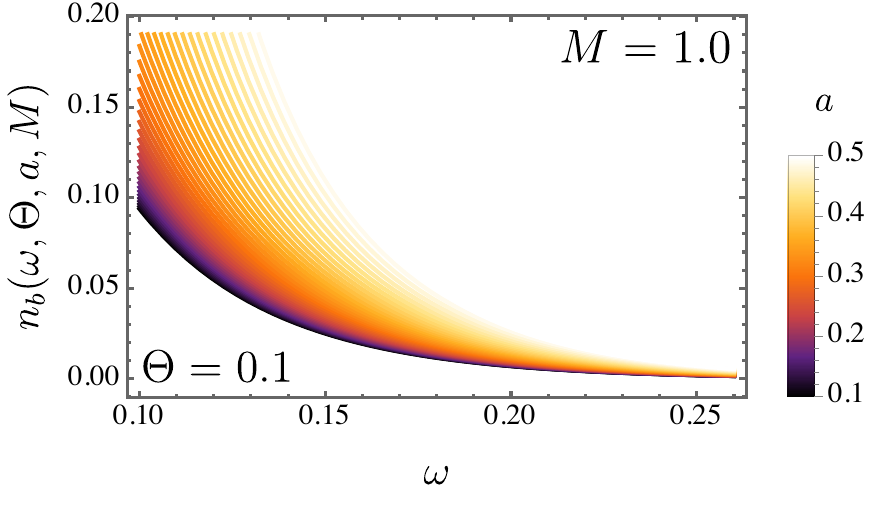}
    \caption{The particle creation rate for bosons, denoted by $n_{b}(\omega,\Theta, a, M)$, is represented as a function of the frequency $\omega$. The left panel illustrates the effect of varying the non--commutative parameter $\Theta$, while the right panel shows the influence of the rotational parameter $a$.}
    \label{nb}
\end{figure}

In Fig. \ref{nb}, the bosonic particle creation rate, $n_{b}(\omega,\Theta, a, M)$, is plotted as a function of the frequency $\omega$. The left panel displays how different values of the non--commutative parameter $\Theta$ affect the rate, whereas the right panel highlights the impact of varying the rotational parameter $a$. In general lines, for a fixed rotational parameter at $a = 0.1$, an increase in the non--commutative parameter $\Theta$ results in an increment in the magnitude of the particle density. Conversely, when $\Theta$ remains constant ($\Theta = 0.1$), varying the rotational parameter $a$ leads to an increase in $n_{b}(\omega,\Theta, a, M)$.

\subsection{Fermionic particle modes}

The emission of Dirac particles through the tunneling formalism is now investigated at the outer event horizon of a Kerr black hole. For simplicity, the focus is restricted to a massless spinor field $\Psi$ governed by the covariant Dirac equation
\ie
-i\hbar\gamma^ae_a^\mu\nabla_\mu\Psi=0 ,
\fe
where $\nabla_\mu$ represents the spinor covariant derivative, expressed as $\nabla_\mu = \partial_\mu + \frac{1}{4}\omega_\mu^{ab}\gamma_{[a}\gamma_{b]}$. Here, $\omega_\mu^{ab}$ denotes the spin connection, which can be written in terms of the tetrad components $e_a^\mu$. The $\gamma$ matrices are defined as
\ie
 \gamma^0=\left(%
 \begin{array}{cc}
  i & 0 \\
  0 & -i \\
 \end{array}\right),\,
 \gamma^1=\left(%
 \begin{array}{cc}
  0 & \sigma^3 \\
  \sigma^3 & 0 \\
 \end{array}\right),\,
 \gamma^2=\left(%
 \begin{array}{cc}
  0 & \sigma^1 \\
  \sigma^1 & 0 \\
 \end{array}%
 \right),\,
 \gamma^3=\left(%
 \begin{array}{cc}
  0 & \sigma^2 \\
  \sigma^2 & 0 \\
 \end{array}%
 \right),
\fe
where the matrices $\sigma^k$ $(k=1,2,3)$ represent the Pauli matrices and the tetrad fields $e_a^\mu$ can be chosen as
\ie
\nonumber
e_0^\mu=\left(\sqrt{-g^{tt}}, 0, 0, \frac{-g^{t\phi}}{\sqrt{-g^{tt}}}\right), \,\,\, e_1^\mu=\left(0, \sqrt{\frac{\Delta(r)}{\Sigma}}, 0, 0\right), 
\fe
\ie
\nonumber
e_2^\mu=\left(0, 0, \frac{1}{\sqrt{\Sigma}}, 0\right), \,\,\, e_3^\mu=\left(0, 0, 0, \frac{1}{\sqrt{g_{\phi\phi}}}\right).
\fe

The spin--up spinor field $\Psi$ is assumed using the following ansatz \cite{kerner2008fermions,di2008fermion}:
\begin{eqnarray}
\Psi=\left(%
\begin{array}{c}
A(t,r,\theta,\phi) \\
0\\
B(t,r,\theta,\phi) \\
0
\end{array}%
\right)\textrm{exp}\left[\frac{i}{\hbar}I(t,r,\theta,\phi)\right] .
\end{eqnarray}

The focus will be restricted to the spin--up case, as the spin--down scenario can be treated in an analogous manner. To implement the WKB approximation, the assumed form of the spinor field $\Psi$ is substituted into the covariant Dirac equation. By isolating the exponential term and omitting terms proportional to $\hbar$, the resulting system simplifies to the following four equations
\begin{eqnarray}
\left\{
\begin{array}{ll}
iA\left(\sqrt{-g^{tt}}\partial_t I-\frac{g^{t\phi}}{\sqrt{-g^{tt}}}\partial_\phi I\right)
+B\sqrt{\frac{\Delta}{\Sigma}}\partial_r I =0,  \\
\left( \frac{1}{\sqrt{\Sigma}}\partial_\theta I
+i\frac{1}{\sqrt{g_{\phi\phi}}}\partial_\phi I\right)B=0,  \\
A\sqrt{\frac{\Delta}{\Sigma}}\partial_r I
-iB\left( \sqrt{-g^{tt}}\partial_t I-\frac{g^{t\phi}}{\sqrt{-g^{tt}}}\partial_\phi I\right)=0,  \\
\left( \frac{1}{\sqrt{\Sigma}}\partial_\theta I
+i\frac{1}{\sqrt{g_{\phi\phi}}}\partial_\phi I\right)A = 0.
\end{array}
\right.
\end{eqnarray}

It should be emphasized that while $A$ and $B$ vary, their derivatives, as well as the components $\omega_\mu$, are proportional to $\hbar$ and can be disregarded in the leading--order WKB approximation. As the analysis is confined to the region outside the event horizon, the condition $\Delta \geq 0$ holds throughout the equations derived above. Moreover, the second and fourth equations imply that
\begin{eqnarray}
\frac{1}{\sqrt{\Sigma}}\partial_\theta I
+i\frac{1}{\sqrt{g_{\phi\phi}}}\partial_\phi I=0.
\end{eqnarray}

The first and third equations reveal that a non-trivial solution for $A$ and $B$ can only arise when the determinant of the coefficient matrix vanishes. This requirement results in the following expression
\ie
\left( \sqrt{-g^{tt}}\partial_t I-\frac{g^{t\phi}}{\sqrt{-g^{tt}}}\partial_\phi I\right)^2
-\frac{\Delta}{\Sigma}(\partial_r I)^2=0.
\fe

Since the Kerr spacetime possesses two Killing vectors, $(\partial/\partial t)^{\mu}$ and $(\partial/\partial \phi)^{\mu}$, the classical action $I(t, r, \theta, \phi)$ can be decomposed into separable variables
\ie
I = - \omega t + J\phi + R(r,\theta) + K,
\fe
where $\omega$ denotes the energy and $J$ the angular momentum of the Dirac particle, with $K$ being a complex--valued constant. In this manner,
\ie
\left(\sqrt{-g^{tt}}\omega+\frac{g^{t\phi}}{\sqrt{-g^{tt}}}J\right)^2
-\frac{\Delta}{\Sigma}(\partial_r R)^2=0.
\fe

Observe that $R(r,\theta)$ is a complex--valued function. By specifying a particular value of $\theta_0$, the expression becomes \cite{5vanzzo,07femions}
\begin{eqnarray}
R_\pm(r,\theta_0)&=&\pm\int \mathrm{d}r\sqrt{\frac{\Sigma(\theta_0)}{\Delta}}
\left[ \sqrt{-g^{tt}(\theta_0)}\omega+\frac{g^{t\phi}(\theta_0)}{\sqrt{-g^{tt}(\theta_0)}}J\right],\nonumber\\
&=&\pm\int \frac{\mathrm{d}r}{\Delta}\left[ \omega\sqrt{(r^2+a^2)^2-\Delta
a^2\textrm{sin}^2\theta_0} - J\frac{a(r^2+a^2-\Delta)}{\sqrt{(r^2+a^2)^2-\Delta
a^2\textrm{sin}^2\theta_0}}\right].\nonumber
\end{eqnarray}

The imaginary part of $R_+$ can be obtained from the expression above. By integrating around the pole near the event horizon, the following result is derived\footnote{For a more detailed treatment, please,  see \cite{07femions,012fermions} and the references therein.}
\ie
\textrm{Im}R_\pm=\pm\frac{\pi(r^2+a^2)}{r_+-r_-}(\omega-J\Omega^{H}),
\fe
where $\Omega_H = \frac{a}{r_+^2 + a^2}$ represents the angular velocity at the event horizon. Notably, this expression shows no dependence on the coordinate $\theta$.

According to the Hamilton--Jacobi method \cite{011fermions,012fermions}, one solution describes Dirac particles escaping from the outer event horizon, while the other corresponds to particles approaching it. The probabilities for the particles crossing the outer horizon in both directions are expressed as
\begin{eqnarray}
\mathcal{P}_{out}&=&\textrm{exp}\left[-\frac{2}{\hbar}\textrm{Im}I\right]
=\textrm{exp}\left[-\frac{2}{\hbar}(\textrm{Im}R_++\textrm{Im}K)\right],\nonumber\\
\mathcal{P}_{in}&=&\textrm{exp}\left[-\frac{2}{\hbar}\textrm{Im}I\right]
=\textrm{exp}\left[-\frac{2}{\hbar}(\textrm{Im}R_-+\textrm{Im}K)\right].
\end{eqnarray}

To ensure that the probabilities are properly normalized, it is important to recognize that the likelihood of any incoming wave crossing the outer event horizon is equal to unity \cite{012fermions}. This condition establishes the relation $\textrm{Im}K = -\textrm{Im}R_-$. Furthermore, since $\textrm{Im}R_+ = -\textrm{Im}R_-$, the tunneling probability for a Dirac particle moving from inside to outside the event horizon reads
\begin{eqnarray}
\hat{\Gamma} &=&\textrm{exp}\left[-\frac{4}{\hbar}\textrm{Im}R_+\right],\nonumber\\ & = & \textrm{exp}\left[\frac{4\pi(r_+^2+a^2)}{(r_+-r_-)}(\omega-J\Omega^H)\right].
\end{eqnarray}

The fermionic spectrum of Hawking radiation for Dirac particles emitted from a Kerr black hole can be derived from the tunneling probability by applying standard methods of Refs. \cite{13fermions,14fermions}
\ie
\begin{split}
& n_{f}(\omega, \Theta, a, M) = \frac{1}{e^{2\pi(\omega-J\Omega^H)/\kappa}+1} \\
& = \frac{1}{\exp \left(\frac{\pi  \left(64 M^2-\Theta ^2\right) \left(\omega  \left(\sqrt{\left(\Theta ^2-64 M^2\right)^2-4096 a^2 M^2}-\Theta ^2+64 M^2\right)-32 a^2 M\right)}{16 M \sqrt{\left(\Theta ^2-64 M^2\right)^2-4096 a^2 M^2}}\right)+1}.
\end{split}
\fe

\begin{figure}
    \centering
     \includegraphics[scale=0.545]{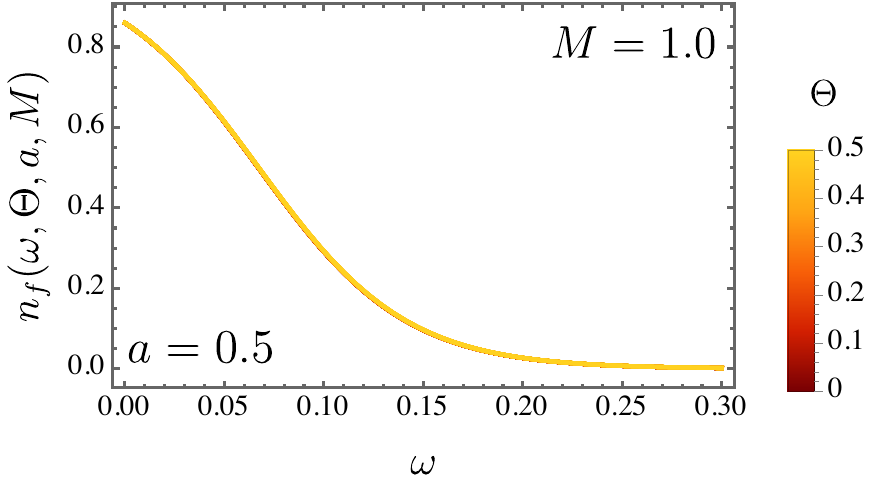}
     \includegraphics[scale=0.545]{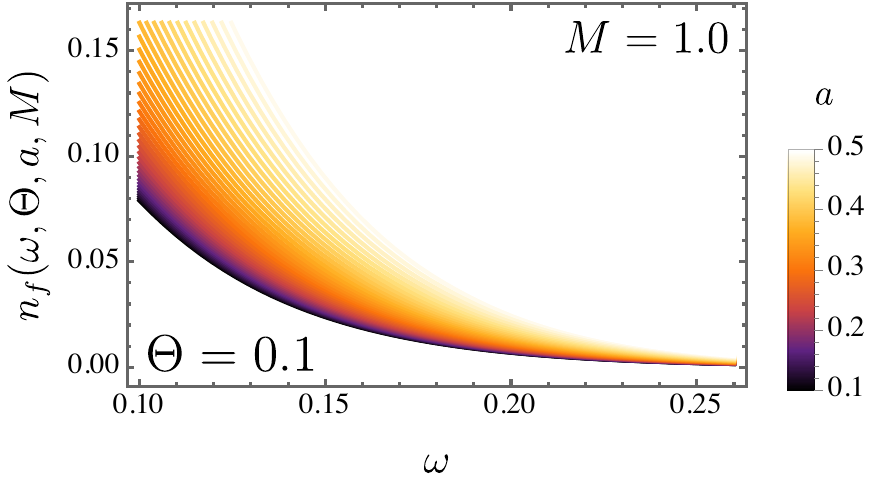}
    \caption{Behavior of the fermionic particle creation rate $n_{f}(\omega, \Theta, a, M)$ with respect to the frequency $\omega$. The left panel depicts the variation with the non--commutative parameter $\Theta$, and the right panel illustrates the influence of the rotational parameter $a$}.
    \label{nf}
\end{figure}

Furthermore, in Fig. \ref{nf}, the fermionic particle creation rate, $n_{f}(\omega,\Theta, a, M)$, is plotted as a function of the frequency $\omega$. The left panel displays how different values of the non--commutative parameter $\Theta$ affect the rate, whereas the right panel highlights the impact of varying the rotational parameter $a$. In general lines, for a fixed rotational parameter at $a = 0.1$, an increase in the non--commutative parameter $\Theta$ results in an increase of the particle density. Conversely, when $\Theta$ remains constant, varying the rotational parameter $a$ leads to an increase in $n_{f}(\Theta, a, M)$.

\section{Geodesics\label{geoss}}

This section investigates how non--commutativity affects the trajectories of particles moving along geodesics in the spacetime of a rotating black hole. The metric's axisymmetry, characterized by the Killing vectors $\partial_t$ and $\partial_\phi$, allows the focus to be narrowed to radial motion, simplifying the analysis. The equations describing the particle dynamics along these geodesics are derived from the Lagrangian formulation, following the methodology presented in \cite{Wald}, which serves as the framework for this analysis
\ie
\mathcal{L} = g_{\mu\nu}\dot{x}^{\mu}\dot{x}^{\nu}.
\fe
The parameter $\mathcal{L}$ can be set to $-1$, $0$, or $1$, representing timelike, null, and spacelike geodesics, respectively. The dot notation signifies differentiation with respect to the affine parameter $\lambda$, with the velocity components defined as $\dot{x}^{\mu} = \frac{\mathrm{d}x^{\mu}}{\mathrm{d}\lambda}$. Using this definition, the equations describing the particle motion take the following form
\ie
\begin{split}
\mathcal{L}  = & -\left[\frac{\Delta(r) -a^2\sin^2{\theta}}{\Sigma}\right]\Dot{t}^2 
   + \frac{\Sigma}{\Delta(r)}  \Dot{r}^2 + \Sigma \, \Dot{\theta}^2  \\
   &  -2a\sin^{2}\theta  \left[1 -\frac{\Delta(r) - a^{2} \sin^{2}\theta  }{\Sigma}   \right]   \Dot{t} \Dot{\phi}   +   \frac{\sin^{2}\theta}{\Sigma} \left[ (r^{2} + a^{2})^{2}  - \Delta a^{2} \sin^{2} \theta   \right]  \Dot{\phi}^2 
    \label{metricr}
    \end{split}
\fe
To simplify the analysis, we confine the particle motion to the equatorial plane by fixing $\theta = \frac{\pi}{2}$. Under this condition, the previous equation simplifies to
\ie
\begin{split}
\mathcal{L}  = & -\left[\frac{\Delta(r) -a^2}{\Sigma}\right]\Dot{t}^2 
   + \frac{\Sigma}{\Delta(r)}  \Dot{r}^2  \\
   &  -2a \left[1 -\frac{\Delta(r) - a^{2}   }{\Sigma}   \right]   \Dot{t} \Dot{\phi}   +   \frac{1}{\Sigma} \left[ (r^{2} + a^{2})^{2}  - \Delta a^{2}  \right]  \Dot{\phi}^2 
    \label{leriril}
    \end{split}
\fe
Since the system conserves two fundamental quantities, the energy $E$ and the angular momentum $L$, the equation turns out to be written
\ie
\label{energy}
E = - g_{t\mu}\dot{x}^{\mu} =  \left(\frac{\Delta(r) - a^{2}}{\Sigma}\right) \Dot{t} + 2a \left[1 -\frac{\Delta(r) - a^{2}   }{\Sigma}   \right] \Dot{\phi},
\fe
and
\ie
\label{angularmomentum}
L = g_{\phi \mu}\dot{x}^{\mu} = - 2a \left[1 -\frac{\Delta(r) - a^{2}   }{\Sigma}   \right] \Dot{t} + \frac{1}{\Sigma} \left[ (r^{2} + a^{2})^{2}  - \Delta a^{2}  \right] \Dot{\phi}.
\fe
To streamline the procedure for solving Eqs. (\ref{energy}) and (\ref{angularmomentum}), a more convenient notation is introduced, defined below:
\ie
E = A \Dot{t} + B \Dot{\phi},
\fe
and
\ie
L = - B \Dot{t} + C \Dot{\phi}, 
\fe
where the shorthand notation is defined by $A \equiv \frac{\Delta(r) - a^{2}}{\Sigma}$, $B \equiv a \left[1 - \frac{\Delta(r) - a^{2}}{\Sigma} \right]$, and $C \equiv \frac{1}{\Sigma} \left[ (r^{2} + a^{2})^{2} - \Delta a^{2} \right]$. It is important to note that
\ie
CE - BL = (AC + B^{2})\Dot{t}  = \Delta(r)\Dot{t}
\fe
and also
\ie
AL + BE= (AC + B^{2})\Dot{\phi}  = \Delta(r) \Dot{\phi}  ,
\fe
in which $ AC + B^{2} = \Delta(r)$. Thereby,
\ie
\Dot{t} = \frac{1}{\Delta(r) } \left[ \frac{1}{\Sigma} \left[ (r^{2} + a^{2})^{2}  - \Delta a^{2}  \right] E -  a \left[1 -\frac{\Delta(r) - a^{2}   }{\Sigma}   \right]L  \right],
\fe
\ie
\Dot{\phi} = \frac{1}{\Delta(r) }  \left[   \left( \frac{\Delta(r) - a^{2}}{\Sigma}   \right) L +  a \left[1 -\frac{\Delta(r) - a^{2}   }{\Sigma}   \right]   E  \right].
\fe
The next step involves deriving the expression for the radial component of the four-velocity, rewritten in terms of the parameters $A$, $B$, and $C$ defined earlier:
\ie
\begin{split}
g_{\mu\nu}\dot{x}^{\mu}\dot{x}^{\nu} & = \mathcal{L} \\
= & - A \Dot{t}^{2} - 2 B \Dot{t} \Dot{\phi} + C \Dot{\phi}^{2} + D \Dot{r}^{2}\\
& = - [A \Dot{t} + B \Dot{\phi}] \Dot{t} + [-B \Dot{t} + C \Dot{\phi}] \Dot{\phi} + \frac{\Sigma}{\Delta(r)} \Dot{r}^{2}\\
& = - E \Dot{t} + L\Dot{\phi} + \frac{D}{\Delta(r)} \Dot{r}^{2},
\end{split}
\fe
where $D \equiv \Sigma$. With this definition, the radial equation can be expressed as
\ie
\begin{split}
\Dot{r}^{2} & = \frac{\Delta(r)}{D} \left( E \Dot{t} - L \Dot{\phi} + \mathcal{L} \right) \\
& = \frac{1}{D} \left[  CE^{2} - 2 BLE -AL^{2}  + \Delta(r)\mathcal{L} \right].
\end{split}
\fe
It is worth poing out that
\ie  
CE^{2} - 2 BLE -AL^{2}  + \mathcal{L} = \left( E - \mathcal{V}_{-} \right)\left( E + \mathcal{V}_{+} \right),
\fe
with $\mathcal{V}_{\pm} = \frac{\pm \sqrt{A C L^2 + B^2 L^2 - C \mathcal{L}} + B L}{C}$, which results in the following expression
\ie
\label{radialequation}
\Dot{r}^{2} = \frac{1}{D}\left[ \left( E - \mathcal{V}_{-} \right)\left( E + \mathcal{V}_{+} \right) \right].
\fe

Explicitly, $\mathcal{V}_{\pm}$ can be expressed as
\ie
\begin{split}
\mathcal{V}_{\pm} = & \frac{ \pm 4 M r \sqrt{\frac{64 M \left(L^2 r \left(a^2+r (r-2 M)\right)-\mathcal{L} \left(a^2 (2 M+r)+r^3\right)\right)+2 \Theta ^2 \left(a^2 \mathcal{L}+L^2 r^2\right)}{M r}}+a L \left(64 M^2-\Theta ^2\right)}{32 M \left(a^2 (2 M+r)+r^3\right)-a^2 \Theta ^2}.
\end{split}
\fe

\begin{figure}
    \centering
     \includegraphics[scale=0.54]{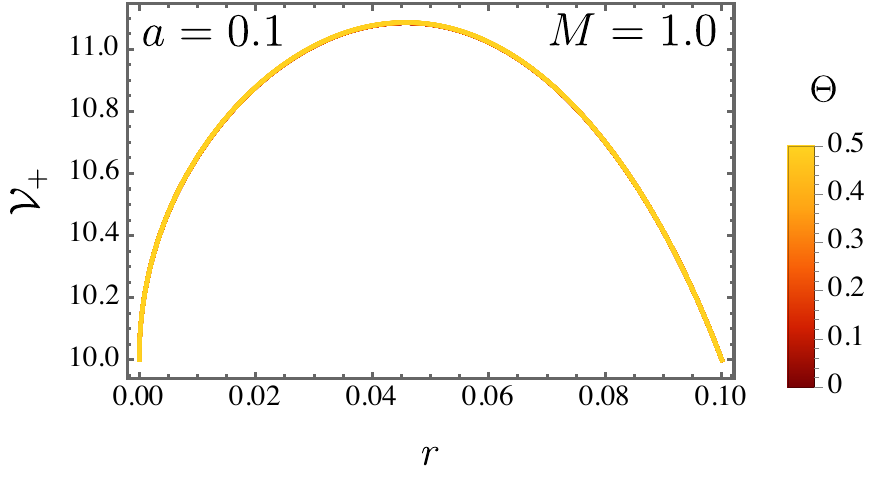}
    \includegraphics[scale=0.53]{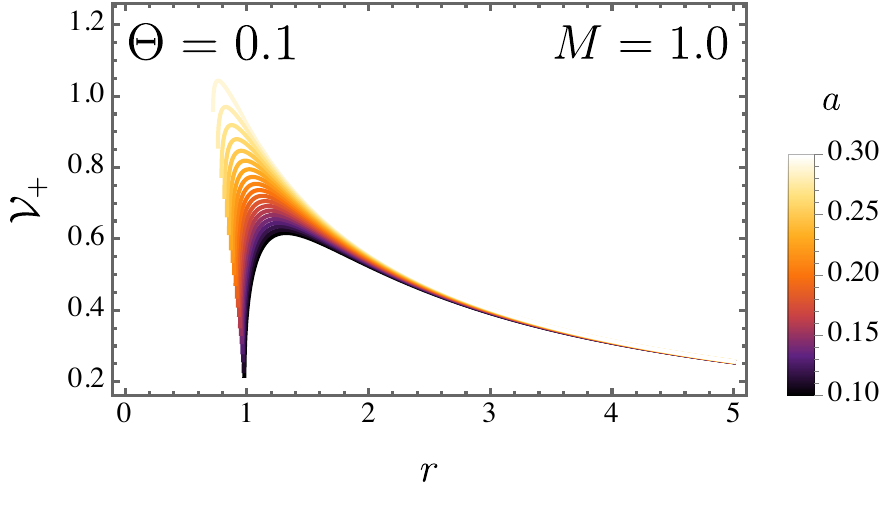}
    \caption{The potential $\mathcal{V}_{+}$ is illustrated for various combinations of the rotational parameter $a$ and the non--commutative parameter $\Theta$.}
    \label{potentialsplus}
\end{figure}

\begin{figure}
    \centering
     \includegraphics[scale=0.54]{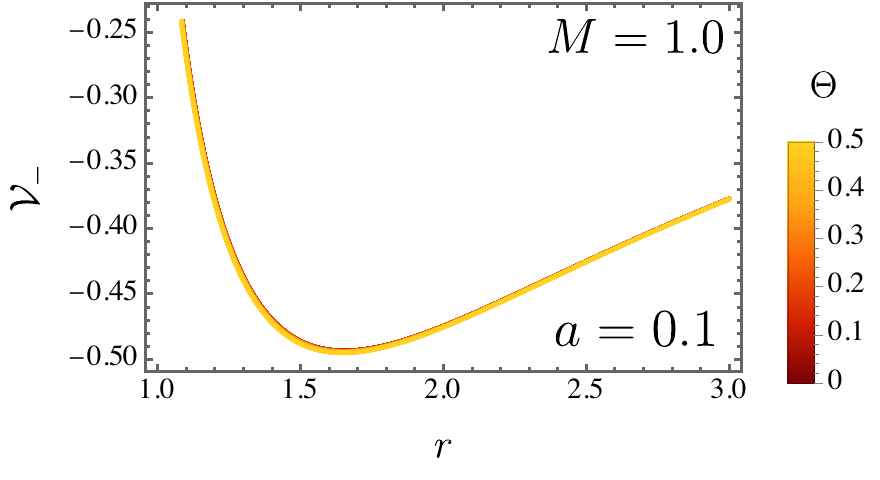}
    \includegraphics[scale=0.53]{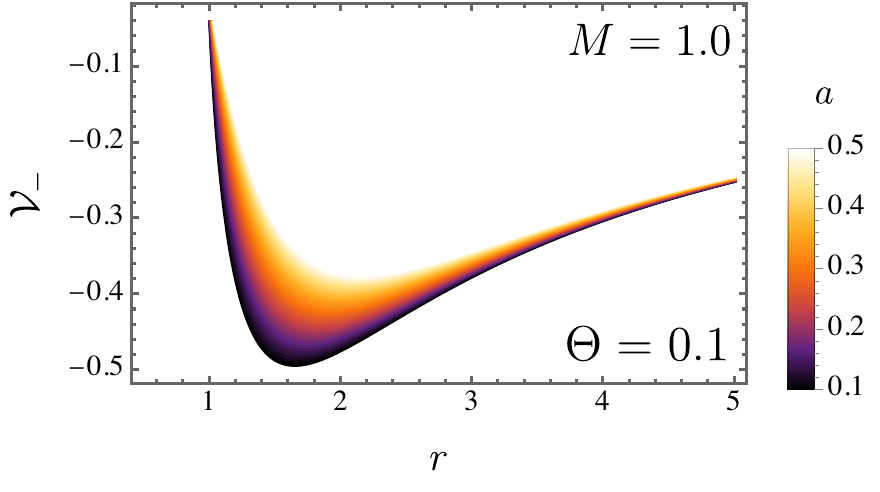}
    \caption{The potential $\mathcal{V}_{-}$ is shown for different values of the rotational parameter $a$ and the non--commutative parameter $\Theta$.}
    \label{potentialsminus}
\end{figure}

\begin{figure}
    \centering
     \includegraphics[scale=0.29]{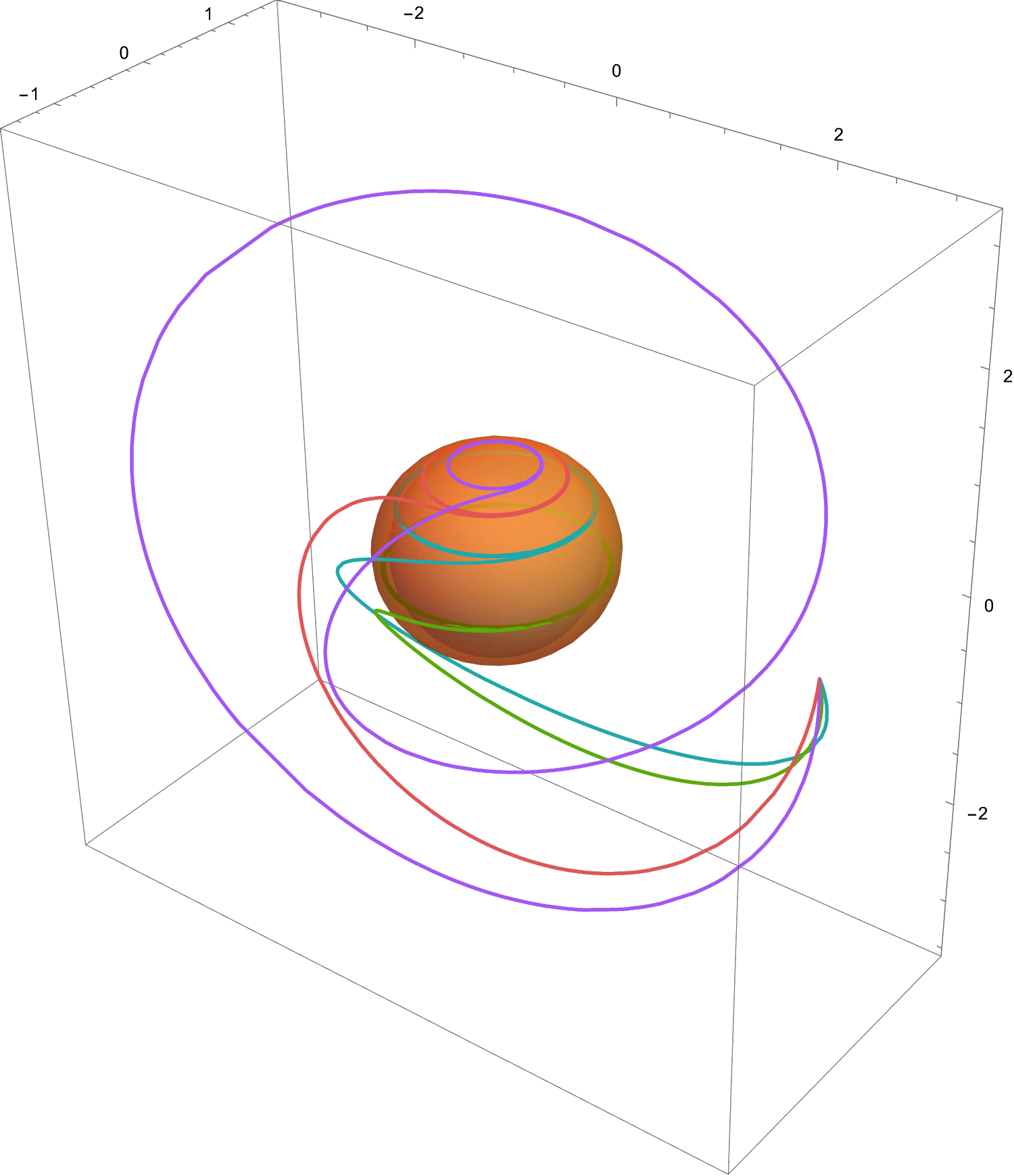}
     \includegraphics[scale=0.29]{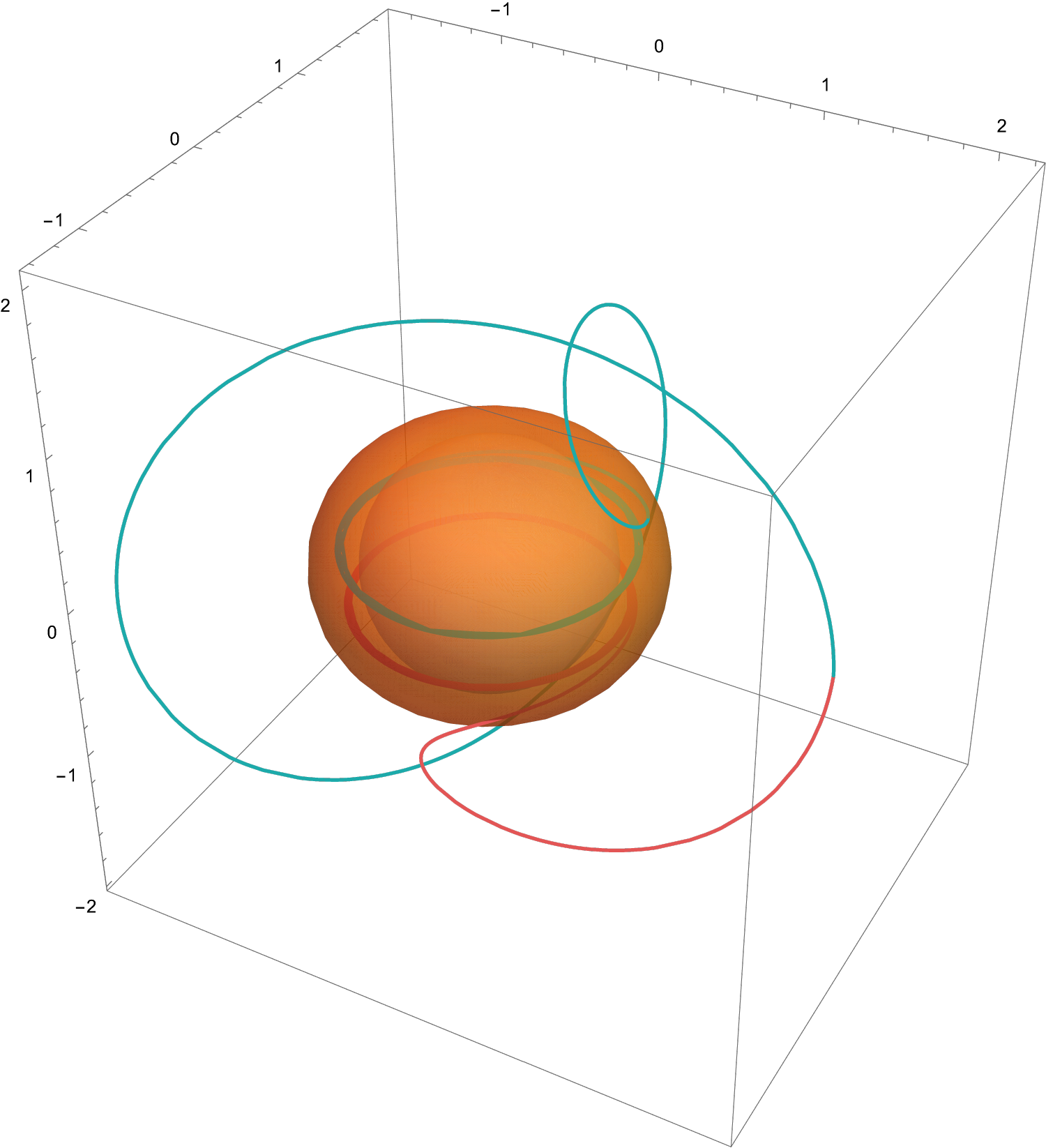}
     \includegraphics[scale=0.26]{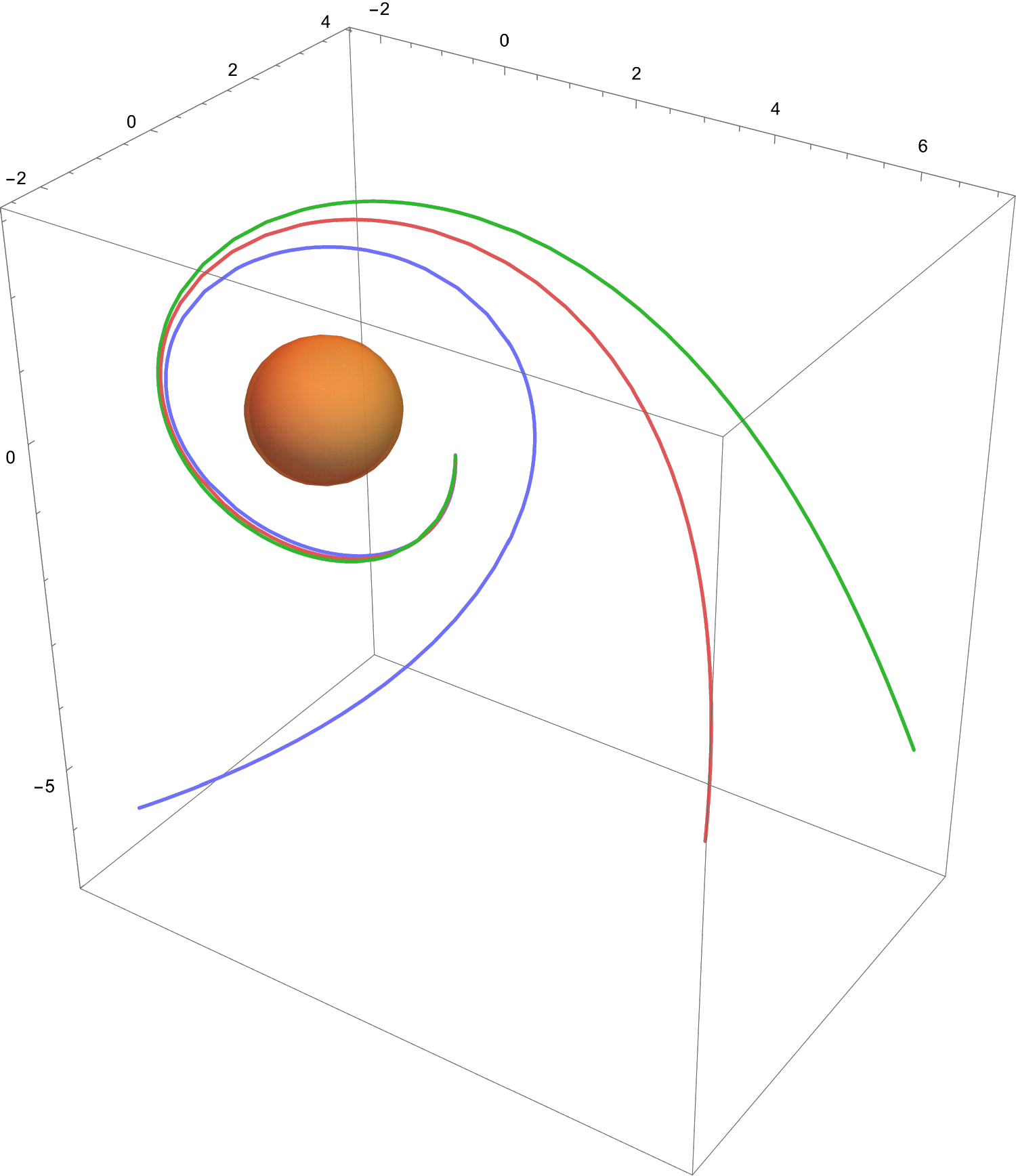}
     \includegraphics[scale=0.33]{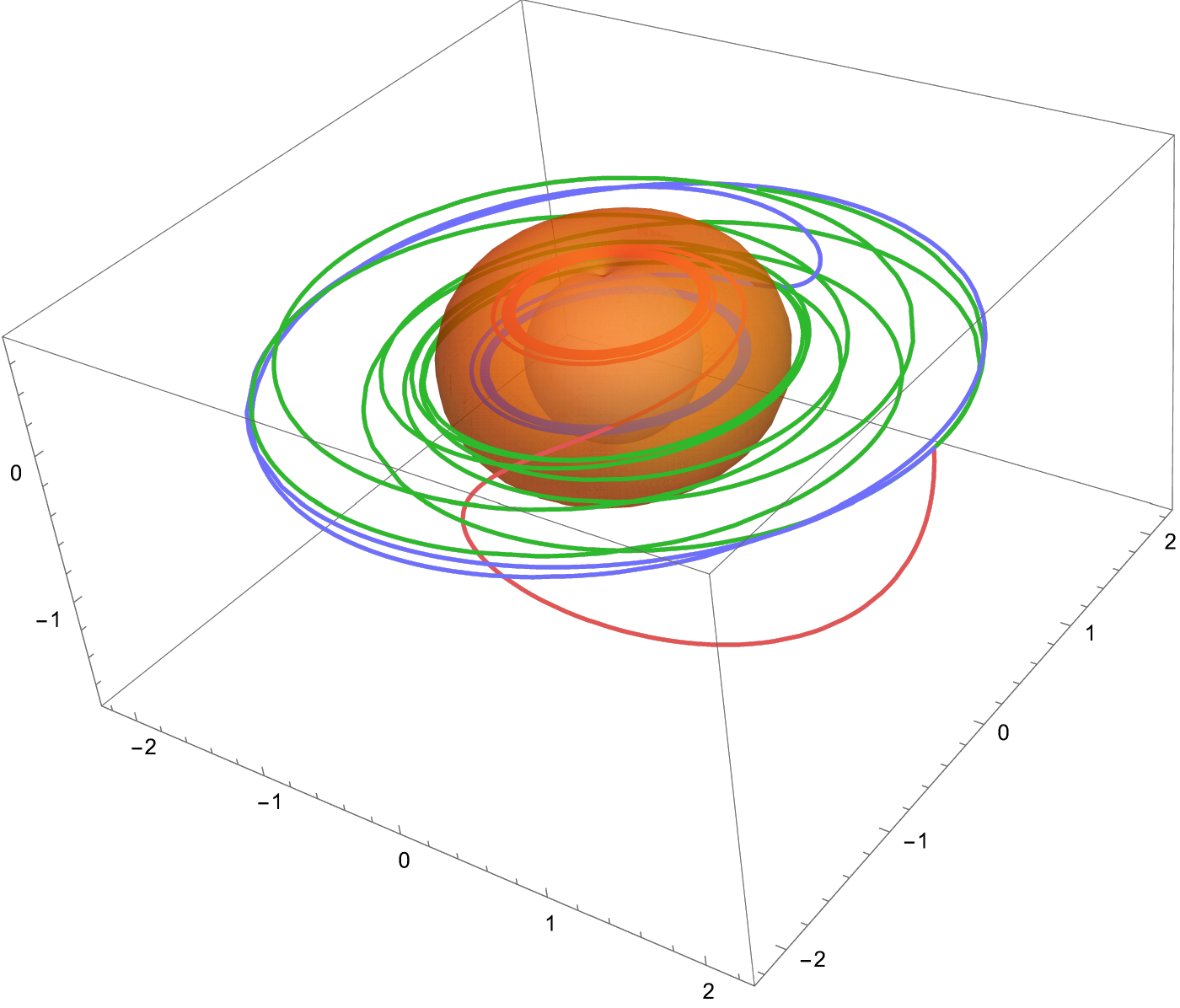}
    \caption{The geodesic trajectories are presented for the configuration with $\Theta = 0.001$.}
    \label{geodesicsfull}
\end{figure}

Figs. \ref{potentialsplus} and \ref{potentialsminus} depict the behavior of the potentials $\mathcal{V}_{\pm}$ in the timelike case, where $\mathcal{L} = -1$. In Fig. \ref{potentialsplus}, the dependence of $\mathcal{V}_{+}$ on the radial coordinate $r$ is shown. The left panel examines how the potential responds to variations in $a$ with $\Theta$ held constant, while the right panel illustrates the influence of different values of $\Theta$ for a fixed $a$. Likewise, Fig. \ref{potentialsminus} presents the radial profile of $\mathcal{V}_{-}$, using identical parameter choices for both $a$ and $\Theta$, enabling a straightforward comparison between the two potentials.  
To further clarify the geodesic motion, Fig. \ref{geodesicsfull} provides a detailed visualization of the particle trajectories.

\subsection{The radial acceleration analysis for null geodesics}

For null geodesics, the radial equation (\ref{radialequation}) simplifies to
\ie
\begin{split}
\nonumber
\Dot{r}^{2} & =  \frac{1}{D} \left( E - \mathcal{V}_{-} \right)\left( E + \mathcal{V}_{+} \right).
\end{split}
\fe
Null geodesics are physically admissible only if the condition $\dot{r}^2 \geq 0$ is satisfied. Evaluating this constraint shows that massless particles can traverse null geodesics when the inequality holds. This requirement is met when the energy constant $E$ lies beyond the bounds set by the potentials, ensuring the existence of the geodesic path. Consequently, null geodesics occur under the condition below:
\ie
\nonumber
E < \mathcal{V}_{-} \quad \text{or} \quad E > \mathcal{V}_{+}.
\fe

The region defined by $\mathcal{V}_{-} < E < \mathcal{V}_{+}$ remains forbidden for particle trajectories. To further explore the orbital behavior, examining the radial acceleration is necessary. It is achieved by taking the derivative of Eq. (\ref{radialequation}) with respect to the affine parameter $s$, resulting in
\ie
\begin{split}
2 \dot{r} \ddot{r} =  \left[ \left( \frac{1}{D} \right)^{\prime} (E- \mathcal{V}_{+})(E-\mathcal{V}_{-}) - \frac{1 }{D}\mathcal{V}_{+}^{\prime}(E-\mathcal{V}_{-})\right. \left.  - \frac{1 }{D}\mathcal{V}_{-}^{\prime}(E-\mathcal{V}_{+}) \right] \Dot{r},
\end{split}
\fe
or, in other words,
\ie
\begin{split}
\Ddot{r} =  \frac{1}{2}\left( \frac{1}{D} \right)^{\prime} (E- \mathcal{V}_{+})(E-\mathcal{V}_{-}) - \frac{1}{2 D} \left[ \mathcal{V}_{+}^{\prime}(E-\mathcal{V}_{-}) - \mathcal{V}_{-}^{\prime}(E-\mathcal{V}_{+}) \right].
\end{split}
\fe

\begin{figure}
    \centering
     \includegraphics[scale=0.54]{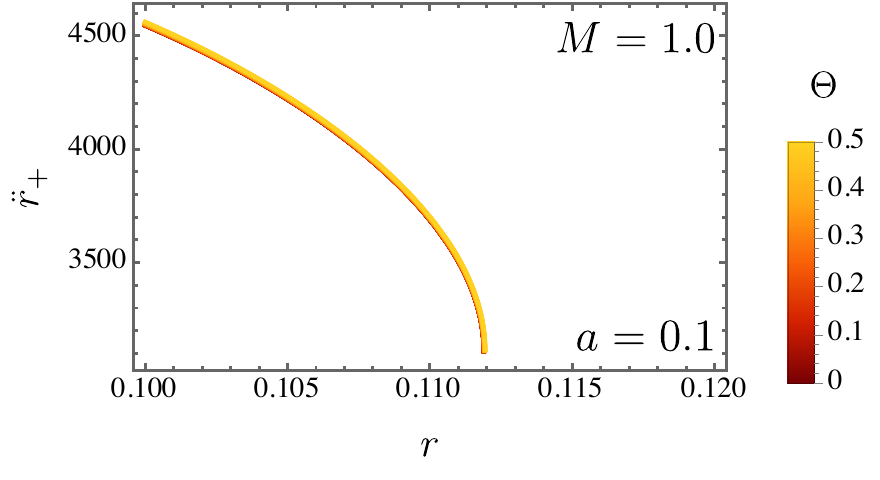}
    \includegraphics[scale=0.54]{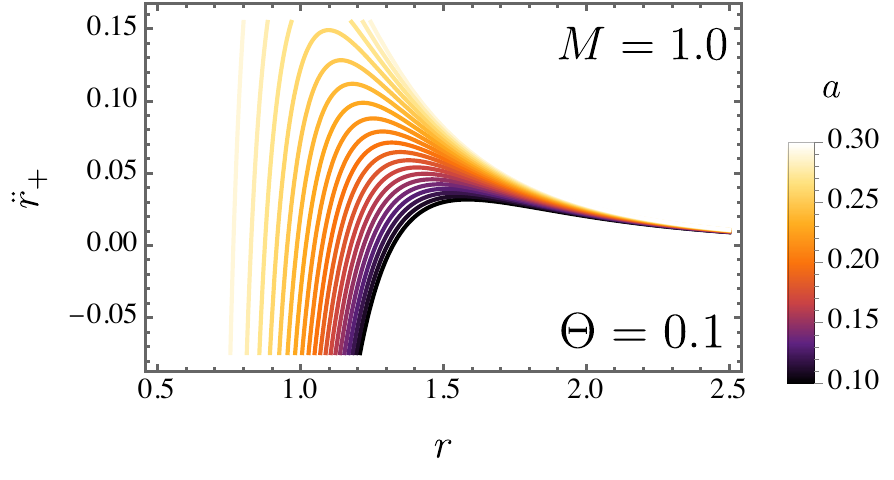}
    \caption{The radial acceleration $\ddot{r}_{+}$ is shown as a function of $r$ for various choices of the parameters $a$ and $\Theta$.}
    \label{rplusa}
\end{figure}

\begin{figure}
    \centering
     \includegraphics[scale=0.54]{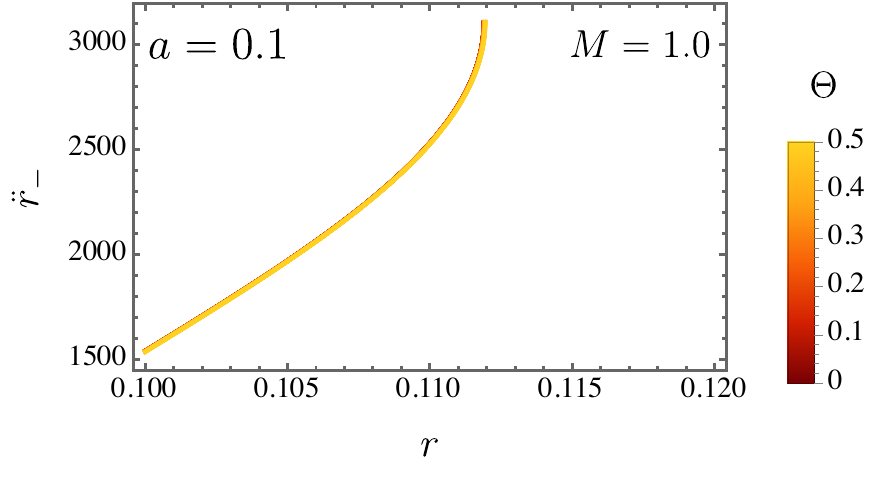}
    \includegraphics[scale=0.54]{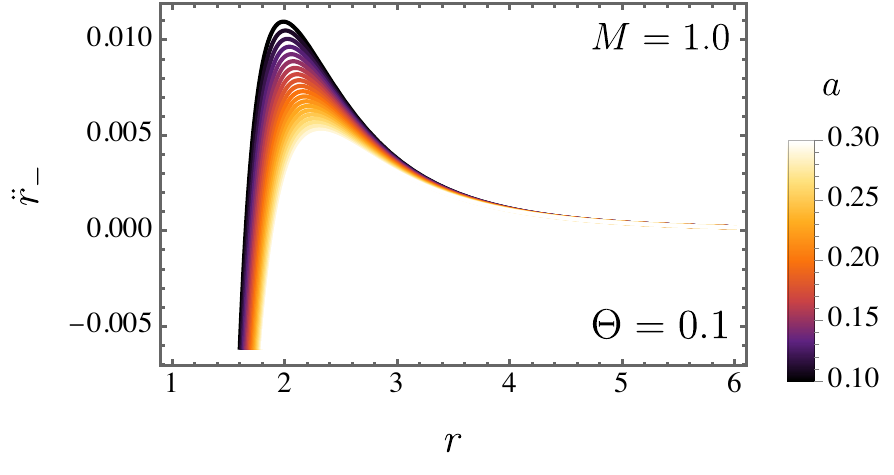}
    \caption{The radial acceleration $\ddot{r}_{-}$ is represented as a function of $r$ for multiple combinations of the parameters $a$ and $\Theta$.}
    \label{rminusa}
\end{figure}

Here, the prime symbol ($'$) represents differentiation with respect to the radial coordinate $r$. The radial acceleration is analyzed at positions where the radial velocity $\dot{r}$ equals zero, which occurs when the energy parameter $E$ coincides with one of the potentials, either $\mathcal{V}_{+}$ or $\mathcal{V}_{-}$
\ie
 \Ddot{r}_{+} = - \frac{1}{2 D}  \mathcal{V}_{+}^{\prime}(\mathcal{V}_{+} -\mathcal{V}_{-}), \,\,\,\,\,\,\text{if} \,\,\,\,\, E = \mathcal{V}_{+},
\fe
and also
\ie
 \Ddot{r}_{-} = - \frac{1}{2 D}  \mathcal{V}_{-}^{\prime}(\mathcal{V}_{-}-\mathcal{V}_{+}), \,\,\,\,\,\,\text{if} \,\,\,\,\, E = \mathcal{V}_{-},
\fe
where $r_{\pm}$ correspond to the radial accelerations. To illustrate the behavior of $\ddot{r}_{\pm}$ more clearly, Figs. \ref{rplusa} and \ref{rminusa} are included for reference.

\section{Shadows }

Photon geodesics are entirely governed by the geometry and symmetries of the spacetime. Consequently, examining null geodesics provides valuable information about the characteristics of the modified gravity theory. In this spacetime, the motion of photons is described by four constants of motion: the vanishing rest mass $m = 0$, the energy $E = -g_{t\phi}\dot{\phi} - g_{tt}\dot{t}$, and the azimuthal angular momentum $L_z = g_{\phi\phi}\dot{\phi} + g_{\phi t}\dot{t}$. Additionally, the Carter constant $Q$, arising from a hidden symmetry, allows the equations for null geodesics to be expressed in a first-order form. These equations can be derived using Carter's formalism through the Hamilton--Jacobi equation, as demonstrated as follows
\ie
\begin{split}
& \Sigma \dot{t} = \frac{r^{2}+a^{2}}{\Delta} \left[ E(r^{2} + a^{2}) - a L_{z}   \right] - a(aE\sin^{2}\theta - L_{z}), \\
& \Sigma \dot{\phi} = \frac{a}{\Delta} [E(r^{2}+a^{2}) - a L_{z}] - \left( aE - \frac{L_{z}}{\sin^{2}\theta} \right), \\
& \Sigma^{2} \dot{r}^{2} = \left[ (r^{2} + a^{2})  - a L_{z}  \right]^{2} - \Delta(r) \left[ \mathcal{K} + (aE - L_{z})^{2}  \right] \equiv \mathcal{R}(r), \\
&\Sigma^{2} \dot{\theta}^{2} = \mathcal{K} - \left(  \frac{L_{z}^{2}}{\sin^{2}\theta} -a^{2}E^{2}   \right) \cos^{2}\theta \equiv \Tilde{\Theta}(r).
\end{split}
\fe

Now, let us re--scaled energy parameters $\xi = L/E$ and $\eta = K/E^2$, referred to as critical impact parameters within the expressions for $R(r)$ and $\Tilde{\Theta}(r)$. The investigation is limited to the region beyond the event horizon, focusing on spherical photon trajectories --- null geodesics confined to specific radii $r_p$. These orbits are described by the conditions $\mathcal{R}'(r_p) = 0$ and $\mathcal{R}''(r_p) = 0$ for all $r_p > r_+$ \cite{afrim}. This formulation allows the determination of critical impact parameters for the non--commutative black hole under study, leading to the following equations:
\ie
\begin{split}
& \xi_{c} = \frac{(a^{2}+r^{2})\Delta^{\prime}(r) - 4 r \Delta(r) }{a \Delta^{\prime}(r)},\\
& \eta_{c}   = \frac{r^{2}(  8 \Delta(r) (2 a^{2} + r\Delta^{\prime} (r)   )   -r^{2} \Delta^{\prime 2}(r) - 16\Delta^{2}(r) )  }{a^{2}\Delta^{\prime 2}(r)}.
\end{split}
\fe

Spherical photon orbits are confined within a three-dimensional region called the photon shell, which defines the bright boundary of the black hole shadow. This region is described by several parameters: the time coordinate $t$, ranging from $-\infty$ to $\infty$; the radial position $r_p$ within the interval $r_{p-}$ to $r_{p+}$; the azimuthal angle $\phi$, varying between $0$ and $2\pi$; and the polar angle $\theta$, which spans from $\theta_{-}$ to $\theta_{+}$, where $\theta_{\mp} = \arccos(\mp\sqrt{\tilde{\omega}})$
\ie
\tilde{\omega} = \frac{a^{2}-\eta_{c} - \xi^{2}_{c} + \sqrt{(a^{2} - \eta_{c} - \xi_{c}^{2})^{2} + 4 a^{2}\eta_{c}  } }{2 a^{2}}.
\fe

\begin{figure}
    \centering
    \includegraphics[scale=0.8]{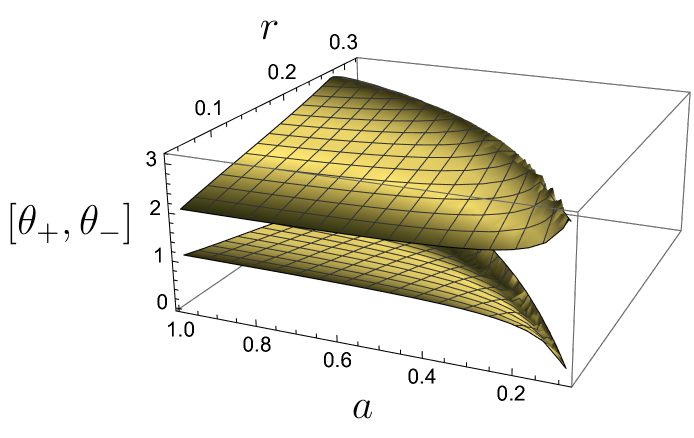}
    \caption{Parameters $\theta_{+}$ and $\theta_{-}$ are shown for different $a$ and for a fixed value of $\Theta =0.1$.}
    \label{thetabehavior}
\end{figure}

To clarify the results further, Fig. \ref{thetabehavior} displays the three--dimensional representation of $\theta_{+}$, $\theta_{-}$, and the combined range $\theta_{+} + \theta_{-}$.

The radii $r_{p\mp}$ correspond to the prograde and retrograde photon orbits, respectively, and are obtained by solving the equation $\eta_c = 0$ under the constraint $\xi_c(r_{p\mp}) \gtrless 0$. The explicit expression for $r_{p}$ is given by
\ie
\begin{split}
r_{p} = & -\frac{\left(-64 M^3+\Theta ^2 M+\sqrt[3]{\psi }\right)^2}{64 M^2 \sqrt[3]{\psi }}
\end{split},
\fe
in which 
\ie
\begin{split}
\tilde{\psi} = & 128 \sqrt{a^2 M^8 \left(\Theta ^2-64 M^2\right)^2 \left(4096 a^2 M^2+\left(\Theta ^2-64 M^2\right)^2\right)}\\
& +8192 a^2 \left(64 M^7-\Theta ^2 M^5\right)+\left(64 M^3-\Theta ^2 M\right)^3.
\end{split}
\fe

The point where photons shift from prograde to retrograde orbits is marked by the intermediate radius $r_{p0}$, derived from the condition $\xi_c = 0$. At this radius, spherical photon orbits occur with zero angular momentum, and it can be described by
\ie
\begin{split}
&r_{p 0} =   \frac{\frac{18 \left(4096 a^2 M^2+\left(\Theta ^2-64 M^2\right)^2\right)}{\sqrt[3]{\lambda^{\dagger} }}+18 \sqrt[3]{\lambda^{\dagger} }+18 \left(64 M^2-\Theta ^2\right)}{1152 M},
\end{split}
\fe
where
\ie
\lambda^{\dagger} =64 \sqrt{-a^2 M^2 \left(4096 a^2 M^2+\left(\Theta ^2-64 M^2\right)^2\right)^2}+4096 a^2 \left(64 M^4-\Theta ^2 M^2\right)+\left(64 M^2-\Theta ^2\right)^3.
\fe

Spherical photon orbits within the photon shell typically exhibit oscillatory motion along the $\theta$-direction, varying between the polar angles $\theta_{\mp}$ \cite{o100}. In contrast, at the outer and inner limits of the shell, located at $r = r_{p\mp}$, the motion becomes restricted to the equatorial plane ($\theta = \pi/2$), where the orbits reduce to planar trajectories \cite{o100, o101}.

The black hole shadow's shape is influenced by its spin, additional spacetime parameters, and the observer's viewing angle $\theta_0$ relative to the spin axis. The overall size of the shadow is determined by the black hole's mass $M$. For an observer located at $r_0 \to \infty$ and viewing from an inclination angle $\theta_0$, the shadow manifests as a dark silhouette against a luminous background, described using the following celestial coordinates
\ie
\left\{ \tilde{\alpha}, \tilde{\beta} \right\} = \left\{ - \xi_{c} \csc \theta_{0}, \pm \sqrt{\eta_{c} + a^{2} \cos^{2} \theta_{0} - \xi^{2}_{c} \cot^{2}\theta_{0}   }    \right\}.
\fe

\begin{figure}
    \centering
     \includegraphics[scale=0.56]{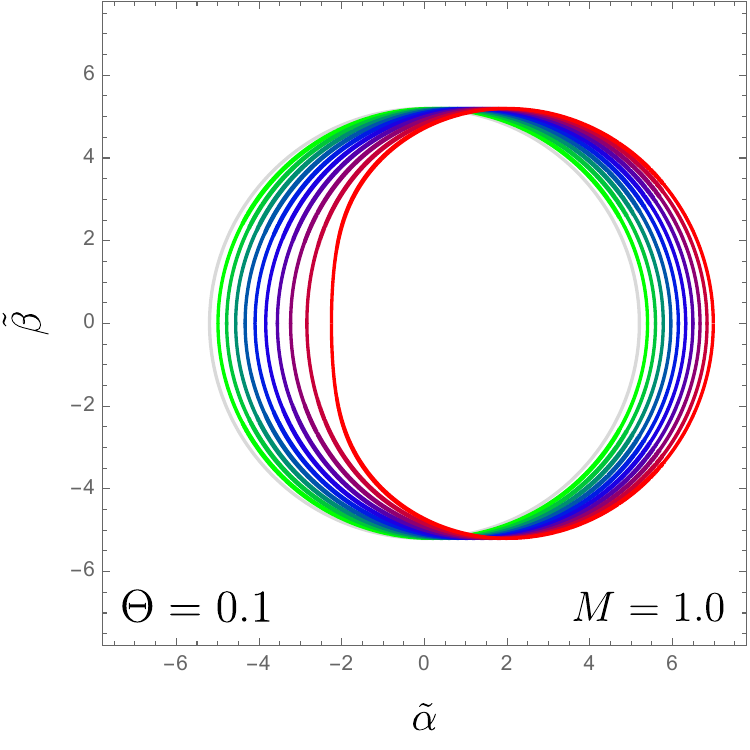}
      \includegraphics[scale=0.56]{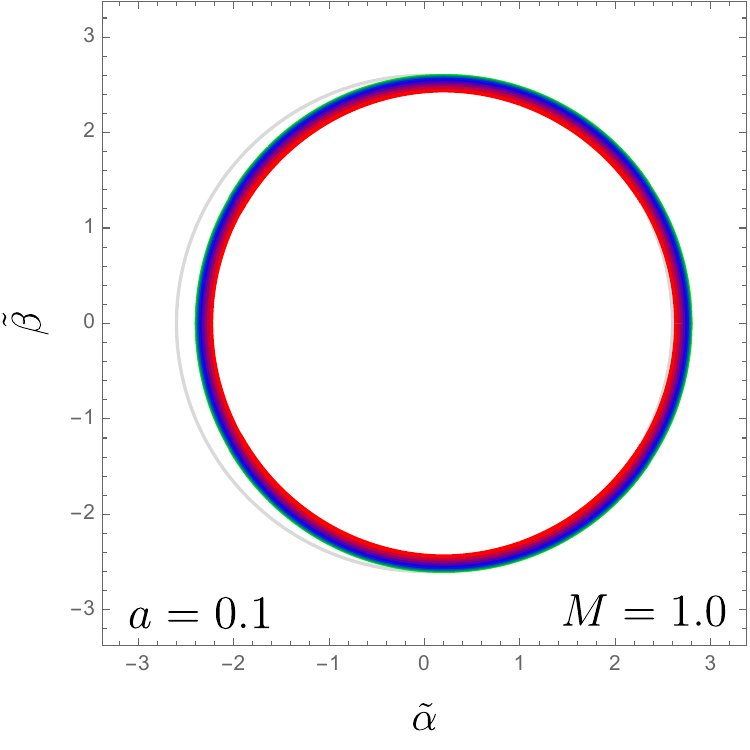}
    \caption{The shadows are presented for various configurations of the non--commutative parameter $\Theta$ and the spin parameter $a$.}
    \label{asdshas}
\end{figure}

To better illustrate the results, Fig. \ref{asdshas} shows the evolution of black hole shadows under different configurations of the spin parameter $a$ and the non--commutative parameter $\Theta$. The left panel explores the impact of varying $a$ while keeping $\Theta$ fixed. The gray curve corresponds to the shadow of a Schwarzschild black hole, serving as a reference. The colored curves represent progressively increasing values of $a$, ranging from $0.1$ to $0.99$ in steps of $0.1$, highlighting how the shadow shape becomes more distorted with higher spin.  

In the right panel, the focus shifts to the effect of varying the non--commutative parameter $\Theta$ while holding the spin parameter constant at $a = 0.1$. The values of $\Theta$ span from $0.1$ to $1.0$, increasing from left to right. Although the shadows may seem almost circular at first glance, they are in fact slightly elliptical. This subtle deviation is particularly noticeable in the red curve, where the difference from the gray reference circle becomes clearer, confirming that the colored curves represent ellipses rather than perfect circles.

\section{Shadow constraints from EHT observation}

The groundbreaking observation of the $M87^*$ black hole shadow by the Event Horizon Telescope (EHT) collaboration \cite{collaboration2019firstL1,akiyama2019firstL6} has opened a new era for probing strong-field gravity. To quantitatively compare theoretical black hole models with these observations, it is essential to define robust observables that characterize the shadow's morphology. Following the approach in Ref. \cite{kumar2020rotating,afrin2022testing}, we employ the shadow area $\mathcal{A}$ as a primary quantifier.

The area of the shadow is calculated via the integral
\begin{equation}
\mathcal{A} = 2 \int Y(r_p)  \mathrm{d}X(r_p) = 2 \int_{r_p^-}^{r_p^+} Y(r_p) \frac{\mathrm{d}X}{\mathrm{d}r_p}  \mathrm{d} r_p,
\end{equation}
where $X$ and $Y$ are the celestial coordinates on the observer's sky, and $r_p$ denotes the photon orbit radius. A derived and observationally pertinent quantity is the angular diameter $\theta_d$ of the shadow. For a characteristic radius defined as $R_a = \sqrt{A/\pi}$, the angular diameter is given by
\begin{equation}
\theta_d = 2 \frac{R_a}{d},
\end{equation}
where $d$ is the distance to the black hole. This angular size depends on the black hole's fundamental parameters: its mass $M$, specific angular momentum $a$, the inclination angle $\theta_0$, and, in our case, the non--commutative parameter $\Theta$. We apply this framework to a non--commutative rotating black hole. For $M87^*$, the EHT collaboration determined a mass of $M = 6.5 \times 10^9 M_\odot$ and a distance of $d = 16.8$ Mpc \cite{akiyama2019firstL6}, reporting an angular diameter for the emission region of $\theta_d = 42 \pm 3~\mu\text{as}$ \cite{AkiyamaL4,akiyama2019firstL6}. We compute the theoretical shadow diameter $\theta_d$ for our model as a function of the spin and the non--commutativity parameter in the mass unit as $a/M$ and $\Theta/M$, respectively.

\begin{figure}
    \centering
     \includegraphics[width=90mm]{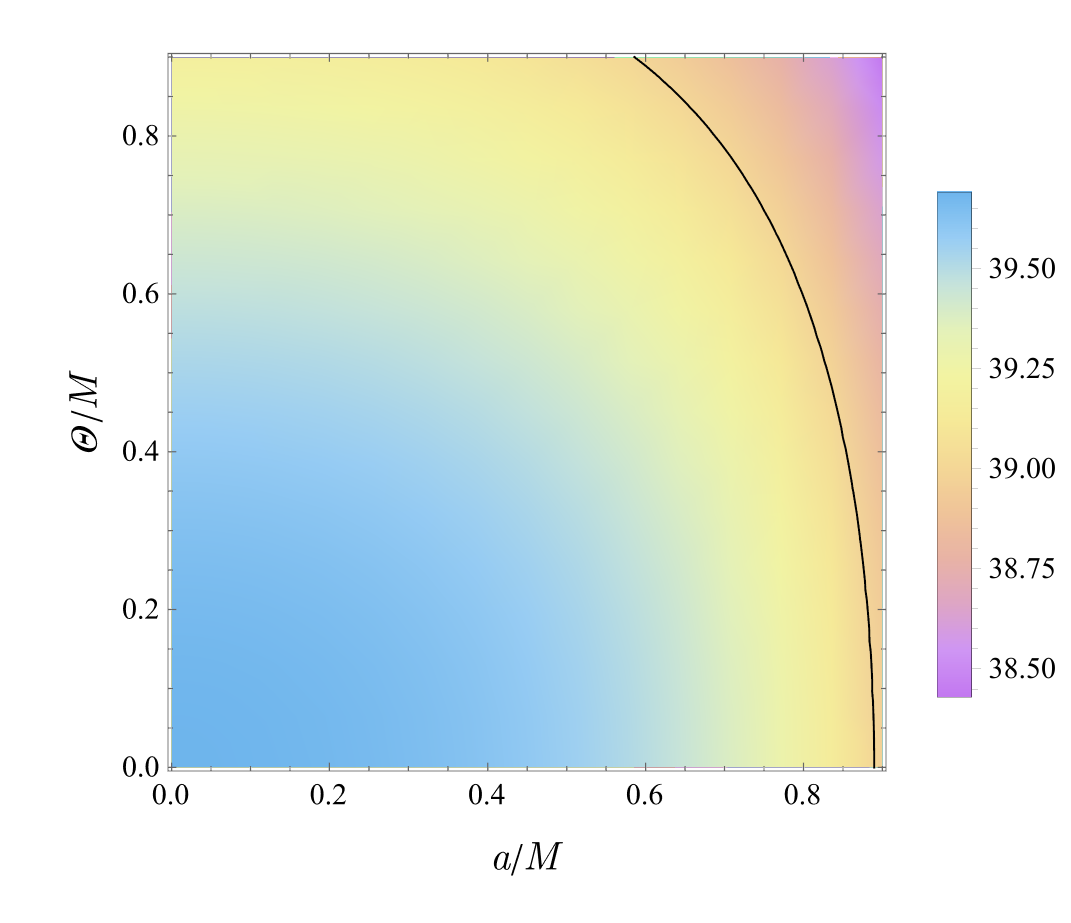}
    \caption{The angular diameter of the shadow $\theta_d$ ($\mu$as) for a modified Kerr black hole with additional non--commutative parameter $\Theta/M$, plotted against the dimensionless spin $a/M$. The black solid line at 39 $\mu$as marks the observed value for M87$^*$ within the EHT's range (42 $\pm$ 3 $\mu$as). Theoretical predictions assume $M = 6.5 \times 10^9 M_\odot$, $d = 16.8$ Mpc, and edge-on inclination ($\theta_0 = 90^\circ$).}
    \label{fig:ConsM87}
\end{figure}

Fig. \ref{fig:ConsM87} shows the result for the inclination angle $\theta_0 = 90^\circ$. The observational constraint from M87$^*$ is indicated by the black solid line, which corresponds to the lower edge of the 1$\sigma$ confidence interval (39 $\mu$as) from the EHT's reported diameter of $42 \pm 3$ $\mu$as. The region in the normalized $\Theta$-$a$ parameter space enclosed by this curve represents the set of black hole configurations whose predicted shadow size is consistent with the $M87^*$ observations. In other words, this result carries an important implication: the spin of $M87^*$ cannot be determined independently without a concurrent constraint on the non–commutative parameter $\Theta$. Consequently, future observations capable of independently measuring the spin could provide a means to test quantum–gravity–inspired modifications to the Kerr solution.

\section{Gravitational lensing in the strong deflection regime}

Let us start with the Lagrangian written below
\ie
\mathcal{L}=\frac{1}{2}g^{\Theta}_{\mu\nu}u^\mu u^\nu.
\fe
A particle’s motion can be described using the $4$--velocity $u^\mu = \mathrm{d}^\mu / \mathrm{d}\lambda$, where $\lambda$ serves as an affine parameter. Due to the stationarity and axisymmetry of our metric displayed in Eq. (\ref{rotatingmetric}), its invariance under translations in $t$ and $\phi$ results in the presence of two Killing vectors, $\xi_{(t)}^\mu$ and $\xi_{(\phi)}^\mu$ as
\ie
\xi_{(t)}^\mu=\delta_t^\mu \; , \quad \xi_{\phi}^\mu=\delta_\phi^\mu.
\fe

The conserved quantities associated with a geodesic can be determined using the Killing vectors in conjunction with the $4$--velocity of light rays. These quantities are expressed as  $\varepsilon = -\xi_{(t)}^\mu u_\mu, \quad \ell = \xi_{(\phi)}^\mu u_\mu$,
where $\varepsilon$ represents the energy of the light ray, and $\ell$ corresponds to its azimuthal angular momentum as measured at spatial infinity. Since light rays follow null geodesics, their $4$--velocity satisfies the constraint $u^\mu u_\mu = 0$. To distinguish whether the motion aligns with or opposes the frame--dragging effect, the impact parameter is introduced as follows:
\ie
b_s=s\left|\frac{\ell}{\varepsilon}\right|\equiv s\,b.
\fe
To characterize the direction of a light ray’s motion relative to the frame-dragging effect, the parameter $s = \text{Sign}(\ell / \varepsilon)$ is introduced, where $b$ represents its absolute value. When $s = +1$ and $b_s > 0$, the trajectory follows a direct orbit, whereas $s = -1$ with $b_s < 0$ corresponds to a retrograde path, following a specific sign convention. For simplicity, attention is restricted to light rays propagating within the equatorial plane of the black hole, imposing the conditions $\theta = \pi / 2$ and $\dot{\theta} = 0$. Under this assumption, the radial equation of motion can be rewritten in the form presented in \cite{lensingmain,lensing23}
\ie
\frac{1}{b^2}  =\frac{\dot{r}^2}{\ell^2} + \tilde{\mathcal{V}}^{\dagger}(r),
\fe
and based on this formulation, the function $\tilde{\mathcal{V}}^{\dagger}$ is introduced as
\ie
\tilde{\mathcal{V}}^{\dagger}(r)=\frac{1}{r^2}\left[1-\frac{a^2}{b^2}-\frac{2M_{\Theta} }{r}\left(1-\frac{a}{b_s}\right)^2\right]. 
\fe

The equation governing the radial motion resembles that of a particle moving under an effective potential $\tilde{\mathcal{V}}^{\dagger}(r)$, where the term ${\dot r}^2 / \ell^2$ plays the role of kinetic energy, and the total energy remains fixed at $1/b^2$. Consider a scenario in which a light ray originates from a distant region, moves toward the black hole, and then returns to the asymptotic region before reaching an observer. Such trajectories exhibit a turning point, corresponding to the closest approach distance $r_0$ to the black hole. This critical distance is dictated by the impact parameter $b$, which is determined through the condition:
\ie
\left.\frac{\dot{r}^2}{\ell^2}\right|_{r=r_0}=\frac{1}{b^2}-\tilde{\mathcal{V}}^{\dagger}(r_0)=0. \label{vdager}
\fe

Using Eq. (\ref{vdager}), as also demonstrated in \cite{lensingmain,lensing22,lensing23}, the impact parameter $b$ can be expressed in terms of the closest approach distance $r_0$. This relation serves as a key input for deriving analytical expressions for the deflection angle within the strong deflection limit (SDL), given by:
\ie \label{bbrr}
b(r_0)=\frac{2sM_{\Theta}a -r_0 \sqrt{a^2-2r_0 M_{\Theta}+r_0^2}}{2M_{\Theta}-r_0}.
\fe

The nature of light ray trajectories is dictated by the relationship between $1/b^2$ and the peak value of the effective potential $\tilde{\mathcal{V}}^{\dagger}(r)$. The innermost paths followed by light rays play a fundamental role in shaping the observed silhouette of the black hole. The smallest possible radius, denoted as $r_{sc}$, corresponds to the turning point $r_0$ being precisely at the maximum of $\tilde{\mathcal{V}}^{\dagger}(r)$. At this critical location, the impact parameter takes the value $b_{sc}$, satisfying the condition:
\ie\label{dfg}
\left.\frac{\mathrm{d}\,\tilde{\mathcal{V}}^{\dagger}(r)}{\mathrm{d}r}\right|_{r=r_{sc} }  = 0.
\fe

The radius at which light follows a circular trajectory, defining the photon sphere, is determined by the condition at which the turning point coincides with the maximum of the effective potential. This characteristic radius is given by (see \cite{lensing22,lensing23,lensingmain})
\ie \label{xcv}
r_{sc}=  2 M_{\Theta}\bigg\{ 1+ \cos\bigg[\frac{2}{3} \cos^{-1} \bigg( \frac{-sa}{M_{\Theta}} \bigg) \bigg]\bigg\}
\fe
where the impact parameter reads
\ie \label{asdasd}
b_{sc}  =-a+ s6 M_{\Theta} \cos\bigg[\frac{1}{3} \cos^{-1} \bigg( \frac{-sa}{M_{\Theta}} \bigg) \bigg] .
\fe

For our rotating non--commutative solution, the black hole’s spin enhances the repulsive effects on light rays in direct orbits more than in retrograde ones due to the $1/r^3$ term in the effective potential. This repulsion prevents direct orbit trajectories from collapsing into the event horizon, shifting the innermost circular paths inward. Consequently, the critical impact parameter in direct orbits, $b_{+c}$, is smaller than $b_{-c}$ for retrograde ones. As the spin parameter $a$ increases, the critical impact parameter $b_{+c}$ decreases, while $\vert b_{-c} \vert$ grows instead \cite{lensing22,lensing23,lensingmain}. Additionally, the black hole’s rotation reduces the deflection angle for direct orbits compared to retrograde ones when both share the same impact parameter $b$ \cite{lensing22,lensing23,lensingmain}.

Next, let us introduce a new variable
\ie
z \equiv 1-\frac{r_0}{r}.
\fe
The geodesic equations governing $r$ and $\phi$, as derived in \cite{lensing23,lensingmain}, can be reformulated using the variable $z$, following the approach in \cite{tsukamoto2017deflection}
\ie\label{cvcvcv}
\frac{\mathrm{d}z}{\mathrm{d}\phi}=\frac{1}{r_0}\frac{1-\frac{2{M_{\Theta}}}{r_0}(1-z)+\frac{a^2}{r_0^2}(1-z)^2}
                  {1-\frac{2{M_{\Theta}}}{r_0}(1-z)(1-\frac{a}{b_s})}\sqrt{B(z,r_0)}.
\fe
Here, the function $B(z, r_0)$ takes the form of a trinomial expression in terms of $z$.
\ie
B(z,r_0)=c_1(r_0)z+c_2(r_0)z^2+c_3(r_0)z^3,
\fe
where
\ie
\begin{split} \label{kkkk}
c_1(r_0)=&-6M_{\Theta}r_0\left(1-\frac{a}{b_s}\right)^2+2r_0^2\left(1-\frac{a^2}{b^2}\right),\\
c_2(r_0)=&6M_{\Theta}r_0\left(1-\frac{a}{b_s}\right)^2-r_0^2\left(1-\frac{a^2}{b^2}\right) ,\\
c_3(r_0)=&-2M_{\Theta}r_0\left(1-\frac{a}{b_s}\right)^2 .
\end{split}
\fe

After that, we consider
\ie
\frac{1-\frac{2{M_{\Theta}}}{r_0}(1-\frac{a}{b_s})+\frac{2{M_{\Theta}}}{r_0}(1-\frac{a}{b_s})z}
{ 1-\frac{2M_{\Theta}}{r_0}+\frac{a^2}{r_0^2}+(\frac{2{M_{\Theta}}}{r_0}-\frac{2a^2}{r_0^2})z+\frac{a^2}{r_0^2}z^2}
=\frac{r_0^2}{a^2}\left(\frac{C_-}{z-z_-}+\frac{C_+}{z-z_+}\right).
\fe
Notice that the roots are represented by $z_-$, $z_+$, and the coefficients $C_-$, $C_+$ are \cite{lensingmain}
\ie
\begin{split} \label{zkzk}
z_-=&1-\frac{r_0r_-}{a^2} ,\\
z_+=&1-\frac{r_0r_+}{a^2} ,
\end{split}
\fe
\ie\label{ccpp}
\begin{split}
C_-=&\frac{a^2-2M_{\Theta}r_-(1-\frac{a}{b_s})}{2r_0 \sqrt{M_{\Theta}^2-a^2}} ,\\
C_+=&\frac{-a^2+2M_{\Theta}r_+(1-\frac{a}{b_s})}{2r_0 \sqrt{M_{\Theta}^2-a^2}}
\end{split}
\fe
with $r_+$ and $r_-$ represent the outer and inner horizons of our black hole solution, as outlined in Sec. \ref{generalfeatures}. Notably, for any value of the spin parameter $a$, the roots satisfy $z_-, z_+ \leq 0$. Using Eq. (\ref{cvcvcv}), the deflection angle can then be expressed as a function of the closest approach distance $r_0$, yielding:
\ie
\hat\alpha(r_0) = I (r_0) -\pi \, , \quad I(r_0)= \int_0^1 f(z,r_0) \mathrm{d}z ,
\fe
so that the integrand turns out to be
\ie
 f(z,r_0)= \frac{r_0^2}{a^2}\left(\frac{C_-}{z-z_-}+\frac{C_+}{z-z_+}\right)\frac{2{r_0}}{\sqrt{c_1(r_0)z+c_2(r_0)z^2+c_3(r_0)z^3}} .
\fe

In SDL under consideration, as the closest approach distance approaches its critical value, i.e., $r_0 \to r_{sc}$, the coefficient $c_1(r_0)$ in (\ref{kkkk}), derived from (\ref{dfg}), vanishes. Consequently, for small $z$, the integrand $f(z, r_0)$ behaves as $\frac{1}{z}$, leading to a logarithmic divergence in the deflection angle as $r_0 \to r_{sc}$. To handle this behavior, we introduce a new function, $f_D(z, r_0)$
\ie
f_D(z,r_0) = \frac{r_0^2}{a^2}\left(\frac{C_-}{z-z_-}+\frac{C_+}{z-z_+}\right)\frac{2{r_0}}{\sqrt{c_1(r_0)z+c_2(r_0)z^2}}.
\fe
Notice that this function $f_D(z, r_0)$ isolates the divergent contribution, allowing the remaining part to be expressed as $f_R(z, r_0) = f(z, r_0) - f_D(z, r_0)$. Since $f_R(z, r_0)$ does not exhibit any singularity, its integral remains finite.

The divergence arises from the integral of the function $f_D(z, r_0)$, which influences both the coefficient $\bar{a}$ in the logarithmic term and $\bar{b}$ in the regular part, leading to
\ie \label{isd}
\begin{split}
I_D(r_0)=&\int_0^1 f_D(z,r_0) \mathrm{d}z\\
=&\frac{2r_0^3}{a^2}\frac{C_-}{\sqrt{c_1(r_0)z_-+c_2(r_0)z_-^2}}
\ln{\left(\frac{\sqrt{c_1(r_0)z_-+c_2(r_0)z_-}+\sqrt{c_1(r_0)+c_2(r_0)z_-}}
{\sqrt{c_1(r_0)z_-+c_2(r_0)z_-}-\sqrt{c_1(r_0)+c_2(r_0)z_-}} \right)} \\
+&\frac{2r_0^3}{a^2}\frac{C_+}{\sqrt{c_1(r_0)z_++c_2(r_0)z_+^2}}
\ln{\left(\frac{\sqrt{c_1(r_0)z_++c_2(r_0)z_+}+\sqrt{c_1(r_0)+c_2(r_0)z_+}}
{\sqrt{c_1(r_0)z_++c_2(r_0)z_+}-\sqrt{c_1(r_0)+c_2(r_0)z_+}} \right)}.
\end{split}
\fe

In SDL, the coefficient $c_1(r_0)$ from (\ref{kkkk}) and the impact parameter $b(r_0)$ can be expanded as power series in terms of the small deviation $r_0 - r_{sc}$ as follows
\ie
c_1(r_0)=c_{1 sc}' (r_0-r_{sc})+{O}(r_0-r_{sc})^2 \, ,\label{dl}
\fe
\ie \label{nnn}
b(r_0) = b_{sc}+\frac{b_{sc}''}{2!}(r_0-r_{sc})^2+{O}(r_0-r_{sc})^3.
\fe
Here, it is important to mention that$c_1(r_{sc}) \equiv c_{1sc} = 0$, while $b(r_{sc}) \equiv b_{sc}$ corresponds to the critical impact parameter given by (\ref{asdasd}). The subscript $sc$ indicates evaluation at $r = r_{sc}$, and the prime denotes differentiation with respect to $r_0$. Substituting $c_{1sc} = 0$ into (\ref{kkkk}), we obtain
\ie\label{ffff}
 c_{3sc}=-\frac{2}{3}c_{2sc}.
\fe

If we combine (\ref{dl}) with (\ref{nnn}), the coefficient $c_1(r_0)$ may be rewritten in terms of the small deviation $b - b_{sc}$ as:
\ie \label{cnn}
\lim_{r_0\to r_{sc}} c_1(r_0)=\lim_{b\to b_{sc}}c_{1 sc}'\sqrt{\frac{2b_{sc}}{b_{sc}''}}\left(\frac{b}{b_{sc}}-1\right)^{1/2}.
\fe

In this manner, substituting (\ref{cnn}) into (\ref{isd}) transforms $I_D$ into \cite{lensingmain}
\ie \label{ID}
\begin{split}
I_D(b)\simeq &-\left(\frac{r_{sc}^3}{a^2}\frac{C_{- sc}}{\sqrt{c_{2 sc} \, z_{- sc}^2}}+\frac{r_{sc}^3}{a^2}\frac{C_{+ sc}}{\sqrt{c_{2 sc}\, z_{+sc}^2}} \right)\ln{\left(\frac{b}{b_{sc}}-1\right)}\\
&+\frac{r_{sc}^3}{a^2}\frac{C_{- sc}}{\sqrt{c_{2 sc}\, z_{-sc}^2}}\ln{\left(\frac{16 \, c^2_{2 sc} \, z_{-sc}^2 b_{sc}''}{c_{1sc}'^2 2b_{sc}(z_{- sc}-1)^2}\right)}
+\frac{r_{sc}^3}{a^2}\frac{C_{+sc}}{\sqrt{c_{2 sc}z_{+sc}^2}}\ln{\left(\frac{16\, c^2_{2sc} \, z_{+sc}^2 \, b_{sc}''}{c_{1sc}'^2 2b_{sc}(z_{+sc}-1)^2}\right)}.
\end{split}
\fe

Finally, the coefficient $\bar{a}$ and the portion of $I_D(b)$ contributing to $\bar{b}$, referred to as $b_D$, are expressed as:
\ie
\begin{split} \label{abak}
\bar{a}=&\frac{r_{sc}^3}{\sqrt{c_{2 sc}
}}\left[ \frac{C_{-sc}}{r_{sc}r_{-}-a^2}+\frac{C_{+sc}}{r_{sc}r_{+}-a^2}\right]
\end{split}
\fe
as well as
\ie
\begin{split} \label{bbdd}
b_D=
\bar{a} \ln{\left[ \frac{8c_{2sc}^2 b_{sc}''}{c_{1sc}'^2 b_{sc}}\right]}
+\frac{ 2 r_{sc}^3}{\sqrt{c_{2sc}}}\left[ \frac{C_{-sc}}{r_{sc}r_{-} -a^2}\ln{\left(1-\frac{a^2}{r_{sc}r_{-}}\right)}+\frac{C_{+sc} }{r_{sc}r_{+}-a^2}\ln{\left(1-\frac{a^2}{r_{sc}r_{+}}\right)} \right].
\end{split}
\fe

By expressing $z_{\pm}$ in terms of $r_{\pm}$ using (\ref{zkzk}), the integration of $f_R(z, r_{sc})$ in SDL yields the leading--order contribution to the coefficient $\bar{b}$. This contribution, labeled as $b_R$, is given by \cite{lensingmain}
\ie
\begin{split} \label{bk}
b_R& =I_R(r_{sc})=\int_0^1 f_R(z,r_{sc}) \mathrm{d}z\\
=&\frac{2 r_{0}^3}{a^2}\frac{C_-}{\sqrt{c_2}z_-}\ln{\left(\frac{z_-}{z_--1} \frac{\sqrt{c_2+c_3}+\sqrt{c_2}}{\sqrt{c_2+c_3}-\sqrt{c_2}} \frac{c_3}{4c_2} \right)}\\
&+\frac{2 r_{0}^3}{a^2}\frac{C_-}{\sqrt{c_2+c_3z_-}z_-}\ln{\left( \frac{\sqrt{c_2+c_3z_-}-\sqrt{c_2+c_3}} {\sqrt{c_2+c_3z_-}+\sqrt{c_2+c_3}} \frac{\sqrt{c_2+c_3z_-}+\sqrt{c_2}}{\sqrt{c_2+c_3z_-}-\sqrt{c_2}}\right)}\\
&+\frac{2 r_{0}^3}{a^2}\frac{C_+}{\sqrt{c_2}z_+}\ln{\left(\frac{z_+}{z_+-1} \frac{\sqrt{c_2+c_3}+\sqrt{c_2}}{\sqrt{c_2+c_3}-\sqrt{c_2}} \frac{c_3}{4c_2} \right)}\\
&+\frac{2 r_{0}^3}{a^2}\frac{C_+}{\sqrt{c_2+c_3z_+}z_+}\ln{\left( \frac{\sqrt{c_2+c_3z_+}-\sqrt{c_2+c_3}} {\sqrt{c_2+c_3z_+}+\sqrt{c_2+c_3}} \frac{\sqrt{c_2+c_3z_+}+\sqrt{c_2}}{\sqrt{c_2+c_3z_+}-\sqrt{c_2}}\right)}\Big\vert_{r_0=r_{sc}}.
\end{split}
\fe

Therefore, the coefficient $\bar{b}$ is determined by the sum of $b_D$ and $b_R$, which reads
\ie
\bar{b}=-\pi+b_D+b_R,
\fe
where by utilizing (\ref{bbdd}) and (\ref{bk}), along with the substitutions from (\ref{ffff}) and (\ref{zkzk}), we express $c_{3sc}$ in terms of $c_{2sc}$ as $c_{2sc} = -\frac{2}{3} c_{3sc}$ and replace $z_{\pm}$ with $r_{\pm}$. After performing the necessary algebraic manipulations, we get \cite{lensingmain}
\ie
\begin{split}\label{bbccc}
\bar{b}=&-\pi+\bar{a} \ln{\left( \frac{36}{7+4\sqrt{3}}\frac{8c_{2sc}^2 b_{sc}''}{c_{1sc}'^2 b_{sc}} \right)}\\
&+\frac{r_{sc}^3}{\sqrt{c_{2 sc}}}\frac{2 a C_{-sc}}{ a^2-r_{sc}r_- } \frac{\sqrt{3}}{\sqrt{a^2+2r_{sc}r_-}} \ln{\left( \frac{\sqrt{a^2+2r_{sc}r_-}-a}{\sqrt{a^2+2r_{sc}r_-}+a} \frac{\sqrt{a^2+2r_{sc}r_-}+\sqrt{3}a}{\sqrt{a^2+2r_{sc}r_-}-\sqrt{3}a}\right)}\\
&+\frac{r_{sc}^3}{\sqrt{c_{2sc}}}\frac{2 a C_{+sc}}{ a^2-r_{sc}r_+ } \frac{\sqrt{3}}{\sqrt{a^2+2r_{sc}r_+}} \ln{\left( \frac{\sqrt{a^2+2r_{sc}r_+}-a}{\sqrt{a^2+2r_{sc}r_+}+a} \frac{\sqrt{a^2+2r_{sc}r_+}+\sqrt{3}a}{\sqrt{a^2+2r_{sc}r_+}-\sqrt{3}a}\right)}.
\end{split}
\fe

By employing the expressions for $r_{sc}$ in (\ref{xcv}), $b_{sc}$ in (\ref{asdasd}), and $b(r_0)$ in (\ref{bbrr}), along with the definitions of $C_\pm$ and $c_2$ from (\ref{ccpp}) and (\ref{kkkk}), the coefficients $\bar{a}$ and $\bar{b}$ from (\ref{abak}) and (\ref{bbccc}) can be determined. For the parameter range considered, $\bar{a}$ remains positive, while $\bar{b}$ is negative. Additionally, both $\bar{a}$ and $\vert \bar{b} \vert$ increase with $a$ in direct orbits and decrease in retrograde ones. Consequently, the deflection angle $\hat{\alpha}$ reduces for direct trajectories and increases for retrograde ones as the black hole’s spin grows, given a fixed impact parameter.

The expressions for $\bar{a}$ and $\bar{b}$ corresponding to the Schwarzschild black hole, as presented in \cite{tsukamoto2017deflection,lensingmain,034}, can be recovered by taking the limit $a \to 0$. In this case, the parameters transform as follows:  
$r_+ \to 2M_{\Theta}$,  
$r_- \to a^2 / 2M_{\Theta}$,  
$C_{+sc} \to 2M_{\Theta} / r_{sc}$,  
$C_{-sc} \to a^3 / (2b_{sc} M_{\Theta} r_{sc})$,  
and $c_{2sc} \to r_{sc}^2$ with $c_{1sc} = 0$.  
It can be verified that in this limit, $\bar{a} = 1$ in (\ref{abak}), and $\bar{b}$ in (\ref{bbccc}) simplifies to an expression proportional to $\bar{a}$, as outlined below
\ie
\begin{split}
\bar{b}=&-\pi+\bar{a} \ln{\left( 36(7-4\sqrt{3})\frac{8c_{2 sc}^2 b_{sc}''}{c_{1sc}'^2 b_{sc}} \right)}\\
=&-\pi+ \ln{\left( 216(7-4\sqrt{3}) \right)}.
\end{split}
\fe

It is worth mentioning that, in the second equality, additional substitutions have been applied based on the Schwarzschild black hole case, where $r_{sc} = 3M_{\Theta}$. These include $b_{sc} \to 3\sqrt{3}M_{\Theta}$, $b_{sc}'' \to \sqrt{3}/M_{\Theta}$, $c_{1sc}' \to 6M_{\Theta}$, and $c_{2sc} \to 9M_{\Theta}^2$.


\section{Conclusion\label{cccon}}

In this work, we began by analyzing a spherically symmetric black hole within the context of non-commutative gauge theory. Using a modified Newman--Janis algorithm, we obtained a new rotating black hole solution. The physical properties of this solution were then investigated, with particular focus on the event horizon structure, which revealed the presence of both an inner horizon $r_{-}$ and an outer horizon $r_{+}$. It was observed that as the non--commutative parameter $\Theta$ and the rotation parameter $a$ increased, the horizon radii diminished.

The analysis also extended to the ergosphere, where both an inner boundary $r_{e_{-}}$ and an outer boundary $r_{e_{+}}$ were identified. Similar to the event horizons, the ergospheres contracted as $\Theta$ and $a$ increased. Additionally, the angular velocity was examined, revealing a reduction in magnitude with increasing $\Theta$, while growing with $a$ up to approximately $r \approx 2$. The influence of non-commutativity on the angular velocity was then compared to the standard Kerr case ($\Theta = 0$).

To conduct a comprehensive thermodynamic analysis, we determined the surface gravity, which was crucial for evaluating key thermodynamic quantities such as the Hawking temperature, entropy, and heat capacity. Both the Hawking temperature and entropy exhibited a decreasing trend with increasing $a$ and, for small values of $M$, also with increasing $\Theta$. The remnant mass was examined by imposing the extremal black hole condition $T \to 0$, yielding $M_{\pm}(a,\Theta) = \frac{1}{8} \left(\sqrt{16 a^2+\Theta ^2} \pm 4 a\right)$. Meanwhile, the heat capacity displayed an overall split as $a$ and $\Theta$ grew, taking on both positive and negative values. All findings were compared to the standard Kerr case ($\Theta = 0$).

Quantum radiation was also examined for our axisymmetric non--commutative black hole solution. Both bosonic and fermionic particle modes were considered, leading to the corresponding particle creation densities $n_{b}(\omega, \Theta, a, M)$ and $n_{f}(\omega, \Theta, a, M)$, respectively. In both cases, the parameters $\Theta$ and $a$ contributed to an overall increase in particle creation densities.

The geodesic motion was also investigated, with numerical simulations in 3D provided for better visualization. Particular attention was given to null geodesics and their associated radial accelerations. Additionally, the photon sphere and black hole shadows were analyzed in detail using analytical methods.

Finally, our investigation focused on gravitational lensing in the strong deflection limit. As a potential extension, applying these analyses to Lorentz--violating scenarios, such as bumblebee \cite{aa2025does} and Kalb--Ramond \cite{araujo2024particle} gravities, appears to be a promising direction for further research. These and other related aspects are currently under development.




\section*{Acknowledgments}
\hspace{0.5cm} A. A. Araújo Filho is supported by Conselho Nacional de Desenvolvimento Cient\'{\i}fico e Tecnol\'{o}gico (CNPq) and Fundação de Apoio à Pesquisa do Estado da Paraíba (FAPESQ), project No. 150891/2023-7. Furthermore, the authors thank M. Ostroff for the valuable discussions and for providing the code used to compute the geodesic segment. A. {\"O}. would like to acknowledge the contribution of the COST Action CA21106 - COSMIC WISPers in the Dark Universe: Theory, astrophysics and experiments (CosmicWISPers), the COST Action CA22113 - Fundamental challenges in theoretical physics (THEORY-CHALLENGES) and CA23130 - Bridging high and low energies in search of quantum gravity (BridgeQG).  N. H. would like to acknowledge networking support of the COST Action CA 22113 - Fundamental challenges in theoretical physics (Theory and Challenges), CA 21106 - COSMIC WISPers in the Dark Universe: Theory, astrophysics and experiments (CosmicWISPers), CA 21136 - Addressing observational tensions in cosmology with systematics and fundamental physics (CosmoVerse), and CA 23130 - Bridging high and low energies in search of quantum gravity (BridgeQG).

\bibliographystyle{ieeetr}
\bibliography{main}

\begin{thebibliography}{100}

\bibitem{ferrari2003finiteness}
A.~F. Ferrari, H.~O. Girotti, M.~Gomes, A.~Y. Petrov, A.~Ribeiro, and
  A.~Da~Silva, ``On the finiteness of noncommutative supersymmetric qed3 in the
  covariant superfield formulation,'' {\em Physics Letters B}, vol.~577,
  no.~1-2, pp.~83--92, 2003.

\bibitem{szabo2003quantum}
R.~J. Szabo, ``Quantum field theory on noncommutative spaces,'' {\em Physics
  Reports}, vol.~378, no.~4, pp.~207--299, 2003.

\bibitem{ferrari2004superfield}
A.~F. Ferrari, H.~O. Girotti, M.~Gomes, A.~Y. Petrov, A.~Ribeiro, V.~O.
  Rivelles, and A.~Da~Silva, ``Superfield covariant analysis of the divergence
  structure of noncommutative supersymmetric qed 4,'' {\em Physical Review D},
  vol.~69, no.~2, p.~025008, 2004.

\bibitem{3}
N.~Seiberg and E.~Witten, ``String theory and noncommutative geometry,'' {\em
  Journal of High Energy Physics}, vol.~1999, no.~09, p.~032, 1999.

\bibitem{szabo2006symmetry}
R.~J. Szabo, ``Symmetry, gravity and noncommutativity,'' {\em Classical and
  Quantum Gravity}, vol.~23, no.~22, p.~R199, 2006.

\bibitem{ferrari2004towards}
A.~F. Ferrari, H.~O. Girotti, M.~Gomes, A.~Y. Petrov, A.~Ribeiro, V.~O.
  Rivelles, and A.~da~Silva, ``Towards a consistent noncommutative
  supersymmetric yang-mills theory: Superfield covariant analysis,'' {\em
  Physical Review D}, vol.~70, no.~8, p.~085012, 2004.

\bibitem{chamseddine2001deforming}
A.~H. Chamseddine, ``Deforming einstein's gravity,'' {\em Physics Letters B},
  vol.~504, no.~1-2, pp.~33--37, 2001.

\bibitem{sharif2011thermodynamics}
M.~Sharif and W.~Javed, ``Thermodynamics of a bardeen black hole in
  noncommutative space,'' {\em Canadian Journal of Physics}, vol.~89, no.~10,
  pp.~1027--1033, 2011.

\bibitem{banerjee2008noncommutative}
R.~Banerjee, B.~R. Majhi, and S.~Samanta, ``Noncommutative black hole
  thermodynamics,'' {\em Physical Review D}, vol.~77, no.~12, p.~124035, 2008.

\bibitem{AraujoFilho:2024rss}
A.~A. Ara\'ujo~Filho, J.~R. Nascimento, A.~Y. Petrov, P.~J. Porf\'\i{}rio, and
  A.~\"Ovg\"un, ``{Properties of an axisymmetric Lorentzian non-commutative
  black hole},'' {\em Phys. Dark Univ.}, vol.~47, p.~101796, 2025.

\bibitem{nozari2007thermodynamics}
K.~Nozari and B.~Fazlpour, ``Thermodynamics of noncommutative schwarzschild
  black hole,'' {\em Modern Physics Letters A}, vol.~22, no.~38,
  pp.~2917--2930, 2007.

\bibitem{t29}
A.~A. Ara{\'u}jo~Filho, S.~Zare, P.~J. Porf{\'\i}rio, J.~K{\v{r}}{\'\i}{\v{z}},
  and H.~Hassanabadi, ``Thermodynamics and evaporation of a modified
  schwarzschild black hole in a non--commutative gauge theory,'' {\em Physics
  Letters B}, vol.~838, p.~137744, 2023.

\bibitem{nozari2006reissner}
K.~Nozari and B.~Fazlpour, ``Reissner-nordstr$\backslash$"$\{$o$\}$ m black
  hole thermodynamics in noncommutative spaces,'' {\em arXiv preprint
  gr-qc/0608077}, 2006.

\bibitem{myung2007thermodynamics}
Y.~S. Myung, Y.-W. Kim, and Y.-J. Park, ``Thermodynamics and evaporation of the
  noncommutative black hole,'' {\em Journal of High Energy Physics}, vol.~2007,
  no.~02, p.~012, 2007.

\bibitem{lopez2006towards}
J.~Lopez-Dominguez, O.~Obregon, M.~Sabido, and C.~Ramirez, ``Towards
  noncommutative quantum black holes,'' {\em Physical Review D}, vol.~74,
  no.~8, p.~084024, 2006.

\bibitem{nicolini2006noncommutative}
P.~Nicolini, A.~Smailagic, and E.~Spallucci, ``Noncommutative geometry inspired
  schwarzschild black hole,'' {\em Physics Letters B}, vol.~632, no.~4,
  pp.~547--551, 2006.

\bibitem{ghosh2018noncommutative}
S.~G. Ghosh, ``Noncommutative geometry inspired einstein--gauss--bonnet black
  holes,'' {\em Classical and Quantum Gravity}, vol.~35, no.~8, p.~085008,
  2018.

\bibitem{nicolini2009noncommutative}
P.~Nicolini, ``Noncommutative black holes, the final appeal to quantum gravity:
  a review,'' {\em International Journal of Modern Physics A}, vol.~24, no.~07,
  pp.~1229--1308, 2009.

\bibitem{nozari2008hawking}
K.~Nozari and S.~H. Mehdipour, ``Hawking radiation as quantum tunneling from a
  noncommutative schwarzschild black hole,'' {\em Classical and Quantum
  Gravity}, vol.~25, no.~17, p.~175015, 2008.

\bibitem{lekbich2024optical}
H.~Lekbich, N.~Parbin, D.~J. Gogoi, A.~E. Boukili, and M.~Sedra, ``The optical
  features of noncommutative charged 4d-ads-einstein--gauss--bonnet black hole:
  shadow and deflection angle,'' {\em The European Physical Journal C},
  vol.~84, no.~4, p.~350, 2024.

\bibitem{wei2015shadow}
S.-W. Wei, P.~Cheng, Y.~Zhong, and X.-N. Zhou, ``Shadow of noncommutative
  geometry inspired black hole,'' {\em Journal of Cosmology and Astroparticle
  Physics}, vol.~2015, no.~08, p.~004, 2015.

\bibitem{sharif2016shadow}
M.~Sharif and S.~Iftikhar, ``Shadow of a charged rotating non-commutative black
  hole,'' {\em The European Physical Journal C}, vol.~76, pp.~1--9, 2016.

\bibitem{ovgun2020shadow}
A.~{\"O}vg{\"u}n, I.~Sakall{\i}, J.~Saavedra, and C.~Leiva, ``Shadow cast of
  noncommutative black holes in rastall gravity,'' {\em Modern Physics Letters
  A}, vol.~35, no.~20, p.~2050163, 2020.

\bibitem{ding2011strong}
C.~Ding, S.~Kang, C.-Y. Chen, S.~Chen, and J.~Jing, ``Strong gravitational
  lensing in a noncommutative black-hole spacetime,'' {\em Physical Review
  D—Particles, Fields, Gravitation, and Cosmology}, vol.~83, no.~8,
  p.~084005, 2011.

\bibitem{saleem2023observable}
R.~Saleem and M.~I. Aslam, ``Observable features of charged kiselev black hole
  with non-commutative geometry under various accretion flow,'' {\em The
  European Physical Journal C}, vol.~83, no.~3, pp.~1--14, 2023.

\bibitem{ding2011probing}
C.~Ding and J.~Jing, ``Probing spacetime noncommutative constant via charged
  astrophysical black hole lensing,'' {\em Journal of High Energy Physics},
  vol.~2011, no.~10, pp.~1--19, 2011.

\bibitem{newcommutativity}
N.~Heidari, H.~Hassanabadi, A.~A.~P. Araújo~Filho, and J.~Kriz, ``Exploring
  non-commutativity as a perturbation in the schwarzschild black hole:
  quasinormal modes, scattering, and shadows,'' {\em The European Physical
  Journal C}, vol.~84, p.~566, 2024.

\bibitem{araujo2025particle}
A.~Ara{\'u}jo~Filho, ``Particle production induced by a lorentzian
  non--commutative spacetime,'' {\em Annals of Physics}, p.~170167, 2025.

\bibitem{campos2022quasinormal}
J.~A.~V. Campos, M.~Anacleto, F.~A. Brito, and E.~Passos, ``Quasinormal modes
  and shadow of noncommutative black hole,'' {\em Scientific Reports}, vol.~12,
  no.~1, p.~8516, 2022.

\bibitem{Anacleto:2022shk}
M.~A. Anacleto, F.~A. Brito, J.~A.~V. Campos, and E.~Passos, ``{Absorption,
  scattering and shadow by a noncommutative black hole with global monopole},''
  {\em Eur. Phys. J. C}, vol.~83, no.~4, p.~298, 2023.

\bibitem{Anacleto:2019tdj}
M.~A. Anacleto, F.~A. Brito, J.~A.~V. Campos, and E.~Passos, ``{Absorption and
  scattering of a noncommutative black hole},'' {\em Phys. Lett. B}, vol.~803,
  p.~135334, 2020.

\bibitem{AraujoFilho:2025rzh}
A.~A. Ara{\'u}jo~Filho, N.~Heidari, and Y.~Shi, ``{Neutrino dynamics in a
  non-commutative spacetime},'' 4 2025.

\bibitem{Pozar:2025yoj}
F.~Po{\v{z}}ar, ``{Corrections to Kerr-Newman black hole from noncommutative
  Einstein-Maxwell equation},'' {\em Phys. Lett. B}, vol.~868, p.~139751, 2025.

\bibitem{araujo2025non}
A.~A. Ara{\'u}jo~Filho, N.~Heidari, and I.~P. Lobo, ``A non-commutative
  kalb-ramond black hole,'' {\em Journal of Cosmology and Astroparticle
  Physics}, vol.~2025, no.~09, p.~076, 2025.

\bibitem{araujo2025gedddodesics}
A.~A. Ara{\'u}jo~Filho, N.~Heidari, and A.~{\"O}vg{\"u}n, ``Geodesics,
  accretion disk, gravitational lensing, time delay, and effects on neutrinos
  induced by a non-commutative black hole,'' {\em Journal of Cosmology and
  Astroparticle Physics}, vol.~2025, no.~06, p.~062, 2025.

\bibitem{heidari2025non}
N.~Heidari, A.~A. Ara{\'u}jo~Filho, and I.~P. Lobo, ``Non-commutativity in
  hayward spacetime,'' {\em Journal of Cosmology and Astroparticle Physics},
  vol.~2025, no.~09, p.~051, 2025.

\bibitem{Juric:2025kjl}
T.~Juri{\'c}, A.~N. Kumara, and F.~Po{\v{z}}ar, ``{Constructing noncommutative
  black holes},'' {\em Nucl. Phys. B}, vol.~1017, p.~116950, 2025.

\bibitem{yakut2005evolution}
K.~Yakut and P.~P. Eggleton, ``Evolution of close binary systems,'' {\em The
  Astrophysical Journal}, vol.~629, no.~2, p.~1055, 2005.

\bibitem{pretorius2005evolution}
F.~Pretorius, ``Evolution of binary black-hole spacetimes,'' {\em Physical
  review letters}, vol.~95, no.~12, p.~121101, 2005.

\bibitem{kjeldsen1994amplitudes}
H.~Kjeldsen and T.~R. Bedding, ``Amplitudes of stellar oscillations: the
  implications for asteroseismology,'' {\em arXiv preprint astro-ph/9403015},
  1994.

\bibitem{dziembowski1992effects}
W.~Dziembowski and P.~R. Goode, ``Effects of differential rotation on stellar
  oscillations-a second-order theory,'' {\em The Astrophysical Journal},
  vol.~394, pp.~670--687, 1992.

\bibitem{unno1979nonradial}
W.~Unno, Y.~Osaki, H.~Ando, and H.~Shibahashi, ``Nonradial oscillations of
  stars,'' {\em Tokyo: University of Tokyo Press}, 1979.

\bibitem{hurley2002evolution}
J.~R. Hurley, C.~A. Tout, and O.~R. Pols, ``Evolution of binary stars and the
  effect of tides on binary populations,'' {\em Monthly Notices of the Royal
  Astronomical Society}, vol.~329, no.~4, pp.~897--928, 2002.

\bibitem{heuvel2011compact}
E.~v.~d. Heuvel, ``Compact stars and the evolution of binary systems,'' in {\em
  Fluid Flows To Black Holes: A Tribute to S Chandrasekhar on His Birth
  Centenary}, pp.~55--73, World Scientific, 2011.

\bibitem{riles2017recent}
K.~Riles, ``Recent searches for continuous gravitational waves,'' {\em Modern
  Physics Letters A}, vol.~32, no.~39, p.~1730035, 2017.

\bibitem{kokkotas1999quasi}
K.~D. Kokkotas and B.~G. Schmidt, ``Quasi-normal modes of stars and black
  holes,'' {\em Living Reviews in Relativity}, vol.~2, no.~1, pp.~1--72, 1999.

\bibitem{roy2020revisiting}
P.~D. Roy, S.~Aneesh, and S.~Kar, ``Revisiting a family of wormholes: geometry,
  matter, scalar quasinormal modes and echoes,'' {\em The European Physical
  Journal C}, vol.~80, no.~9, pp.~1--17, 2020.

\bibitem{oliveira2019quasinormal}
R.~Oliveira, D.~Dantas, V.~Santos, and C.~Almeida, ``Quasinormal modes of
  bumblebee wormhole,'' {\em Classical and Quantum Gravity}, vol.~36, no.~10,
  p.~105013, 2019.

\bibitem{berti2009quasinormal}
E.~Berti, V.~Cardoso, and A.~O. Starinets, ``Quasinormal modes of black holes
  and black branes,'' {\em Classical and Quantum Gravity}, vol.~26, no.~16,
  p.~163001, 2009.

\bibitem{horowitz2000quasinormal}
G.~T. Horowitz and V.~E. Hubeny, ``Quasinormal modes of ads black holes and the
  approach to thermal equilibrium,'' {\em Physical Review D}, vol.~62, no.~2,
  p.~024027, 2000.

\bibitem{heidari2024impact}
N.~Heidari, J.~Reis, H.~Hassanabadi, {\em et~al.}, ``The impact of an
  antisymmetric tensor on charged black holes: evaporation process, geodesics,
  deflection angle, scattering effects and quasinormal modes,'' {\em arXiv
  preprint arXiv:2404.10721}, 2024.

\bibitem{Hamil:2024ppj}
B.~Hamil and B.~C. L\"utf\"uo\u{g}lu, ``{Noncommutative Schwarzschild black
  hole surrounded by quintessence: Thermodynamics, Shadows and Quasinormal
  modes},'' {\em Phys. Dark Univ.}, vol.~44, p.~101484, 2024.

\bibitem{nollert1999quasinormal}
H.-P. Nollert, ``Quasinormal modes: the characteristicsound'of black holes and
  neutron stars,'' {\em Classical and Quantum Gravity}, vol.~16, no.~12,
  p.~R159, 1999.

\bibitem{ferrari1984new}
V.~Ferrari and B.~Mashhoon, ``New approach to the quasinormal modes of a black
  hole,'' {\em Physical Review D}, vol.~30, no.~2, p.~295, 1984.

\bibitem{santos2016quasinormal}
V.~Santos, R.~V. Maluf, and C.~A.~S. Almeida, ``Quasinormal frequencies of
  self-dual black holes,'' {\em Physical Review D}, vol.~93, no.~8, p.~084047,
  2016.

\bibitem{ovgun2018quasinormal}
A.~{\"O}vg{\"u}n, I.~Sakall{\i}, and J.~Saavedra, ``Quasinormal modes of a
  schwarzschild black hole immersed in an electromagnetic universe,'' {\em
  Chinese Physics C}, vol.~42, no.~10, p.~105102, 2018.

\bibitem{jusufi2024charged}
K.~Jusufi, B.~Cuadros-Melgar, G.~Leon, A.~Jawad, {\em et~al.}, ``Charged black
  holes with yukawa potential,'' {\em arXiv preprint arXiv:2401.15211}, 2024.

\bibitem{rincon2020greybody}
{\'A}.~Rinc{\'o}n and V.~Santos, ``Greybody factor and quasinormal modes of
  regular black holes,'' {\em The European Physical Journal C}, vol.~80,
  no.~10, pp.~1--7, 2020.

\bibitem{araujo2024dark}
A.~A. Ara{\'u}jo~Filho, K.~Jusufi, B.~Cuadros-Melgar, and G.~Leon, ``Dark
  matter signatures of black holes with yukawa potential,'' {\em Physics of the
  Dark Universe}, p.~101500, 2024.

\bibitem{london2014modeling}
L.~London, D.~Shoemaker, and J.~Healy, ``Modeling ringdown: Beyond the
  fundamental quasinormal modes,'' {\em Physical Review D}, vol.~90, no.~12,
  p.~124032, 2014.

\bibitem{maggiore2008physical}
M.~Maggiore, ``Physical interpretation of the spectrum of black hole
  quasinormal modes,'' {\em Physical Review Letters}, vol.~100, no.~14,
  p.~141301, 2008.

\bibitem{flachi2013quasinormal}
A.~Flachi and J.~P. Lemos, ``Quasinormal modes of regular black holes,'' {\em
  Physical Review D}, vol.~87, no.~2, p.~024034, 2013.

\bibitem{blazquez2018scalar}
J.~L. Bl{\'a}zquez-Salcedo, X.~Y. Chew, and J.~Kunz, ``Scalar and axial
  quasinormal modes of massive static phantom wormholes,'' {\em Physical Review
  D}, vol.~98, no.~4, p.~044035, 2018.

\bibitem{konoplya2011quasinormal}
R.~Konoplya and A.~Zhidenko, ``Quasinormal modes of black holes: From
  astrophysics to string theory,'' {\em Reviews of Modern Physics}, vol.~83,
  no.~3, p.~793, 2011.

\bibitem{araujo2024exploring}
A.~A. Ara{\'u}jo~Filho, J.~A. A.~S. Reis, and H.~Hassanabadi, ``Exploring
  antisymmetric tensor effects on black hole shadows and quasinormal
  frequencies,'' {\em Journal of Cosmology and Astroparticle Physics},
  vol.~2024, no.~05, p.~029, 2024.

\bibitem{lee2020quasi}
C.~O. Lee, J.~Y. Kim, and M.-I. Park, ``Quasi-normal modes and stability of
  einstein--born--infeld black holes in de sitter space,'' {\em The European
  Physical Journal C}, vol.~80, no.~8, pp.~1--21, 2020.

\bibitem{jawad2020quasinormal}
A.~Jawad, S.~Chaudhary, M.~Yasir, A.~{\"O}vg{\"u}n, and {\.I}.~Sakall{\i},
  ``Quasinormal modes of extended gravity black holes through higher order wkb
  method,'' {\em International Journal of Geometric Methods in Modern Physics},
  p.~2350129, 2023.

\bibitem{Fernando:2012yw}
S.~Fernando and J.~Correa, ``{Quasinormal Modes of Bardeen Black Hole: Scalar
  Perturbations},'' {\em Phys. Rev. D}, vol.~86, p.~064039, 2012.

\bibitem{jcap5}
L.~Hui, D.~Kabat, and S.~S. Wong, ``Quasinormal modes, echoes and the causal
  structure of the green's function,'' {\em Journal of Cosmology and
  Astroparticle Physics}, vol.~2019, no.~12, p.~020, 2019.

\bibitem{maluf2014einstein}
R.~Maluf, C.~Almeida, R.~Casana, and M.~Ferreira~Jr, ``Einstein-hilbert
  graviton modes modified by the lorentz-violating bumblebee field,'' {\em
  Physical Review D}, vol.~90, no.~2, p.~025007, 2014.

\bibitem{jcap4}
M.~Cadoni, M.~Oi, and A.~P. Sanna, ``Quasi-normal modes and microscopic
  description of 2d black holes,'' {\em Journal of High Energy Physics},
  vol.~2022, no.~1, pp.~1--23, 2022.

\bibitem{JCAP1}
M.~Okyay and A.~{\"O}vg{\"u}n, ``Nonlinear electrodynamics effects on the black
  hole shadow, deflection angle, quasinormal modes and greybody factors,'' {\em
  Journal of Cosmology and Astroparticle Physics}, vol.~2022, no.~01, p.~009,
  2022.

\bibitem{Daghigh:2008jz}
R.~G. Daghigh and M.~D. Green, ``{Highly Real, Highly Damped, and Other
  Asymptotic Quasinormal Modes of Schwarzschild-Anti De Sitter Black Holes},''
  {\em Class. Quant. Grav.}, vol.~26, p.~125017, 2009.

\bibitem{JCAP2}
Y.~Zhao, X.~Ren, A.~Ilyas, E.~N. Saridakis, and Y.-F. Cai, ``Quasinormal modes
  of black holes in f (t) gravity,'' {\em Journal of Cosmology and
  Astroparticle Physics}, vol.~2022, no.~10, p.~087, 2022.

\bibitem{Gogoi:2023kjt}
D.~J. Gogoi, A.~\"Ovg\"un, and M.~Koussour, ``{Quasinormal modes of black holes
  in f(Q) gravity},'' {\em Eur. Phys. J. C}, vol.~83, no.~8, p.~700, 2023.

\bibitem{JCAP3}
S.~Boudet, F.~Bombacigno, G.~J. Olmo, and P.~J. Porfirio, ``Quasinormal modes
  of schwarzschild black holes in projective invariant chern-simons modified
  gravity,'' {\em Journal of Cosmology and Astroparticle Physics}, vol.~2022,
  no.~05, p.~032, 2022.

\bibitem{Daghigh:2020fmw}
R.~G. Daghigh, M.~D. Green, and G.~Kunstatter, ``{Scalar Perturbations and
  Stability of a Loop Quantum Corrected Kruskal Black Hole},'' {\em Phys. Rev.
  D}, vol.~103, no.~8, p.~084031, 2021.

\bibitem{mmm2}
A.~A. Ara{\'u}jo~Filho, ``Implications of a simpson--visser solution in
  verlinde’s framework,'' {\em The European Physical Journal C}, vol.~84,
  no.~1, pp.~1--22, 2024.

\bibitem{Daghigh:2022uws}
R.~G. Daghigh, M.~D. Green, and J.~C. Morey, ``{Calculating quasinormal modes
  of Schwarzschild anti\textendash{}de Sitter black holes using the continued
  fraction method},'' {\em Phys. Rev. D}, vol.~107, no.~2, p.~024023, 2023.

\bibitem{Daghigh:2020jyk}
R.~G. Daghigh, M.~D. Green, and J.~C. Morey, ``{Significance of Black Hole
  Quasinormal Modes: A Closer Look},'' {\em Phys. Rev. D}, vol.~101, no.~10,
  p.~104009, 2020.

\bibitem{mmm1}
A.~A. Ara{\'u}jo~Filho, ``Analysis of a regular black hole in verlinde’s
  gravity,'' {\em Classical and Quantum Gravity}, vol.~41, no.~1, p.~015003,
  2023.

\bibitem{Fernando:2016ftj}
S.~Fernando, ``{Quasinormal modes of dilaton-de Sitter black holes: scalar
  perturbations},'' {\em Gen. Rel. Grav.}, vol.~48, no.~3, p.~24, 2016.

\bibitem{Heidari:2023bww}
N.~Heidari, H.~Hassanabadi, A.~A. Ara{\'u}jo~Filho, J.~K{\v{r}}{\'\i}{\v{z}},
  S.~Zare, and P.~J. Porf{\'\i}rio, ``Gravitational signatures of a
  non--commutative stable black hole,'' {\em Physics of the Dark Universe},
  vol.~43, p.~101382, 2023.

\bibitem{Daghigh:2020mog}
R.~G. Daghigh, M.~D. Green, J.~C. Morey, and G.~Kunstatter, ``{Scalar
  Perturbations of a Single-Horizon Regular Black Hole},'' {\em Phys. Rev. D},
  vol.~102, no.~10, p.~104040, 2020.

\bibitem{liu2022quasinormal}
D.~Liu, Y.~Yang, A.~{\"O}vg{\"u}n, Z.-W. Long, and Z.~Xu, ``The quasinormal
  modes and greybody bounds of rotating black holes in a dark matter halo,''
  {\em arXiv preprint arXiv:2204.11563}, 2022.

\bibitem{yang2023probing}
Y.~Yang, D.~Liu, A.~{\"O}vg{\"u}n, Z.-W. Long, and Z.~Xu, ``Probing hairy black
  holes caused by gravitational decoupling using quasinormal modes and greybody
  bounds,'' {\em Physical Review D}, vol.~107, no.~6, p.~064042, 2023.

\bibitem{Gogoi:2023fow}
D.~J. Gogoi, A.~\"Ovg\"un, and D.~Demir, ``{Quasinormal modes and greybody
  factors of symmergent black hole},'' {\em Phys. Dark Univ.}, vol.~42,
  p.~101314, 2023.

\bibitem{maluf2013matter}
R.~Maluf, V.~Santos, W.~Cruz, and C.~Almeida, ``Matter-gravity scattering in
  the presence of spontaneous lorentz violation,'' {\em Physical Review D},
  vol.~88, no.~2, p.~025005, 2013.

\bibitem{kim2018quasi}
J.~Y. Kim, C.~O. Lee, and M.-I. Park, ``Quasi-normal modes of a natural ads
  wormhole in einstein--born--infeld gravity,'' {\em The European Physical
  Journal C}, vol.~78, no.~12, pp.~1--15, 2018.

\bibitem{Daghigh:2005ph}
R.~G. Daghigh and G.~Kunstatter, ``{Highly damped quasinormal modes of generic
  single horizon black holes},'' {\em Class. Quant. Grav.}, vol.~22,
  pp.~4113--4128, 2005.

\bibitem{lambiase2023investigating}
G.~Lambiase, R.~C. Pantig, D.~J. Gogoi, and A.~{\"O}vg{\"u}n, ``Investigating
  the connection between generalized uncertainty principle and asymptotically
  safe gravity in black hole signatures through shadow and quasinormal modes,''
  {\em arXiv preprint arXiv:2304.00183}, 2023.

\bibitem{hassanabadi2023gravitational}
A.~A. Araújo~Filho, H.~Hassanabadi, N.~Heidari, J.~Kr{\'\i}z, and S.~Zare,
  ``Gravitational traces of bumblebee gravity in metric-affine formalism,''
  {\em Class. Quant. Grav.}, vol.~41, p.~055003, 2024.

\bibitem{Yang:2022ifo}
Y.~Yang, D.~Liu, A.~\"Ovg\"un, Z.-W. Long, and Z.~Xu, ``{Probing hairy black
  holes caused by gravitational decoupling using quasinormal modes and greybody
  bounds},'' {\em Phys. Rev. D}, vol.~107, no.~6, p.~064042, 2023.

\bibitem{gogoi2023quasinormal}
D.~J. Gogoi, A.~{\"O}vg{\"u}n, and M.~Koussour, ``Quasinormal modes of black
  holes in $ f (q) $ gravity,'' {\em arXiv preprint arXiv:2303.07424}, 2023.

\bibitem{023}
F.~Eisenhauer, R.~Genzel, T.~Alexander, R.~Abuter, T.~Paumard, T.~Ott,
  A.~Gilbert, S.~Gillessen, M.~Horrobin, S.~Trippe, {\em et~al.}, ``Sinfoni in
  the galactic center: young stars and infrared flares in the central
  light-month,'' {\em The Astrophysical Journal}, vol.~628, no.~1, p.~246,
  2005.

\bibitem{024}
E.~H.~T. Collaboration {\em et~al.}, ``First m87 event horizon telescope
  results. iv. imaging the central supermassive black hole,'' {\em arXiv
  preprint arXiv:1906.11241}, 2019.

\bibitem{025}
K.~Akiyama, A.~Alberdi, W.~Alef, K.~Asada, R.~Azulay, A.-K. Baczko, D.~Ball,
  M.~Balokovi{\'c}, J.~Barrett, D.~Bintley, {\em et~al.}, ``First m87 event
  horizon telescope results. ii. array and instrumentation,'' {\em The
  Astrophysical Journal Letters}, vol.~875, no.~1, p.~L2, 2019.

\bibitem{026}
K.~Akiyama, A.~Alberdi, W.~Alef, K.~Asada, R.~Azulay, A.-K. Baczko, D.~Ball,
  M.~Balokovi{\'c}, J.~Barrett, D.~Bintley, {\em et~al.}, ``First m87 event
  horizon telescope results. v. physical origin of the asymmetric ring,'' {\em
  The Astrophysical Journal Letters}, vol.~875, no.~1, p.~L5, 2019.

\bibitem{027}
E.~H.~T. Collaboration {\em et~al.}, ``First m87 event horizon telescope
  results. iv. imaging the central supermassive black hole,'' {\em arXiv
  preprint arXiv:1906.11241}, 2019.

\bibitem{028}
K.~Akiyama, A.~Alberdi, W.~Alef, K.~Asada, R.~Azulay, A.-K. Baczko, D.~Ball,
  M.~Balokovi{\'c}, J.~Barrett, D.~Bintley, {\em et~al.}, ``First m87 event
  horizon telescope results. v. physical origin of the asymmetric ring,'' {\em
  The Astrophysical Journal Letters}, vol.~875, no.~1, p.~L5, 2019.

\bibitem{029}
D.~Ball, C.-K. Chan, P.~Christian, B.~T. Jannuzi, J.~Kim, D.~P. Marrone,
  L.~Medeiros, F.~Ozel, D.~Psaltis, M.~Rose, {\em et~al.}, ``First m87 event
  horizon telescope results. vi. the shadow and mass of the central black
  hole,'' 2019.

\bibitem{pantig2023testing}
R.~C. Pantig, A.~{\"O}vg{\"u}n, and D.~Demir, ``Testing symmergent gravity
  through the shadow image and weak field photon deflection by a rotating black
  hole using the m87 and sgr. a results,'' {\em The European Physical Journal
  C}, vol.~83, no.~3, p.~250, 2023.

\bibitem{ccimdiker2021black}
{\.I}.~{\c{C}}imdiker, D.~Demir, and A.~{\"O}vg{\"u}n, ``Black hole shadow in
  symmergent gravity,'' {\em Physics of the Dark Universe}, vol.~34, p.~100900,
  2021.

\bibitem{pantig2022shadow}
R.~C. Pantig, L.~Mastrototaro, G.~Lambiase, and A.~{\"O}vg{\"u}n, ``Shadow,
  lensing and neutrino propagation by dyonic modmax black holes,'' {\em arXiv
  preprint arXiv:2208.06664}, 2022.

\bibitem{030}
K.~S. Virbhadra and G.~F. Ellis, ``Schwarzschild black hole lensing,'' {\em
  Phys. Rev. D}, vol.~62, no.~8, p.~084003, 2000.

\bibitem{031}
V.~Perlick, ``Theoretical gravitational lensing--beyond the weak-field
  small-angle approximation,'' in {\em The Eleventh Marcel Grossmann Meeting:
  On Recent Developments in Theoretical and Experimental General Relativity,
  Gravitation and Relativistic Field Theories (In 3 Volumes)}, pp.~680--699,
  World Scientific, 2008.

\bibitem{032}
S.~Frittelli, T.~P. Kling, and E.~T. Newman, ``Spacetime perspective of
  schwarzschild lensing,'' {\em Phys. Rev. D}, vol.~61, no.~6, p.~064021, 2000.

\bibitem{033}
V.~Bozza, S.~Capozziello, G.~Iovane, and G.~Scarpetta, ``Strong field limit of
  black hole gravitational lensing,'' {\em General Relativity and Gravitation},
  vol.~33, pp.~1535--1548, 2001.

\bibitem{035}
N.~Tsukamoto, ``Deflection angle in the strong deflection limit in a general
  asymptotically flat, static, spherically symmetric spacetime,'' {\em Phys.
  Rev. D}, vol.~95, no.~6, p.~064035, 2017.

\bibitem{036}
N.~Tsukamoto, Y.~Gong, {\em et~al.}, ``Retrolensing by a charged black hole,''
  {\em Phys. Rev. D}, vol.~95, no.~6, p.~064034, 2017.

\bibitem{aa2024remarks}
A.~A. Ara{\'u}jo~Filho, ``Remarks on a nonlinear electromagnetic extension in
  ads reissner-nordstr$\backslash$" om spacetime,'' {\em arXiv preprint
  arXiv:2410.23165}, 2024.

\bibitem{036.2}
E.~F. Eiroa, G.~E. Romero, and D.~F. Torres, ``Reissner-nordstr{\"o}m black
  hole lensing,'' {\em Physical Review D}, vol.~66, no.~2, p.~024010, 2002.

\bibitem{aa2024static}
A.~A. Ara{\'u}jo~Filho, ``Static limit analysis of a nonlinear electromagnetic
  generalization of the kerr-newman black hole,'' {\em arXiv preprint
  arXiv:2410.12060}, 2024.

\bibitem{036.1}
E.~F. Eiroa and D.~F. Torres, ``Strong field limit analysis of gravitational
  retrolensing,'' {\em Phys. Rev. D}, vol.~69, no.~6, p.~063004, 2004.

\bibitem{37.5}
V.~Bozza, F.~De~Luca, and G.~Scarpetta, ``Kerr black hole lensing for generic
  observers in the strong deflection limit,'' {\em Phys. Rev. D}, vol.~74,
  no.~6, p.~063001, 2006.

\bibitem{37.2}
S.~E. Vazquez and E.~P. Esteban, ``Strong field gravitational lensing by a kerr
  black hole,'' {\em arXiv preprint gr-qc/0308023}, 2003.

\bibitem{37.4}
A.~B. Aazami, C.~R. Keeton, and A.~Petters, ``Lensing by kerr black holes. ii:
  Analytical study of quasi-equatorial lensing observables,'' {\em J. Math.
  Phys.}, vol.~52, no.~10, 2011.

\bibitem{37.3}
V.~Bozza, ``Quasiequatorial gravitational lensing by spinning black holes in
  the strong field limit,'' {\em Physical Review D}, vol.~67, no.~10,
  p.~103006, 2003.

\bibitem{37.6}
V.~Bozza and G.~Scarpetta, ``Strong deflection limit of black hole
  gravitational lensing with arbitrary source distances,'' {\em Phys. Rev. D},
  vol.~76, no.~8, p.~083008, 2007.

\bibitem{37.1}
V.~Bozza, F.~De~Luca, G.~Scarpetta, and M.~Sereno, ``Analytic kerr black hole
  lensing for equatorial observers in the strong deflection limit,'' {\em Phys.
  Rev. D}, vol.~72, no.~8, p.~083003, 2005.

\bibitem{38.4}
N.~Tsukamoto, ``Retrolensing by a wormhole at deflection angles $\pi$ and 3
  $\pi$,'' {\em Phys. Rev. D}, vol.~95, no.~8, p.~084021, 2017.

\bibitem{38.5}
R.~Shaikh, P.~Banerjee, S.~Paul, and T.~Sarkar, ``Strong gravitational lensing
  by wormholes,'' {\em JCAP}, vol.~2019, no.~07, p.~028, 2019.

\bibitem{38.3}
N.~Tsukamoto, ``Strong deflection limit analysis and gravitational lensing of
  an ellis wormhole,'' {\em Phys. Rev. D}, vol.~94, no.~12, p.~124001, 2016.

\bibitem{38.2}
G.~W. Gibbons and M.~Vyska, ``The application of weierstrass elliptic functions
  to schwarzschild null geodesics,'' {\em Class. Quant. Grav.}, vol.~29, no.~6,
  p.~065016, 2012.

\bibitem{38.1}
N.~Tsukamoto, T.~Harada, and K.~Yajima, ``Can we distinguish between black
  holes and wormholes by their einstein-ring systems?,'' {\em Phys. Rev. D},
  vol.~86, no.~10, p.~104062, 2012.

\bibitem{40}
R.~Shaikh and S.~Kar, ``Gravitational lensing by scalar-tensor wormholes and
  the energy conditions,'' {\em Phys. Rev. D}, vol.~96, no.~4, p.~044037, 2017.

\bibitem{Donmez:2024lfi}
O.~Donmez, ``{Bondi-Hoyle-Lyttleton accretion around the rotating hairy
  Horndeski black hole},'' {\em JCAP}, vol.~09, p.~006, 2024.

\bibitem{Koyuncu:2014nga}
F.~Koyuncu and O.~D\"onmez, ``{Numerical simulation of the disk dynamics around
  the black hole: Bondi Hoyle accretion},'' {\em Mod. Phys. Lett. A}, vol.~29,
  p.~1450115, 2014.

\bibitem{Donmez:2023egk}
O.~Donmez, ``{Perturbing the Stable Accretion Disk in Kerr and 4D
  Einstein\textendash{}Gauss\textendash{}Bonnet Gravities: Comprehensive
  Analysis of Instabilities and Dynamics},'' {\em Res. Astron. Astrophys.},
  vol.~24, no.~8, p.~085001, 2024.

\bibitem{o11}
S.~W. Hawking, ``Black hole explosions?,'' {\em Nature}, vol.~248, no.~5443,
  pp.~30--31, 1974.

\bibitem{o1}
S.~W. Hawking, ``Particle creation by black holes,'' {\em Communications in
  mathematical physics}, vol.~43, no.~3, pp.~199--220, 1975.

\bibitem{o111}
S.~W. Hawking, ``Black holes and thermodynamics,'' {\em Physical Review D},
  vol.~13, no.~2, p.~191, 1976.

\bibitem{eeeOvgun:2019ygw}
A.~\"Ovg\"un and I.~Sakall\i{}, ``{Hawking Radiation via Gauss-Bonnet
  Theorem},'' {\em Annals Phys.}, vol.~413, p.~168071, 2020.

\bibitem{gibbons1977cosmological}
G.~W. Gibbons and S.~W. Hawking, ``Cosmological event horizons, thermodynamics,
  and particle creation,'' {\em Physical Review D}, vol.~15, no.~10, p.~2738,
  1977.

\bibitem{eeeKuang:2018goo}
X.-M. Kuang, B.~Liu, and A.~\"Ovg\"un, ``{Nonlinear electrodynamics AdS black
  hole and related phenomena in the extended thermodynamics},'' {\em Eur. Phys.
  J. C}, vol.~78, no.~10, p.~840, 2018.

\bibitem{eeeKuang:2017sqa}
X.-M. Kuang, J.~Saavedra, and A.~\"Ovg\"un, ``{The Effect of the Gauss-Bonnet
  term to Hawking Radiation from arbitrary dimensional Black Brane},'' {\em
  Eur. Phys. J. C}, vol.~77, no.~9, p.~613, 2017.

\bibitem{eeeOvgun:2015box}
A.~\"Ovg\"un and K.~Jusufi, ``{Massive vector particles tunneling from
  noncommutative charged black holes and their GUP-corrected thermodynamics},''
  {\em Eur. Phys. J. Plus}, vol.~131, no.~5, p.~177, 2016.

\bibitem{eeeOvgun:2019jdo}
A.~\"Ovg\"un, I.~Sakall\i{}, J.~Saavedra, and C.~Leiva, ``{Shadow cast of
  noncommutative black holes in Rastall gravity},'' {\em Mod. Phys. Lett. A},
  vol.~35, no.~20, p.~2050163, 2020.

\bibitem{sedaghatnia2023thermodynamical}
P.~Sedaghatnia, H.~Hassanabadi, J.~Porf{\'\i}rio, W.~Chung, {\em et~al.},
  ``Thermodynamical properties of a deformed schwarzschild black hole via dunkl
  generalization,'' {\em arXiv preprint arXiv:2302.11460}, 2023.

\bibitem{o3}
B.~Harms and Y.~Leblanc, ``Statistical mechanics of black holes,'' {\em
  Physical Review D}, vol.~46, no.~6, p.~2334, 1992.

\bibitem{o9}
A.~Jawad and A.~Khawer, ``Thermodynamic consequences of well-known regular
  black holes under modified first law,'' {\em The European Physical Journal
  C}, vol.~78, pp.~1--10, 2018.

\bibitem{o6}
D.~Hansen, D.~Kubiz{\v{n}}{\'a}k, and R.~B. Mann, ``Criticality and surface
  tension in rotating horizon thermodynamics,'' {\em Classical and Quantum
  Gravity}, vol.~33, no.~16, p.~165005, 2016.

\bibitem{araujo2023analysis}
A.~A. Ara{\'u}jo~Filho, ``Analysis of a regular black hole in verlinde’s
  gravity,'' {\em Classical and Quantum Gravity}, vol.~41, no.~1, p.~015003,
  2023.

\bibitem{o8}
D.~Hansen, D.~Kubiz{\v{n}}{\'a}k, and R.~B. Mann, ``Universality of p- v
  criticality in horizon thermodynamics,'' {\em Journal of High Energy
  Physics}, vol.~2017, no.~1, pp.~1--24, 2017.

\bibitem{aa2024implications}
A.~A. Ara{\'u}jo~Filho, ``Implications of a simpson--visser solution in
  verlinde’s framework,'' {\em The European Physical Journal C}, vol.~84,
  no.~1, pp.~1--22, 2024.

\bibitem{o4}
C.~Vaz, ``Canonical quantization and the statistical entropy of the
  schwarzschild black hole,'' {\em Physical Review D}, vol.~61, no.~6,
  p.~064017, 2000.

\bibitem{o7}
D.~Chen, J.~Tao, {\em et~al.}, ``The modified first laws of thermodynamics of
  anti-de sitter and de sitter space--times,'' {\em Nuclear Physics B},
  vol.~918, pp.~115--128, 2017.

\bibitem{o10}
P.~Kraus and F.~Wilczek, ``Self-interaction correction to black hole
  radiance,'' {\em Nuclear Physics B}, vol.~433, no.~2, pp.~403--420, 1995.

\bibitem{013}
K.~Jusufi, {\.I}.~Sakall{\i}, and A.~{\"O}vg{\"u}n, ``Effect of lorentz
  symmetry breaking on the deflection of light in a cosmic string spacetime,''
  {\em Phys. Rev. D}, vol.~96, no.~2, p.~024040, 2017.

\bibitem{o12}
M.~Parikh, ``A secret tunnel through the horizon,'' {\em International Journal
  of Modern Physics D}, vol.~13, no.~10, pp.~2351--2354, 2004.

\bibitem{anacleto2015quantum}
M.~Anacleto, F.~Brito, and E.~Passos, ``Quantum-corrected self-dual black hole
  entropy in tunneling formalism with gup,'' {\em Physics Letters B}, vol.~749,
  pp.~181--186, 2015.

\bibitem{medved2002radiation}
A.~Medved, ``Radiation via tunneling from a de sitter cosmological horizon,''
  {\em Physical Review D}, vol.~66, no.~12, p.~124009, 2002.

\bibitem{mirekhtiary2024tunneling}
F.~Mirekhtiary, A.~Abbasi, K.~Hosseini, and F.~Tulucu, ``Tunneling of
  rotational black string with nonlinear electromagnetic fields,'' {\em Physica
  Scripta}, vol.~99, no.~3, p.~035005, 2024.

\bibitem{silva2013quantum}
C.~Silva and F.~Brito, ``Quantum tunneling radiation from self-dual black
  holes,'' {\em Physics Letters B}, vol.~725, no.~4-5, pp.~456--462, 2013.

\bibitem{del2024tunneling}
F.~Del~Porro, S.~Liberati, and M.~Schneider, ``Tunneling method for hawking
  quanta in analogue gravity,'' {\em arXiv preprint arXiv:2406.14603}, 2024.

\bibitem{calmet2023quantum}
X.~Calmet, S.~D. Hsu, and M.~Sebastianutti, ``Quantum gravitational corrections
  to particle creation by black holes,'' {\em Physics Letters B}, vol.~841,
  p.~137820, 2023.

\bibitem{johnson2020hawking}
G.~Johnson and J.~March-Russell, ``Hawking radiation of extended objects,''
  {\em Journal of High Energy Physics}, vol.~2020, no.~4, pp.~1--16, 2020.

\bibitem{vanzo2011tunnelling}
L.~Vanzo, G.~Acquaviva, and R.~Di~Criscienzo, ``Tunnelling methods and
  hawking's radiation: achievements and prospects,'' {\em Classical and Quantum
  Gravity}, vol.~28, no.~18, p.~183001, 2011.

\bibitem{mitra2007hawking}
P.~Mitra, ``Hawking temperature from tunnelling formalism,'' {\em Physics
  Letters B}, vol.~648, no.~2-3, pp.~240--242, 2007.

\bibitem{zhang2005new}
J.~Zhang and Z.~Zhao, ``New coordinates for kerr--newman black hole
  radiation,'' {\em Physics Letters B}, vol.~618, no.~1-4, pp.~14--22, 2005.

\bibitem{touati2024quantum}
A.~Touati and Z.~Slimane, ``Quantum tunneling from schwarzschild black hole in
  non-commutative gauge theory of gravity,'' {\em Physics Letters B}, vol.~848,
  p.~138335, 2024.

\bibitem{senjaya2024bocharova}
D.~Senjaya, ``The bocharova--bronnikov--melnikov--bekenstein black hole’s
  exact quasibound states and hawking radiation,'' {\em The European Physical
  Journal C}, vol.~84, no.~6, p.~607, 2024.

\bibitem{n57i}
A.~I. Janis and E.~T. Newman, ``Structure of gravitational sources,'' {\em
  Journal of Mathematical Physics}, vol.~6, no.~6, pp.~902--914, 1965.

\bibitem{n57}
E.~T. Newman and A.~Janis, ``Note on the kerr spinning-particle metric,'' {\em
  Journal of Mathematical Physics}, vol.~6, no.~6, pp.~915--917, 1965.

\bibitem{afrim}
M.~Afrin, S.~G. Ghosh, and A.~Wang, ``Testing egb gravity coupled to bumblebee
  field and black hole parameter estimation with eht observations,'' {\em
  Physics of the Dark Universe}, vol.~46, p.~101642, 2024.

\bibitem{n59}
M.~Azreg-A{\"\i}nou, ``Generating rotating regular black hole solutions without
  complexification,'' {\em Physical Review D}, vol.~90, no.~6, p.~064041, 2014.

\bibitem{n58}
M.~Azreg-A{\"\i}nou, ``From static to rotating to conformal static solutions:
  rotating imperfect fluid wormholes with (out) electric or magnetic field,''
  {\em The European Physical Journal C}, vol.~74, pp.~1--11, 2014.

\bibitem{n63}
S.~G. Ghosh and S.~D. Maharaj, ``Radiating kerr-like regular black hole,'' {\em
  The European Physical Journal C}, vol.~75, pp.~1--9, 2015.

\bibitem{n62}
C.~Bambi and L.~Modesto, ``Rotating regular black holes,'' {\em Physics Letters
  B}, vol.~721, no.~4-5, pp.~329--334, 2013.

\bibitem{n60}
T.~Johannsen and D.~Psaltis, ``Metric for rapidly spinning black holes suitable
  for strong-field tests of the no-hair theorem,'' {\em Physical Review
  D—Particles, Fields, Gravitation, and Cosmology}, vol.~83, no.~12,
  p.~124015, 2011.

\bibitem{n64}
S.~G. Ghosh, ``Rotating black hole and quintessence,'' {\em The European
  Physical Journal C}, vol.~76, no.~4, p.~222, 2016.

\bibitem{n65}
S.~G. Ghosh, ``A nonsingular rotating black hole,'' {\em The European Physical
  Journal C}, vol.~75, no.~11, p.~532, 2015.

\bibitem{n61}
K.~Jusufi, M.~Jamil, H.~Chakrabarty, Q.~Wu, C.~Bambi, and A.~Wang, ``Rotating
  regular black holes in conformal massive gravity,'' {\em Physical Review D},
  vol.~101, no.~4, p.~044035, 2020.

\bibitem{1}
M.~Chaichian, A.~Tureanu, and G.~Zet, ``Corrections to schwarzschild solution
  in noncommutative gauge theory of gravity,'' {\em Physics Letters B},
  vol.~660, no.~5, pp.~573--578, 2008.

\bibitem{2}
G.~Zet, V.~Manta, and S.~Babeti, ``Desitter gauge theory of gravitation,'' {\em
  International Journal of Modern Physics C}, vol.~14, no.~01, pp.~41--48,
  2003.

\bibitem{6}
O.~Bertolami and J.~Paramos, ``Vacuum solutions of a gravity model with
  vector-induced spontaneous lorentz symmetry breaking,'' {\em Phys. Rev. D},
  vol.~72, no.~4, p.~044001, 2005.

\bibitem{4}
A.~H. Chamseddine, ``Deforming einstein's gravity,'' {\em Physics Letters B},
  vol.~504, no.~1-2, pp.~33--37, 2001.

\bibitem{5}
B.~Jurco, S.~Schraml, P.~Schupp, and J.~Wess, ``Enveloping algebra-valued gauge
  transformations for non-abelian gauge groups on non-commutative spaces,''
  {\em The European Physical Journal C-Particles and Fields}, vol.~17, no.~3,
  pp.~521--526, 2000.

\bibitem{heidari2023gravitational}
N.~Heidari, H.~Hassanabadi, A.~A. Ara{\'u}jo~Filho, J.~Kriz, S.~Zare, and P.~J.
  Porf{\'\i}rio, ``Gravitational signatures of a non--commutative stable black
  hole,'' {\em Physics of the Dark Universe}, p.~101382, 2023.

\bibitem{heidari2024quantum}
N.~Heidari, A.~{\"O}vg{\"u}n, {\em et~al.}, ``Quantum gravity effects on
  particle creation and evaporation in a non-commutative black hole via mass
  deformation,'' {\em arXiv preprint arXiv:2409.03566}, 2024.

\bibitem{azreg2014generating}
M.~Azreg-A{\"\i}nou, ``Generating rotating regular black hole solutions without
  complexification,'' {\em Physical Review D}, vol.~90, no.~6, p.~064041, 2014.

\bibitem{kumar2020rotating}
R.~Kumar and S.~G. Ghosh, ``Rotating black holes in 4d einstein-gauss-bonnet
  gravity and its shadow,'' {\em Journal of Cosmology and Astroparticle
  Physics}, vol.~2020, no.~07, p.~053, 2020.

\bibitem{brahma2021testing}
S.~Brahma, C.-Y. Chen, and D.-h. Yeom, ``Testing loop quantum gravity from
  observational consequences of nonsingular rotating black holes,'' {\em
  Physical Review Letters}, vol.~126, no.~18, p.~181301, 2021.

\bibitem{islam2023investigating}
S.~U. Islam, J.~Kumar, R.~K. Walia, and S.~G. Ghosh, ``Investigating loop
  quantum gravity with event horizon telescope observations of the effects of
  rotating black holes,'' {\em The Astrophysical Journal}, vol.~943, no.~1,
  p.~22, 2023.

\bibitem{afrin2022testing}
M.~Afrin and S.~G. Ghosh, ``Testing horndeski gravity from eht observational
  results for rotating black holes,'' {\em The Astrophysical Journal},
  vol.~932, no.~1, p.~51, 2022.

\bibitem{visser2007kerr}
M.~Visser, ``The kerr spacetime: A brief introduction,'' {\em arXiv preprint
  arXiv:0706.0622}, 2007.

\bibitem{grumiller2022black}
D.~Grumiller and M.~M. Sheikh-Jabbari, {\em Black hole physics}.
\newblock Springer, 2022.

\bibitem{christodoulou1971reversible}
D.~Christodoulou and R.~Ruffini, ``Reversible transformations of a charged
  black hole,'' {\em Physical Review D}, vol.~4, no.~12, p.~3552, 1971.

\bibitem{ruiz2019thermodynamic}
O.~Ruiz, U.~Molina, and P.~Viloria, ``Thermodynamic analysis of kerr-newman
  black holes,'' in {\em Journal of Physics: Conference Series}, vol.~1219,
  p.~012016, IOP Publishing, 2019.

\bibitem{wald2010general}
R.~M. Wald, {\em General relativity}.
\newblock University of Chicago press, 2010.

\bibitem{bardeen1973four}
J.~M. Bardeen, B.~Carter, and S.~W. Hawking, ``The four laws of black hole
  mechanics,'' {\em Communications in mathematical physics}, vol.~31,
  pp.~161--170, 1973.

\bibitem{page2005hawking}
D.~N. Page, ``Hawking radiation and black hole thermodynamics,'' {\em New
  Journal of Physics}, vol.~7, no.~1, p.~203, 2005.

\bibitem{carlip2014black}
S.~Carlip, ``Black hole thermodynamics,'' {\em International Journal of Modern
  Physics D}, vol.~23, no.~11, p.~1430023, 2014.

\bibitem{davies1978thermodynamics}
P.~C. Davies, ``Thermodynamics of black holes,'' {\em Reports on Progress in
  Physics}, vol.~41, no.~8, p.~1313, 1978.

\bibitem{hawking1976black}
S.~W. Hawking, ``Black holes and thermodynamics,'' {\em Physical Review D},
  vol.~13, no.~2, p.~191, 1976.

\bibitem{christodoulou1970reversible}
D.~Christodoulou, ``Reversible and irreversible transformations in black-hole
  physics,'' {\em Physical Review Letters}, vol.~25, no.~22, p.~1596, 1970.

\bibitem{1bekenstein2020black}
J.~D. Bekenstein, ``Black holes and the second law,'' in {\em JACOB BEKENSTEIN:
  The Conservative Revolutionary}, pp.~303--306, World Scientific, 2020.

\bibitem{2bekenstein1974generalized}
J.~D. Bekenstein, ``Generalized second law of thermodynamics in black-hole
  physics,'' {\em Physical Review D}, vol.~9, no.~12, p.~3292, 1974.

\bibitem{hrelja2024entropy}
A.~Hrelja, T.~Juri{\'c}, and F.~Po{\v{z}}ar, ``Entropy of black holes, charged
  probes and noncommutative generalization,'' {\em arXiv preprint
  arXiv:2407.13233}, 2024.

\bibitem{ciric2018noncommutative}
M.~D. {\'C}iri{\'c}, N.~Konjik, and A.~Samsarov, ``Noncommutative scalar
  quasinormal modes of the reissner--nordstr{\"o}m black hole,'' {\em Classical
  and quantum gravity}, vol.~35, no.~17, p.~175005, 2018.

\bibitem{ciric2024noncommutative}
M.~D. {\'C}iri{\'c}, T.~Juri{\'c}, N.~Konjik, A.~Samsarov, and I.~Smoli{\'c},
  ``Noncommutative reissner-nordstr$\backslash$" om black hole from
  noncommutative charged scalar field,'' {\em arXiv preprint arXiv:2404.03755},
  2024.

\bibitem{dimitrijevic2020noncommutative}
M.~Dimitrijevi{\'c}~{\'C}iri{\'c}, N.~Konjik, and A.~Samsarov, ``Noncommutative
  scalar field in the nonextremal reissner-nordstr{\"o}m background:
  Quasinormal mode spectrum,'' {\em Physical Review D}, vol.~101, no.~11,
  p.~116009, 2020.

\bibitem{juric2023noncommutative}
T.~Juri{\'c} and F.~Po{\v{z}}ar, ``Noncommutative correction to the entropy of
  charged btz black hole,'' {\em Symmetry}, vol.~15, no.~2, p.~417, 2023.

\bibitem{kerr1963gravitational}
R.~P. Kerr, ``Gravitational field of a spinning mass as an example of
  algebraically special metrics,'' {\em Physical review letters}, vol.~11,
  no.~5, p.~237, 1963.

\bibitem{jiang2006hawking}
Q.-Q. Jiang, S.-Q. Wu, and X.~Cai, ``Hawking radiation as tunneling from the
  kerr and kerr-newman black holes,'' {\em Physical Review D—Particles,
  Fields, Gravitation, and Cosmology}, vol.~73, no.~6, p.~064003, 2006.

\bibitem{zhang2005hawking}
J.~Zhang and Z.~Zhao, ``Hawking radiation via tunneling from kerr black
  holes,'' {\em Modern Physics Letters A}, vol.~20, no.~22, pp.~1673--1681,
  2005.

\bibitem{li2008hawking}
R.~Li, J.-R. Ren, and S.-W. Wei, ``Hawking radiation of dirac particles via
  tunneling from the kerr black hole,'' {\em Classical and Quantum Gravity},
  vol.~25, no.~12, p.~125016, 2008.

\bibitem{agullo2010hawking}
I.~Agullo, J.~Navarro-Salas, G.~J. Olmo, and L.~Parker, ``Hawking radiation by
  kerr black holes and conformal symmetry,'' {\em Physical review letters},
  vol.~105, no.~21, p.~211305, 2010.

\bibitem{murata2006hawking}
K.~Murata and J.~Soda, ``Hawking radiation from rotating black holes and
  gravitational anomalies,'' {\em Physical Review D—Particles, Fields,
  Gravitation, and Cosmology}, vol.~74, no.~4, p.~044018, 2006.

\bibitem{corda2013effective}
C.~Corda, S.~Hendi, R.~Katebi, and N.~Schmidt, ``Effective state, hawking
  radiation and quasi-normal modes for kerr black holes,'' {\em Journal of High
  Energy Physics}, vol.~2013, no.~6, pp.~1--12, 2013.

\bibitem{xu2007hawking}
Z.~Xu and B.~Chen, ``Hawking radiation from general kerr-(anti) de sitter black
  holes,'' {\em Physical Review D—Particles, Fields, Gravitation, and
  Cosmology}, vol.~75, no.~2, p.~024041, 2007.

\bibitem{umetsu2010hawking}
K.~Umetsu, ``Hawking radiation from kerr--newman black hole and tunneling
  mechanism,'' {\em International journal of modern physics A}, vol.~25,
  no.~21, pp.~4123--4140, 2010.

\bibitem{arbey2020evolution}
A.~Arbey, J.~Auffinger, and J.~Silk, ``Evolution of primordial black hole spin
  due to hawking radiation,'' {\em Monthly Notices of the Royal Astronomical
  Society}, vol.~494, no.~1, pp.~1257--1262, 2020.

\bibitem{wang2024entanglement}
L.~Wang and R.~Li, ``Entanglement islands and the page curve of hawking
  radiation for rotating kerr black holes,'' {\em Physical Review D}, vol.~110,
  no.~6, p.~066012, 2024.

\bibitem{mcmaken2024hawking}
T.~McMaken and A.~J. Hamilton, ``Hawking radiation inside a rotating black
  hole,'' {\em Physical Review D}, vol.~109, no.~6, p.~065023, 2024.

\bibitem{senjaya2024kerr}
D.~Senjaya, ``The kerr--bumblebee exact massive and massless scalar quasibound
  states and hawking radiation,'' {\em The European Physical Journal C},
  vol.~84, no.~4, p.~424, 2024.

\bibitem{5vanzzo}
M.~Angheben, M.~Nadalini, L.~Vanzo, and S.~Zerbini, ``Hawking radiation as
  tunneling for extremal and rotating black holes,'' {\em Journal of High
  Energy Physics}, vol.~2005, no.~05, p.~014, 2005.

\bibitem{07femions}
R.~Kerner and R.~B. Mann, ``Tunnelling, temperature, and taub-nut black
  holes,'' {\em Physical Review D—Particles, Fields, Gravitation, and
  Cosmology}, vol.~73, no.~10, p.~104010, 2006.

\bibitem{kerner2008fermions}
R.~Kerner and R.~B. Mann, ``Fermions tunnelling from black holes,'' {\em
  Classical and Quantum Gravity}, vol.~25, no.~9, p.~095014, 2008.

\bibitem{di2008fermion}
R.~Di~Criscienzo and L.~Vanzo, ``Fermion tunneling from dynamical horizons,''
  {\em Europhysics Letters}, vol.~82, no.~6, p.~60001, 2008.

\bibitem{012fermions}
P.~Mitra, ``Hawking temperature from tunnelling formalism,'' {\em Physics
  Letters B}, vol.~648, no.~2-3, pp.~240--242, 2007.

\bibitem{011fermions}
E.~T. Akhmedov, V.~Akhmedova, and D.~Singleton, ``Hawking temperature in the
  tunneling picture,'' {\em Physics Letters B}, vol.~642, no.~1-2,
  pp.~124--128, 2006.

\bibitem{13fermions}
T.~Damour and R.~Ruffini, ``Black-hole evaporation in the
  klein-sauter-heisenberg-euler formalism,'' {\em Physical Review D}, vol.~14,
  no.~2, p.~332, 1976.

\bibitem{14fermions}
S.~Sannan, ``Heuristic derivation of the probability distributions of particles
  emitted by a black hole,'' {\em General Relativity and Gravitation}, vol.~20,
  pp.~239--246, 1988.

\bibitem{Wald}
R.~M. Wald, {\em {General Relativity}}.
\newblock Chicago, USA: Chicago Univ. Pr., 1984.

\bibitem{o100}
A.~Anjum, M.~Afrin, and S.~G. Ghosh, ``Investigating effects of dark matter on
  photon orbits and black hole shadows,'' {\em Physics of the Dark Universe},
  vol.~40, p.~101195, 2023.

\bibitem{o101}
E.~Teo, ``Spherical orbits around a kerr black hole,'' {\em General Relativity
  and Gravitation}, vol.~53, no.~1, p.~10, 2021.

\bibitem{collaboration2019firstL1}
E.~H.~T. Collaboration, K.~Akiyama, A.~Alberdi, W.~Alef, K.~Asada, R.~Azuly,
  {\em et~al.}, ``First m87 event horizon telescope results. i. the shadow of
  the supermassive black hole,'' {\em Astrophys. J. Lett}, vol.~875, no.~1,
  p.~L1, 2019.

\bibitem{akiyama2019firstL6}
K.~Akiyama, A.~Alberdi, W.~Alef, K.~Asada, R.~Azulay, A.-K. Baczko, D.~Ball,
  M.~Balokovi{\'c}, J.~Barrett, D.~Bintley, {\em et~al.}, ``First m87 event
  horizon telescope results. vi. the shadow and mass of the central black
  hole,'' {\em The Astrophysical Journal Letters}, vol.~875, no.~1, p.~L6,
  2019.

\bibitem{AkiyamaL4}
E.~H.~T. Collaboration {\em et~al.}, ``First m87 event horizon telescope
  results. iv. imaging the central supermassive black hole,'' {\em The
  Astrophysical Journal Letters}, vol.~875, no.~1, p.~L4, 2019.

\bibitem{lensingmain}
T.~Hsieh, D.-S. Lee, and C.-Y. Lin, ``Strong gravitational lensing by kerr and
  kerr-newman black holes,'' {\em Physical Review D}, vol.~103, no.~10,
  p.~104063, 2021.

\bibitem{lensing23}
Y.-W. Hsiao, D.-S. Lee, and C.-Y. Lin, ``Equatorial light bending around
  kerr-newman black holes,'' {\em Physical Review D}, vol.~101, no.~6,
  p.~064070, 2020.

\bibitem{lensing22}
S.~V. Iyer and E.~C. Hansen, ``Light’s bending angle in the equatorial plane
  of a kerr black hole,'' {\em Physical Review D—Particles, Fields,
  Gravitation, and Cosmology}, vol.~80, no.~12, p.~124023, 2009.

\bibitem{tsukamoto2017deflection}
N.~Tsukamoto, ``Deflection angle in the strong deflection limit in a general
  asymptotically flat, static, spherically symmetric spacetime,'' {\em Phys.
  Rev. D}, vol.~95, no.~6, p.~064035, 2017.

\bibitem{034}
V.~Bozza, ``Gravitational lensing in the strong field limit,'' {\em Phys. Rev.
  D}, vol.~66, no.~10, p.~103001, 2002.

\bibitem{aa2025does}
A.~A. Ara{\'u}jo~Filho, ``How does non-metricity affect particle creation and
  evaporation in bumblebee gravity?,'' {\em arXiv preprint arXiv:2501.00927},
  2025.

\bibitem{araujo2024particle}
A.~A. Ara{\'u}jo~Filho, ``Particle creation and evaporation in kalb-ramond
  gravity,'' {\em arXiv preprint arXiv:2411.06841}, 2024.

\end{thebibliography}

\end{document}